\documentclass[12pt]{article}   % class instead of style due to graphics
\usepackage{graphicx}  % for graphics
\usepackage{latexsym}  % for \Box

\large
\makeatletter
\@addtoreset{equation}{section}
\makeatother

\newtheorem{Theorem}{Theorem}[section]
\newtheorem{Definition}{Definition}[section]
\newtheorem{Lemma}{Lemma}[section]

\def\be{\begin{equation}}
\def\ee{\end{equation}}
\def\ba{\begin{eqnarray}}
\def\ea{\end{eqnarray}}

\def\a{{\cal A}}
\def\g{{\gamma_0}}

\def\ab{{\overline \a}}

\def\Nl{{\mathchoice
{\setbox0=\hbox{$\displaystyle\rm N$}\hbox{\hbox to0pt
{\kern0.4\wd0\vrule height0.9\ht0\hss}\box0}}
{\setbox0=\hbox{$\textstyle\rm N$}\hbox{\hbox to0pt
{\kern0.4\wd0\vrule height0.9\ht0\hss}\box0}}
{\setbox0=\hbox{$\scriptstyle\rm N$}\hbox{\hbox to0pt
{\kern0.4\wd0\vrule height0.9\ht0\hss}\box0}}
{\setbox0=\hbox{$\scriptscriptstyle\rm N$}\hbox{\hbox to0pt
{\kern0.4\wd0\vrule height0.9\ht0\hss}\box0}}}}

\def\Zl{{\mathchoice
{\setbox0=\hbox{$\displaystyle\rm Z$}\hbox{\hbox to0pt
{\kern0.4\wd0\vrule height0.9\ht0\hss}\box0}}
{\setbox0=\hbox{$\textstyle\rm Z$}\hbox{\hbox to0pt
{\kern0.4\wd0\vrule height0.9\ht0\hss}\box0}}
{\setbox0=\hbox{$\scriptstyle\rm Z$}\hbox{\hbox to0pt
{\kern0.4\wd0\vrule height0.9\ht0\hss}\box0}}
{\setbox0=\hbox{$\scriptscriptstyle\rm Z$}\hbox{\hbox to0pt
{\kern0.4\wd0\vrule height0.9\ht0\hss}\box0}}}}

\def\Ql{{\mathchoice
{\setbox0=\hbox{$\displaystyle\rm Q$}\hbox{\hbox to0pt
{\kern0.4\wd0\vrule height0.9\ht0\hss}\box0}}
{\setbox0=\hbox{$\textstyle\rm Q$}\hbox{\hbox to0pt
{\kern0.4\wd0\vrule height0.9\ht0\hss}\box0}}
{\setbox0=\hbox{$\scriptstyle\rm Q$}\hbox{\hbox to0pt
{\kern0.4\wd0\vrule height0.9\ht0\hss}\box0}}
{\setbox0=\hbox{$\scriptscriptstyle\rm Q$}\hbox{\hbox to0pt
{\kern0.4\wd0\vrule height0.9\ht0\hss}\box0}}}}

\def\Rl{{\mathchoice
{\setbox0=\hbox{$\displaystyle\rm R$}\hbox{\hbox to0pt
{\kern0.4\wd0\vrule height0.9\ht0\hss}\box0}}
{\setbox0=\hbox{$\textstyle\rm R$}\hbox{\hbox to0pt
{\kern0.4\wd0\vrule height0.9\ht0\hss}\box0}}
{\setbox0=\hbox{$\scriptstyle\rm R$}\hbox{\hbox to0pt
{\kern0.4\wd0\vrule height0.9\ht0\hss}\box0}}
{\setbox0=\hbox{$\scriptscriptstyle\rm R$}\hbox{\hbox to0pt
{\kern0.4\wd0\vrule height0.9\ht0\hss}\box0}}}}

\def\Co{{\mathchoice
{\setbox0=\hbox{$\displaystyle\rm C$}\hbox{\hbox to0pt
{\kern0.4\wd0\vrule height0.9\ht0\hss}\box0}}
{\setbox0=\hbox{$\textstyle\rm C$}\hbox{\hbox to0pt
{\kern0.4\wd0\vrule height0.9\ht0\hss}\box0}}
{\setbox0=\hbox{$\scriptstyle\rm C$}\hbox{\hbox to0pt
{\kern0.4\wd0\vrule height0.9\ht0\hss}\box0}}
{\setbox0=\hbox{$\scriptscriptstyle\rm C$}\hbox{\hbox to0pt
{\kern0.4\wd0\vrule height0.9\ht0\hss}\box0}}}}

\def\Hl{{\mathchoice
{\setbox0=\hbox{$\displaystyle\rm H$}\hbox{\hbox to0pt
{\kern0.4\wd0\vrule height0.9\ht0\hss}\box0}}
{\setbox0=\hbox{$\textstyle\rm H$}\hbox{\hbox to0pt
{\kern0.4\wd0\vrule height0.9\ht0\hss}\box0}}
{\setbox0=\hbox{$\scriptstyle\rm H$}\hbox{\hbox to0pt
{\kern0.4\wd0\vrule height0.9\ht0\hss}\box0}}
{\setbox0=\hbox{$\scriptscriptstyle\rm H$}\hbox{\hbox to0pt
{\kern0.4\wd0\vrule height0.9\ht0\hss}\box0}}}}

\def\Ol{{\mathchoice
{\setbox0=\hbox{$\displaystyle\rm O$}\hbox{\hbox to0pt
{\kern0.4\wd0\vrule height0.9\ht0\hss}\box0}}
{\setbox0=\hbox{$\textstyle\rm O$}\hbox{\hbox to0pt
{\kern0.4\wd0\vrule height0.9\ht0\hss}\box0}}
{\setbox0=\hbox{$\scriptstyle\rm O$}\hbox{\hbox to0pt
{\kern0.4\wd0\vrule height0.9\ht0\hss}\box0}}
{\setbox0=\hbox{$\scriptscriptstyle\rm O$}\hbox{\hbox to0pt
{\kern0.4\wd0\vrule height0.9\ht0\hss}\box0}}}}

\def\ch{\cosh}
\def\sh{\sinh}
\def\ach{\mbox{arcosh}}
\def\acs{\arccos}
\def\sach{\mbox{{\tiny arcosh}}}

\title{Gauge Field Theory Coherent States (GCS) : II.\\
Peakedness Properties}
\author{T. Thiemann\thanks{thiemann@aei-potsdam.mpg.de}, 
O. Winkler\thanks{winkler@aei-potsdam.mpg.de} \\
MPI f. Gravitationsphysik, Albert-Einstein-Institut,\\ 
Am M\"uhlenberg 1, 14476 Golm near Potsdam, Germany}
\date{{\small Preprint AEI-2000-028}} %\\
\begin{document}

\maketitle

\begin{abstract}
In this article we apply the methods outlined in the previous paper of this
series to the particular set of states obtained by 
choosing the complexifier to be a Laplace operator for each edge 
of a graph. The corresponding coherent state transform was
introduced by Hall for one edge and  
generalized by Ashtekar, Lewandowski, Marolf, Mour\~ao and Thiemann to 
arbitrary, finite, piecewise analytic graphs. 

However, both of these works were incomplete with respect to the 
following two issues :\\
(a) The focus was on the unitarity of the transform and left
the properties of the corresponding coherent states themselves untouched.\\
(b) While these states depend in some sense on complexified connections, 
it remained 
unclear what the complexification was in terms of the coordinates of the 
underlying real phase space.

In this paper we complement these results : First, we explicitly
derive the complexification of the configuration space underlying these 
heat kernel 
coherent states and, secondly, prove that this family of states satisfies 
all the usual properties :\\
i) Peakedness in the configuration, momentum and phase space 
(or Bargmann-Segal) representation.\\
ii) Saturation of the unquenched Heisenberg uncertainty bound.\\
iii) (Over)completeness.

These states therefore comprise a candidate family for the semi-classical
analysis of canonical quantum gravity and quantum gauge theory coupled to
quantum gravity. They also enable
error-controlled approximations to difficult analytical
calculations and therefore set a new starting point for {\it numerical 
canonical quantum general relativity and gauge theory}. 

The text is supplemented by an appendix which contains extensive graphics 
in order to give a feeling for the so far unknown peakedness properties of 
the states constructed.
\end{abstract}

\section{Introduction}
\label{s0}

Quantum General Relativity (QGR) has matured over the past decade to a 
mathematically well-defined theory of quantum gravity. 
In contrast to string theory, by definition QGR is a
manifestly background independent, diffeomorphism 
invariant and non-perturbative theory.
The obvious advantage is that one will never have to postulate the
existence of a non-perturbative extension of the theory,
which in string theory has been called the still unknown 
M(ystery)-Theory.

The disadvantage of a non-perturbative and background independent
formulation is, of course, that one is faced with new and interesting 
mathematical problems so that one cannot just go ahead and 
``start calculating scattering amplitudes'': 
As there is no background around which one could perturb, rather the full 
metric is fluctuating, one is not
doing quantum field theory on a spacetime but only on a differential
manifold. Once there is no (Minkowski) metric at our disposal, one loses
familiar notions such as causality structure, locality, Poincar\'e group 
and so forth, in other words, the theory is not a theory to which
the Wightman axioms apply. Therefore, one must build an entirely
new mathematical apparatus to treat the resulting quantum field theory 
which is {\it drastically different from the Fock space picture 
to which particle physicists are used to}.

As a consequence, the mathematical formulation of the theory was the main 
focus of research in the field over the past decade. The main 
achievements to date are the following (more or less in chronological 
order) : 
\begin{itemize}
\item[i)] {\it Kinematical Framework}\\
The starting 
point was the introduction of new field variables \cite{1} for the 
gravitational field which are better suited to a background  
independent formulation of the quantum theory than the ones employed
until that time. In its original version these variables were
complex valued, however, currently their real valued version,  
considered first in \cite{1a} for {\it classical} Euclidean gravity and 
later in \cite{1b} for {\it classical} Lorentzian gravity, is preferred 
because 
to date it seems that it is only with these variables that one can rigorously
define the kinematics and dynamics of Euclidean or Lorentzian  {\it quantum} 
gravity \cite{1c}. \\
These variables are coordinates for the infinite dimensional phase
space of an $SU(2)$ gauge theory subject to further constraints 
besides the Gauss law, that is, a connection and a canonically
conjugate electric field. As such, it is very natural to introduce
smeared functions of these variables, specifically Wilson loop and 
electric flux functions. (Notice that one does not need a metric 
to define these functions, that is, they are background independent).
This had been done for ordinary gauge fields already before in \cite{2} 
and was then reconsidered for gravity (see e.g. \cite{3}).\\
The next step was the choice of a representation of the canonical
commutation relations between the electric and magnetic degrees
of freedom. This involves the choice of a suitable space of 
distributional connections \cite{4} and a faithful measure thereon \cite{5}
which, as one can show \cite{6}, is $\sigma$-additive.
The proof that the resulting Hilbert space indeed solves the adjointness 
relations induced by the reality structure of the classical theory
as well as the canonical commutation relations induced by the symplectic 
structure of the classical theory can be found in \cite{7}.
Independently, a second representation of the canonical commutation
relations, called the loop representation, 
had been advocated (see e.g. \cite{8} and especially \cite{8a} and 
references therein)
but both representations were shown to be unitarily equivalent in
\cite{9} (see also \cite{10} for a different method of proof).\\
This is then the first major achievement : The theory is based on
a rigorously defined kinematical framework.
\item[ii)] {\it Geometrical Operators}\\
The second major achievement concerns the spectra of positive 
semi-definite, self-adjoint geometrical
operators measuring lengths \cite{11}, areas \cite{12,13}
and volumes \cite{12,14,15,16,8} of curves, surfaces and regions
in spacetime. These spectra are pure point (discete) and imply a discrete
Planck scale structure. It should be pointed out that the discreteness
is, in contrast to other approaches to quantum gravity, not put in
by hand but it is a {\it prediction} !
\item[iii)] {\it Regularization- and Renormalization Techniques}\\
The third major achievement is that there is a new 
regularization and renormalization technique \cite{17,18}
for diffeomorphism covariant, density-one-valued operators at our disposal
which was successfully tested in model theories \cite{19}. This
technique can be applied, in particular, to the standard model
coupled to gravity \cite{20,21} and to the Poincar\'e generators at 
spatial infinity \cite{22}. In particular, it works for {\it Lorentzian}
gravity while all earlier proposals could at best work in the Euclidean 
context only (see, e.g. \cite{8a} and references therein). 
The algebra of important operators of the
resulting quantum field theories was shown to be consistent \cite{23}. 
Most surprisingly, these operators are {\it UV and IR finite} !
Notice that, at least as far as these operators are concerned, this 
result is stronger 
than the believed but unproved finiteness of scattering amplitudes
order by order in perturbation theory of the five critical
string theories, in a sense we claim that the perturbation series converges.
The absence of the divergences that usually plague interacting quantum fields
propagating on a Minkowski background can be understood intuitively
from the diffeomorphism invariance of the theory : ``short and long distances
are gauge equivalent''. We will elaborate more on this point in future 
publications. 
\item[iv)] {\it Spin Foam Models}\\
After the construction of the densely defined Hamiltonian constraint
operator of \cite{17,18}, a formal, Euclidean functional integral was
constructed in \cite{23a} and gave rise to the so-called spin foam 
models   
(a spin foam is a history of a graph with faces as the history of edges)
\cite{23b}. Spin foam models are in close connection with causal
spin-network evolutions \cite{23c}, state sum models \cite{23d} and
topological quantum field theory, in particular BF theory \cite{23e}. To
date most results are at a formal level and for the Euclidean version of the
theory only but the programme is exciting since it may restore manifest
four-dimensional diffeomorphism invariance which in the Hamiltonian
formulation is somewhat hidden.
\item[v)]
Finally, the fifth major achievement is the existence of a rigorous and 
satisfactory framework \cite{24,25,26,27,28,29,30} for the quantum 
statistical description of black holes
which reproduces the Bekenstein-Hawking Entropy-Area relation and applies,
in particular, to physical Schwarzschild black holes while stringy black 
holes so far are under control only for extremal charged black holes.
\end{itemize}
Summarizing, the work of the past decade has now 
culminated in a promising starting point for a quantum theory of the 
gravitational field plus matter and the stage is set to pose and answer 
physical questions. 

The most basic and most important question that one should ask is :
{\it Does the theory have classical general relativity as its classical
limit ?} Notice that even if the answer is negative, the existence
of a consistent, interacting, diffeomorphism invariant quantum field theory 
in four dimensions is already a quite non-trivial result. However, we can 
claim to have a satisfactory quantum theory of Einstein's theory
only if the answer is positive. 

It seems that the most natural framework for deriving the classical limit
of a theory is based on coherent states or best approximation states.
Coherent states have a long history and an extensive literature exists
in a vast range of applications (see e.g. \cite{30a,30b}
and references therein). It has been pointed out by many 
(see e.g. \cite{31}) that they are 
best suited for the analysis of the semi-classical behaviour of any
given system because, among other things, in contrast to the WKB-methods 
more familiar to physicists they avoid the discussion of the critical turning 
points and it is much more natural to ask questions which address regions
in the classical phase space rather than in configuration and momentum space
only.

Surprisingly, the vast majority of coherent states have been constructed 
for systems with only a finite number of degrees of freedom. This is 
astonishing because in the course of constructions of (interacting) 
quantum field theories from given classical ones one is almost always 
forced to regularize and renormalize the operators in that theory and these
are operations which have no classical counterpart. Thus, it would be no 
surprise if it turned out that the classical limit of such quantum field 
theories is {\it not} the classical field theory that one started from.
Just to give an example, even if one could rigorously show that the 
continuum limit of lattice QCD exists, to the best of the knowledge 
of the authors it is at present unclear whether the classical limit of that
continuum quantum field theory would give us back classical $SU(3)$
Yang-Mills theory coupled to quarks.

This paper is the second one in a series of papers \cite{32,33,34,35,36,37}
entitled ``Gauge Field Theory Coherent States'' which are geared at shedding
light at these questions. Specifically, we are interested in the question 
whether the non-perturbative
quantization of continuum Lorentzian general relativity in four dimensions 
with and without matter advertized in \cite{17,18,20} has the correct 
classical limit. In fact we eliminate the
criticism stated in \cite{31a} and show in \cite{31b} that
quantum general relativity as presently formulated {\it does} admit 
graviton states which would then presumably also 
enable us to make contact with results from perturbation theory. 

The general outline of our programme was given in 
\cite{32} where a huge family of coherent states, based on the phase space
complexifier method \cite{38}, was introduced. 
Here we specialize to the ``heat kernel family'' of coherent states
that results by choosing the square of electric flux variables as the 
complexifier which, upon quantization, becomes a Laplacian. 
This choice is motivated, on the one hand by the beautiful analysis of Hall
\cite{39a,39b} who established a unitary transfomation between 
square integrable functions on a compact gauge group with respect to the
Haar measure and square integrable, holomorphic functions on the 
complexified group with respect to the so-called heat kernel measure.
On the other hand, it is convenient since an application of this framework 
to diffeomorphism invariant gauge theories hs already been started in
in \cite{40}. 

The original purpose of \cite{40}
was to solve the reality conditions of quantum general relativity written
in terms of the complex valued Ashtekar connection and therefore the 
properties of the states that came with that heat kernel transform 
remained untouched. Moreover, the heat kernel transform of \cite{40}
obviously complexifies the real connection but it remained unclear
how that complex valued connection is expressed in terms of the coordinates
of the real phase space. Without that knowledge there is obviously
no interpretation of that complex valued connection possible.
In this paper we will fill both of these gaps. Namely, using the 
classical framework of \cite{41} and the complexifier method of 
\cite{38} we explicitly construct the complex connection out of 
the real phase space variables. Secondly, we analyze in detail the 
semi-classical properties of the coherent states so obtained, most
importantly their peakedness properties. 

This we do in great detail for the compact gauge groups of rank one,
that is, $U(1)$ and $SU(2)$, and sketch how the proofs extend to
compact groups of higher rank. Details will appear in 
the forthcoming paper \cite{42}. Coherent states for  
Higgs fields are completely analogous to the coherent states constructed 
here because one can describe them by so-called ``point-holonomies"
\cite{21} which are a special case of the holonomies considered here.
Details and coherent states for fermions are treated in \cite{35}. 

As it will become obvious, the states constructed in this paper can serve as 
a tool to perform error-controlled rigorous approximations in
quantum general relativity and quantum gauge theory coupled to 
quantum gravity and therefore as a starting point for 
{\it numerical canonical quantum general relativity and 
numerical canonical quantum gauge theory coupled to quantum gravity}.\\ 
\\
The present article is organized as follows : \\
\\
Section two is an account of the relevant notions and techniques of
non-perturbative classical and quantum general relativity.

Section three explicitly derives the particular complexification of
the real phase space of gauge theories or real general relativity 
based on heat kernel generators as complexifiers. This
section depends on the recently constructed theory of symplectic manifolds 
of quantum general relativity 
and quantum gauge theory labelled by graphs \cite{41}.
 
Section four introduces the heat kernel family of gauge-non-invariant states 
for a general gauge theory
without fermions in any spacetime dimension and we prove that they satisfy
all the properties that one is used to from the classical harmonic 
oscillator coherent states. That is, these states are labelled by a classical
connection and a classical electric field (a point in phase space) and
we show that these states are peaked on these values in the connection-,
momentum- and Segal-Bargmann representation. Furthermore, we show that 
the system of states is overcomplete, saturates the {\it unquenched} 
Heisenberg 
uncertainty bound with respect to certain complexified holonomy operators
and that each state labelled by a point in phase space can be associated
with a phase space cell with a volume whose size is controlled 
by $\hbar^d$. We do all this for the gauge group $SU(2)$ and point out 
how to generalize to an arbitrary compact gauge group. 

In section five the analysis of section four 3 is generalized 
to the gauge invariant heat kernel family. The proofs follow essentially 
from the proofs derived in section four by employing the group averaging
method of refined algebraic quantization (RAQ) \cite{7}. However,
the results stated in section five are somewhat less complete than those for
section four due to the difficulty to do the group averaging explicitly
which makes it hard to establish sharp peakedness. Fortunately, the results 
of section four are completely sufficient in order to study the semi-classical
behaviour of the theory.

Finally in Appendix A we repeat our analysis for the technically much 
simpler case of $G=U(1)$ and 
in Appendix B we display the peakedness properties of the states 
constructed in the configuration and Bargmann-Segal representation
graphically, both for $SU(2)$ and $U(1)$. 
All graphics have been obtained by means of Mathematica and the admittedly
large amount of plots is justified by the fact that, to the best of our 
knowledge, the behaviour of these states has not been studied 
numerically before.

\section{Kinematical Structure of Diffeomorphism Invariant Quantum
Gauge Theories}
\label{s1}

In this section we will recall the main ingredients of the mathematical
formulation of (Lorentzian) diffeomorphism invariant classical and quantum 
field theories of 
connections with local degrees of freedom in any dimension and for
any compact gauge group. See 
\cite{41,7} and references therein for more details.
Also, in this section we will take all quantities to be dimensionless 
for simplicity, the incoporation of dimensionful parameters will be 
discussed in the next section.

\subsection{Classical Theory}
\label{s1.1}

Let $G$ be a compact gauge group, $\Sigma$ a $D-$dimensional manifold 
admitting a principal $G-$bundle with connection over $\Sigma$.
Let us denote the pull-back to $\Sigma$ of the connection 
by local sections by $A_a^i$
where $a,b,c,..=1,..,D$ denote tensorial indices and $i,j,k,..=1,..,
\dim(G)$ denote indices for the Lie algebra of $G$. 
Likewise, consider a vector bundle of electric fields, whose
projection to $\Sigma$ is a Lie algebra valued 
vector density of weight one. We will denote the set of generators
of the rank $N-1$ Lie algebra of $G$ by $\tau_i$ which are normalized
according to $\mbox{tr}(\tau_i\tau_j)=-N\delta_{ij}$ and 
$[\tau_i,\tau_j]=2f_{ij}\;^k\tau_k$ defines the structure constants 
of $Lie(G)$. 

Let $F^a_i$ be a Lie algebra valued vector density test field of weight one 
and let $f_a^i$ be a Lie algebra valued covector test field. 
We consider the smeared quantities
\be \label{1.1}
F(A):=\int_\Sigma d^Dx F^a_i A_a^i\mbox{ and } 
E(f):=\int_\Sigma d^Dx E^a_i f_a^i 
\ee
While both are diffeomorphism covariant, it is only the latter which is 
gauge covariant, one reason to consider the singular smearings discussed 
below. The choice of the space of pairs of test fields $(F,f)\in{\cal S}$ 
depends on the boundary conditions on
the space of connections and electric fields which in turn depends on the 
topology of $\Sigma$ and will not be specified in what follows. 

Consider the set $M$
of all pairs of smooth functions $(A,E)$ on $\Sigma$ such that 
(\ref{1.1}) is 
well defined for any $(F,f)\in {\cal S}$. 
We define a topology on $M$ through the following globally defined metric :
\ba \label{1.2}
&& d_{\rho,\sigma}[(A,E),(A',E')] \\
&:=& \sqrt{-\frac{1}{N}\int_\Sigma d^Dx 
[\sqrt{\det(\rho)} \rho^{ab} \mbox{tr}([A_a-A'_a][A_b-A'_b])+
\frac{[\sigma_{ab} \mbox{tr}([E^a-E^{a\prime}][E^b-E^{b\prime}])}
{\sqrt{\det(\sigma)}}]} \nonumber
\ea
where $\rho_{ab},\sigma_{ab}$ are fiducial metrics on $\Sigma$ of 
everywhere Euclidean signature. Their fall-off behaviour has to be suited
to the boundary conditions of the fields $A,E$ at spatial infinity.
Notice that the metric (\ref{1.2}) on $M$ is gauge invariant. It can be 
used   in the usual way to equip $M$ with the structure of a smooth,
infinite dimensional differential
manifold modelled on a Banach (in fact Hilbert) space $\cal E$
where ${\cal S}\times {\cal S}\subset {\cal E}$. (It is the 
weighted Sobolev space $H_{0,\rho}^2\times H_{0,\sigma^{-1}}^2$ in the 
notation of \cite{43}). 

Finally, we equip $M$ with the structure of an infinite dimensional 
symplectic manifold through the following strong (in the sense of 
\cite{44}) symplectic structure 
\be \label{1.3}
\Omega((f,F),(f',F'))_m:=\int_\Sigma d^Dx [F^a_i f^{i\prime}_a
-F^{a\prime}_i f_a^i](x)
\ee
for any $(f,F),(f',F')\in {\cal E}$. We have abused the notation by 
identifying the tangent space to $M$ at $m$ with $\cal E$. To prove 
that $\Omega$ is a strong symplectic structure one uses standard 
Banach space techniques. Computing the Hamiltonian vector fields
(with respect to $\Omega$) of the functions $E(f),F(A)$ we obtain the
following elementary Poisson brackets
\be \label{1.4}
\{E(f),E(f')\}=\{F(A),F'(A)\}=0,\;\{E(f),A(F)\}=F(f)
\ee
As a first step towards quantization of the symplectic manifold
$(M,\Omega)$ one must choose a polarization. As usual in gauge theories,
we will use connections as the configuration variables and electric fields 
as canonically conjugate momenta. As a second step one must decide
on a complete set of coordinates of $M$ which are to become the elementary
quantum operators. The analysis just outlined suggests to use the 
coordinates $E(f),F(A)$. However, the well-known immediate problem is that 
these coordinates are not gauge covariant. Thus, we proceed as follows :

Let $\Gamma^\omega_0$ be the set
of all piecewise analytic, finite, oriented graphs $\gamma$ embedded into 
$\Sigma$ 
and denote by $E(\gamma)$ and $V(\gamma)$ respectively its sets of oriented
edges $e$ and vertices $v$ respectively. Here finite means that 
$E(\gamma)$ is a finite set. (One can extend the framework to 
$\Gamma^\infty_0$, the restriction to webs of the set of
piecewise smooth graphs \cite{45,46} but the description becomes more 
complicated and we refrain from doing this here). 
It is possible to consider the set $\Gamma^\omega_\sigma$ of piecewise 
analytic, infinite graphs with
an additional regularity property \cite{34} but for the purpose of this
paper it will be sufficient to stick to $\Gamma^\omega_0$. The subscript
$_0$ as usual denotes ``of compact support'' while $_\sigma$ denotes
``$\sigma$-finite''.

We denote by $h_e(A)$ the holonomy
of $A$ along $e$ and say that a function $f$ on $\a$ is cylindrical with 
respect to $\gamma$ if there exists a function $f_\gamma$ on 
$G^{|E(\gamma)|}$ such that $f=p_\gamma^\ast f_\gamma=f_\gamma\circ 
p_\gamma$ 
where $p_\gamma(A)=\{h_e(A)\}_{e\in E(\gamma)}$. 
Holonomies are invariant under
reparameterizations of the edge and in this article we assume that
the edges are always analyticity preserving diffeomorphic images from 
$[0,1]$ to a
one-dimensional submanifold of $\Sigma$. Gauge transformations are functions
$g:\;\Sigma\mapsto G;\;x\mapsto g(x)$ and they act on
holonomies as $h_e\mapsto g(e(0))h_e g(e(1))^{-1}$. 

Next, given a graph $\gamma$ we choose a polyhedronal decomposition
$P_\gamma$ of $\Sigma$ dual to $\gamma$. The precise definition
of a dual polyhedronal decomposition can be found in \cite{41} but
for the purposes of the present paper it is sufficient to know that
$P_\gamma$ assigns to each edge $e$ of $\gamma$ an open ``face''
$S_e$ (a polyhedron of codimension one embedded into $\Sigma$) with 
the following properties :\\ 
(1) the surfaces $S_e$ are mutually non-intersecting,\\ 
(2) only the edge $e$ intersects $S_e$, the intersection is transversal
and consists only of one point which is an interiour point of both
$e$ and $S_e$,\\
(3) $S_e$ carries the orientation which agrees with the orientation 
of $e$.\\
Furthermore, we choose a system $\Pi_\gamma$ of paths $\rho_e(x) \subset
S_e,\; x\in S_e,\; e\in E(\gamma)$ connecting the intersection point 
$p_e=e\cap S_e$ with $x$. The paths vary smoothly with
$x$ and the triples $(\gamma,P_\gamma,\Pi_\gamma)$
have the property that if $\gamma,\gamma'$ are diffeomorphic, so
are $P_\gamma,P_{\gamma'}$ and $\Pi_\gamma,\Pi_{\gamma'}$.

With these structures we define the following function on $(M,\Omega)$
\be \label{1.5}
P^e_i(A,E):=-\frac{1}{N}
\mbox{tr}(\tau_i h_e(0,1/2)[\int_{S_e} h_{\rho_e(x)} \ast E(x) 
h_{\rho_e(x)}^{-1}] h_e(0,1/2)^{-1})
\ee
where $h_e(s,t)$ denotes the holonomy of $A$ along $e$ between the 
parameter values $s<t$, $\ast$ denotes the Hodge dual, that is,
$\ast E$ is a $(D-1)-$form on $\Sigma$, $E^a:=E^a_i\tau_i$ and
we have chosen a parameterization of $e$ such that $p_e=e(1/2)$.

Notice that in contrast to similar variables used earlier in the literature
the function $P^e_i$ is {\it gauge covariant}. Namely, under gauge 
transformations it transforms as $P^e\mapsto g(e(0)) P^e g(e(0))^{-1}$,
the price to pay being that $P^e$ depends on both $A$ and $E$ and not 
only on $E$. The idea is therefore to use the variables $h_e,P^e_i$
for all possible graphs $\gamma$ as the coordinates of $M$.

The problem with the functions $h_e(A)$ and $P^e_i(A,E)$ on $M$ is that 
they are not differentiable on $M$, that is, $Dh_e, DP^e_i$ are nowhere  
bounded operators on $\cal E$ as one can easily see. The reason for this is,
of course, that these are functions on $M$ which are not properly smeared 
with functions from $\cal S$, rather they are smeared with distributional
test functions with support on $e$ or $S_e$ respectively. Nevertheless
one would like to base the quantization of the theory on these functions 
as basic variables because of their gauge and diffeomorphism covariance.
Indeed, under diffeomorphisms $h_e\mapsto h_{\varphi^{-1}(e)},
P^e_j\mapsto P^{\varphi^{-1}(e)}_j$ where we abuse notation since
$P^e$ depends also explicitly on the $S_e,\rho_e$, see \cite{41} for details.
We proceed as follows. 
\begin{Definition} \label{def1.1}
By $\bar{M}_\gamma$ we denote the direct product 
$[G\times Lie(G)]^{|E(\gamma)|}$. 
The subset of $\bar{M}_\gamma$ of pairs $(h_e(A),P^e_i(A,E))_{e\in 
E(\gamma)}$ as 
$(A,E)$ varies over $M$ will be denoted by $(\bar{M}_\gamma)_{|M}$. We 
have a corresponding map $p_\gamma:\;M\mapsto \bar{M}_\gamma$ which
maps $M$ onto $(\bar{M}_{\gamma})_{|M}$.
\end{Definition}
Notice that the set $(\bar{M}_\gamma)_{|M}$ is in general a proper subset of 
$M_\gamma$,
depending on the boundary conditions on $(A,E)$, the topology of $\Sigma$ 
and the ``size'' of $e,S_e$. For instance, in the limit of $e,S_e\to  
e\cap S_e$ but holding the number of edges fixed, $(\bar{M}_\gamma)_{|M}$  
will consist of only one point in $\bar{M}_\gamma$. This follows from the 
smoothness of the $(A,E)$. 

We equip a subset $M_\gamma$ of $\bar{M}_\gamma$ with the structure of a 
differentiable manifold 
modelled on the Banach space ${\cal E}_\gamma=\Rl^{2\dim(G)|E(\gamma)|}$
by using the natural direct product manifold structure of
$[G\times Lie(G)]^{|E(\gamma)|}$. While $\bar{M}_\gamma$ is a kind of 
distributional
phase space, $M_\gamma$ satisfies appropriate regularity properties similar
to $M$.

In order to proceed and to give $M_\gamma$ a symplectic structure 
{\it derived from $(M,\Omega)$} one must 
regularize the elementary functions $h_e, P^e_i$ by writing them as limits  
(in which the regulator vanishes) of functions which can be expressed  
in terms of the $F(A),E(f)$. Then one can compute their Poisson
brackets with respect to the symplectic structure $\Omega$ at finite
regulator and then take the limit pointwise on $M$. The result is the 
following  
well-defined strong symplectic structure $\Omega_\gamma$ on $M_\gamma$. 
\ba \label{1.6}
\{h_e,h_{e'}\}_\gamma &=& 0\nonumber\\
\{P^e_i,h_{e'}\}_\gamma &=&
\delta^e_{e'} \frac{\tau_i}{2}h_e\nonumber\\
\{P^e_i,P^{e'}_j\}_\gamma &=&
-\delta^{ee'}f_{ij}\;^k P^e_k
\ea
Since $\Omega_\gamma$ is obviously block diagonal, each block standing
for one copy of $G\times Lie(G)$, to check that $\Omega_\gamma$ is 
non-degenerate and closed reduces to doing it for each factor together
with an appeal to well-known Hilbert space techniques to establish that
$\Omega_\gamma$ is a surjection of ${\cal E}_\gamma$.
This is done in \cite{41} where it is shown that each copy is isomorphic
with the cotangent bundle $T^\ast G$ equipped with the symplectic structure
(\ref{1.6}) (choose $e=e'$ and delete the label $e$). \\
\\
Now that we have managed to assign to each graph $\gamma$ a symplectic
manifold $(M_\gamma,\Omega_\gamma)$ we can quantize it by using geometric
quantization. This can be done in a well-defined way because the relations
(\ref{1.6}) show that the corresponding operators are non-distributional.
This is therefore a clean starting point for the regularization of any 
operator
of quantum gauge field theory which can always be written in terms 
of the $\hat{h}_e,\hat{P}^e,\;e\in E(\gamma)$ if we apply this operator to
a function which depends only on the $h_e,\; e\in E(\gamma)$. 

As an example \cite{41}, recall that $(M_\gamma,\Omega_\gamma)$ 
is subject to a coisotropic constraint, the Gauss
constraint, which in terms of the quantities defined above can be written
\be \label{1.7}
G(\Lambda)=\sum_{v\in V(\gamma)}\Lambda^i
[\sum_{e\in E(\gamma),e(0)=v} P^e_i 
-\sum_{e\in E(\gamma),e(1)=v} O_{ij}(h_e) P^e_j]
\ee
where the smooth, Lie-algebra valued function of rapid decrease $\Lambda$ 
is a test function on $\Sigma$ enforcing the local constraint
\be \label{1.8}
G_i(v)=\sum_{e\in E(\gamma),e(0)=v} P^e_i 
-\sum_{e\in E(\gamma),e(1)=v} O_{ij}(h_e) P^e_j
\ee
where $O_{ij}(h)=-\mbox{tr}(h\tau_i h^{-1}\tau_j)/N$.
Since $G(\Lambda)$ is coisotropic, specifically
\be \label{1.9}
\{G(\Lambda),G(\Lambda')\}=-G([\Lambda,\Lambda'])
\ee
the dimension of the physical configuration space 
equals half the dimension of $M_\gamma$ (which is $E\dim(G)$)
minus $V\dim(G)$, the number of constraints.
The question is what $(M_\gamma,\Omega_\gamma)$ has to do with $M,\Omega$.
In \cite{41} it is shown that there exists a partial order $\prec$ on the 
set $\cal L$ of triples $l=(\gamma,P_\gamma,\Pi_\gamma)$. 
In particular, $\gamma\prec\gamma'$ means $\gamma\subset\gamma'$
and $\cal L$ is a directed set so that one can form
a generalized projective limit $M_\infty$ of the $M_\gamma$ (we abuse 
notation in 
displaying the dependence of $M_\gamma$ on $\gamma$ only rather than on
$l$). For this one verifies that the family 
of symplectic structures $\Omega_\gamma$ is self-consistent
in the sense that if 
$(\gamma,P_\gamma,\Pi_\gamma)\prec (\gamma',P_{\gamma'},\Pi_{\gamma'})$ 
then $p_{\gamma'\gamma}^\ast\{f,g\}_\gamma
=\{p_{\gamma'\gamma}^\ast f,p_{\gamma'\gamma}^\ast g\}_{\gamma'}$
for any $f,g\in C^\infty(M_\gamma)$ and 
$p_{\gamma'\gamma}:\;M_{\gamma'}\mapsto M_\gamma$ is a system
of natural projections, more precisely, of (non-invertible) 
symplectomorphisms. 

Now, via the maps $p_\gamma$ of definition \ref{def1.1} we can identify
$M$ with a subset of $M_\infty$. Moreover, in \cite{41} it is shown that
there is a generalized projective sequence $(\gamma_n,P_{\gamma_n},
\Pi_{\gamma_n})$
such that $\lim_{n\to\infty}p_{\gamma_n}^\ast\Omega_{\gamma_n}=\Omega$
pointwise in $M$. This displays $(M,\Omega)$ as embedded into a 
generalized projective
limit of the $(M_\gamma,\Omega_\gamma)$, intuitively speaking, as $\gamma$ 
fills all of $\Sigma$, we recover $(M,\Omega)$ from the 
$(M_\gamma,\Omega_\gamma)$. Of course, this works with $\Gamma^\omega_0$ 
only if $\Sigma$ is compact, otherwise we need the extension to 
$\Gamma^\omega_\sigma$.

It follows that quantization of $(M,\Omega)$, and conversely taking the 
classical limit, can be studied purely in terms of $(M_\gamma,\Omega_\gamma)$
for {\it all} $\gamma$. The quantum kinematical framework for this will be 
given in the next subsection.

\subsection{Quantum Theory}
\label{s1.2}

Let us denote the set of all smooth connections by $\a$. This is our
classical configuration space and we will choose for its coordinates the
holonomies $h_e(A),\;e\in\gamma,\;\gamma\in\Gamma^\omega_0$. 
$\a$ is naturally equipped with a metric topology induced by (\ref{1.2}). 

Recall the notion of a function cylindrical over a graph from the 
previous subsection.
A particularly useful set of cylindrical functions are the so-called 
spin-netwok functions \cite{47,48,9}. A spin-network function is 
labelled by a graph $\gamma$, a set of non-trivial irreducible 
representations 
$\vec{\pi}=\{\pi_e\}_{e\in E(\gamma)}$ (choose from each equivalence 
class of equivalent
representations once and for all a fixed representant), one for each 
edge of $\gamma$, and a set $\vec{c}=\{c_v\}_{v\in V(\gamma)}$ of
contraction matrices, one for each vertex of $\gamma$, which 
contract the indices of the tensor product 
$\otimes_{e\in E(\gamma)} \pi_e(h_e)$ in such a way that the resulting
function is gauge invariant. We denote spin-network functions as
$T_I$ where $I=\{\gamma,\vec{\pi},\vec{c}\}$ is a compound label.
One can show that these functions are linearly independent.
From now on we denote by $\tilde{\Phi}_\gamma$ finite linear combinations of
spin-network functions over $\gamma$, by $\Phi_\gamma$ the finite linear 
combinations of elements from any possible $\tilde{\Phi}_{\gamma'},\;
\gamma'\subset\gamma$ a subgraph of $\gamma$  
and by $\Phi$ the finite linear 
combinations of spin-network functions over an arbitrary collection 
of graphs. Clearly $\tilde{\Phi}_{\gamma}$ is a subspace of 
$\Phi_\gamma$.  To express this distinction we will say that functions 
in $\tilde{\Phi}_\gamma$ are labelled by the ``coloured graphs'' $\gamma$
while functions in $\Phi_\gamma$ are labelled simply by graphs $\gamma$
where we abuse notation by using the same symbol $\gamma$.

The set $\Phi$ of finite linear combinations of spin-network functions 
forms an Abelian $^\ast$ algebra 
of functions on $\a$. By completing it with respect to the sup-norm 
topology it 
becomes an Abelian C$^\ast$ algebra $\cal B$
(here the compactness of $G$ is crucial). The spectrum $\ab$ of this algebra, 
that is, the set of all algebraic homomorphisms ${\cal B}\mapsto\Co$
is called the quantum configuration space. This space is equipped with
the Gel'fand topology, that is, the space of continuous functions
$C^0(\ab)$
on $\ab$ is given by the Gel'fand transforms of elements of $\cal B$.
Recall that the Gel'fand transform is given by $\tilde{f}(\bar{A}):=
\bar{A}(f)\;\forall \bar{A}\in \ab$. It is a general result that $\ab$ with 
this topology is a compact Hausdorff space. Obviously, the elements of
$\a$ are contained in $\ab$ and one can show that $\a$ is even dense
\cite{49}. Generic elements of $\ab$ are, however, distributional.

The idea is now to construct a Hilbert space consisting of square
integrable functions on $\ab$ with respect to some measure $\mu$. Recall 
that one can define a measure on a locally compact Hausdorff space 
by prescribing a positive linear functional $\chi_\mu$ on the space 
of continuous functions thereon. The particular measure
we choose is given by $\chi_{\mu_0}(\tilde{T}_I)=1$ if $I=\{\{p\},
\vec{0},\vec{1}\}$ and $\chi_{\mu_0}(\tilde{T}_I)=0$ otherwise. Here
$p$ is any point in $\Sigma$, $0$ denotes the 
trivial representation and $1$ the trivial contraction matrix. In other 
words, (Gel'fand transforms of) spin-network functions play the same role 
for $\mu_0$ as 
Wick-polynomials do for Gaussian measures and like those they form
an orthonormal basis in the Hilbert space ${\cal H}:=L_2(\ab,d\mu_0)$ 
obtained by completing their finite linear span $\Phi$.\\
An equivalent definition of $\ab,\mu_0$ is as follows :\\ 
$\ab$ is in one to one correspondence, via the surjective map $H$ defined 
below, with the set $\ab':=\mbox{Hom}({\cal X},G)$
of homomorphisms from the groupoid $\cal X$ of composable, holonomically
independent, analytical paths
into the gauge group. The correspondence is explicitly given by
$\ab\ni\bar{A}\mapsto H_{\bar{A}}\in\mbox{Hom}({\cal X},G)$
where ${\cal X}\ni e\mapsto H_{\bar{A}}(e):=\bar{A}(h_e)=
\tilde{h}_e(\bar{A})\in G$ and $\tilde{h}_e$ is the Gel'fand transform
of the function $\a\ni A\mapsto h_e(A)\in G$. Consider now the restriction
of $\cal X$ to ${\cal X}_\gamma$, the groupoid of composable edges of  
the graph $\gamma$. One can then show that the projective limit of the 
corresponding {\it cylindrical sets} 
$\ab'_\gamma:=\mbox{Hom}({\cal X}_\gamma,G)$ coincides with $\ab'$.
Moreover, we have $\{\{H(e)\}_{e\in E(\gamma)};\;H\in\ab'_\gamma\}=
\{\{H_{\bar{A}}(e)\}_{e\in E(\gamma)};\;\bar{A}\in\ab\}=
G^{|E(\gamma)|}$.
Let now $f\in{\cal B}$ be a function cylindrical over $\gamma$ then 
$$
\chi_{\mu_0}(\tilde{f})=\int_{\ab} d\mu_0(\bar{A}) \tilde{f}(\bar{A})
=\int_{G^{|E(\gamma)|}} \otimes_{e\in E(\gamma)} d\mu_H(h_e)
f_\gamma(\{h_e\}_{e\in E(\gamma)})
$$
where $\mu_H$ is the Haar measure on $G$.
As usual, $\a$ turns out to be contained in a measurable subset of 
$\ab$ which has measure zero with respect to $\mu_0$.

Let $\Phi_\gamma$, as before, be the finite linear span of spin-network 
functions
over $\gamma$ and ${\cal H}_\gamma$ its completion with respect to
$\mu_0$. Clearly, $\cal H$ itself is the completion of the finite linear
span $\Phi$ of vectors from the mutually orthogonal $\tilde{\Phi}_\gamma$. 
Our 
basic coordinates of $M_\gamma$ are promoted to operators on ${\cal H}$ with 
dense domain $\Phi$. As $h_e$ is group-valued and $P^e$ is real-valued
we must check that the adjointness relations coming from these reality 
conditions as well as the Poisson brackets (\ref{1.6}) are implemented on
our ${\cal H}$. This turns out to be precisely the case if we choose
$\hat{h}_e$ to be a multiplication operator and $\hat{P}^e_j=i
\hbar\kappa X^e_j/2$
where $X^e_j=X_j(h_e)$ and $X^j(h),\;h\in G$ is the vector field on $G$
generating left translations into the $j-th$ coordinate direction of 
$Lie(G)\equiv T_h(G)$ (the tangent space of $G$ at $h$ can be identified 
with the Lie algebra of $G$) and $\kappa$ is the coupling constant
of the theory. For details see \cite{7,41}.

\section{The Heat Kernel Complexifier}
\label{s2}

The results of this section hold for arbitrary compact, semisimple connected 
gauge groups and direct products of such with Abelian ones. We will be as 
explicit as in \cite{32} in order to make this paper self-contained.\\
\\
As we want to bring in Planck's constant $\hbar$ as a measure of closeness
to classical physics, we need to spend a few moments on dimensionalities, see 
\cite{32} for a general discussion. The dimension of the time coordinate 
$x^0$ 
is taken to be the same as that of the spatial coordinates $x^a$,
namely $[x^0]=[x^a]=$cm$^1$ which can always be achieved by absorbing 
an appropriate power of the speed of light into the coupling constant 
$\kappa$ of the theory.

We will take take our connection one-form to be of dimension $[A]=$cm$^{-1}$
so that its holonomy is dimensionless. In $D+1$ spacetime dimensions
the kinetic term of the classical action is given by
$$
A_{kin}=\frac{1}{\kappa}\int_\Rl dt\int_{\Sigma} d^Dx\; E^a_i(x) 
\dot{A}_a^i(x) $$
and its dimension is that of an action, that is, $[A_{kin}]=[\hbar]$. 
In Yang-Mills theories the electric field is a first derivative of 
$A_a^i$ and thus has dimension $[E^a_i]=$cm$^{-2}$. In general relativity
the metric components, the D-beins and also $[E^a_i]=$cm$^0$ are 
dimensionfree. It follows that in Yang-Mills (YM) theory the Feinstruktur
constant 
\be \label{2.1}
\alpha:=\hbar\kappa 
\ee
has dimension $[\alpha]:=$cm$^{D-3}$ and in general relativity (GR) 
$[\alpha]=$cm$^{D-1}$. 

Let now $\gamma$ be a graph and consider the symplectic manifold
$(M_\gamma,\Omega_\gamma)$ introduced in section \ref{1.1}
with its canonical coordinates $h_e,P^e_i:\;e\in E(\gamma)$. 
The electric flux variable (\ref{1.5}) then 
has dimension $[P^e_i]=$cm$^{D-3}$ in YM and cm$^{D-1}$ in GR respectively
and in general let $[P^e_i]=$cm$^{n'_D}$.
Let now $a$ be an arbitrary but fixed constant with the dimension of a 
length, $[a]=$cm$^1$, say $a=1$cm if $n'_D\not=0$ and let $a$ be 
dimensionfree otherwise. Then we introduce the dimensionfree
quantity
\be \label{2.2}
p^e_i:=\frac{P^e_i}{a^{n_D}} 
\ee
where $n_D=n'_D$ if $n'_D\not=0$ and $n_D=1$ otherwise.
Notice that a natural choice for a dimensionful constant 
in general relativity in any $D$ would
be $a=1/\sqrt{|\Lambda|}$ where $\Lambda$ is the (supposed to be 
non-vanishing) cosmological constant.

On the other hand, it is $E^a_i/\kappa$ which is canonically conjugate 
to $A_a^i$ rather than $E^a_i$ itself, therefore the brackets
(\ref{1.6}) get modified into
\ba \label{2.3}
\{h_e,h_{e'}\}_\gamma &=& 0\nonumber\\
\{\frac{P^e_i}{\kappa},h_{e'}\}_\gamma &=&
\delta^e_{e'} \frac{\tau_i}{2}h_e\nonumber\\
\{\frac{P^e_i}{\kappa},\frac{P^{e'}_j}{\kappa}\}_\gamma &=&
-\delta^{ee'}f_{ij}\;^k \frac{P^e_k}{\kappa}
\ea
We are now ready to define the complexifier for the symplectic manifold 
$M_\gamma$, it is given by
\be \label{2.4}
C_\gamma:=\frac{1}{2\kappa a^{n_D}}\sum_{e\in E(\gamma)}\delta^{ij}
P^e_i P^e_j
\ee
and since $C_\gamma$ is gauge invariant it will pass to the reduced phase 
space. Using the partial order $\prec$ of \cite{41} or section 
\ref{s1.1} it is immediately
clear that $C_\gamma$ defines a self-consistently defined function on
the $M_\gamma$, that is, for $\gamma\prec\gamma'$ we have 
$\{p_{\gamma'\gamma}^\ast C_\gamma, p_{\gamma'\gamma}^\ast 
f_\gamma\}_{\gamma'}=p_{\gamma'\gamma}^\ast \{C_\gamma,f_\gamma\}_\gamma $
for any $f_\gamma\in C^\infty(M_\gamma)$.

We can explicitly compute the complexified holonomy and complexified 
momenta for any compact, semi-simple gauge group $G$. Since 
$\{P^e_i,C_\gamma\}=0$ (gauge invariance of $C_\gamma$) we have 
\ba \label{2.5}
\{h_e,C_\gamma\}_\gamma &=& -P^e_i\frac{\tau_i}{2 a^{n_D}} 
h_e=-p^e_i\frac{\tau_i}{2} h_e 
\nonumber\\ 
\{h_e,C_\gamma\}_{\gamma(2)}&=& \frac{1}{a^{2n_D}}P^e_i P^e_j
\frac{\tau_i\tau_j}{4} h_e
=(-p^e_j\frac{\tau_j}{2})^2 h_e 
\nonumber\\
&& (=-\frac{p_e^2}{4} h_e)
\ea 
where in the last line we have displayed a simplification that results for
$G=SU(2)$ upon using the Clifford 
algebra relation $\tau_i\tau_j=-\delta_{ij}1_G+f_{ij}\;^k\tau_k$ for the 
Pauli matrices and we define generally $p^e:=\sqrt{p^e_j p^e_j}$.
In the second line of (\ref{2.5}) we have made us of the fact that $G$ is semi-simple
so that the structure constants are completely skew and so $\{p^e_j,C_\gamma\}=0$.

We therefore conclude that the complexification of 
$h_e$ is given by (see \cite{38} for full details)
\ba \label{2.6}
h^\Co_e &:=& g_e =\sum_{n=0}^\infty \frac{i^n}{n!}\{h_e,C\}_{(n)}
\nonumber\\
&=& [\sum_{n=0}^\infty \frac{i^n}{n!} (-p^e_j\frac{\tau_j}{2})^n] h_e 
\nonumber\\
&=& e^{-i\tau_j p^e_j/2} h_e=:H_e h_e \nonumber\\
&& (=[\ch(\frac{p^e}{2})1_G-i\frac{p^e_j}{p^e}\tau_j\sh(\frac{p^e}{2})]h_e)
\ea
and similarly $P^{e\Co}_i=P^e_i$
where we follow the notation of \cite{39a} to denote elements of 
$G^\Co$ by $g$ while elements of $G$ are denoted by $h$. In the last line of 
(\ref{2.6}) we have again displayed the formula for the special case of $G=SU(2)$.
Thus we have established the following.
\begin{Lemma} \label{la2.1}~\\
The complexification of the holonomy for compact and semisimple $G$ is given 
directly as a left 
polar decomposition, where the right unitary factor is the holonomy of the 
compact gauge group while the left positive definite hermitean factor is
just the exponential of $-i p^e_j\tau_j/2$.
\end{Lemma}
For $G=U(1)$ the generator $\tau_j/2$ has to be 
replaced by the imaginary unit $i$.

Notice that (\ref{2.6}) makes sense since $p^e_j$ is dimensionless. 
Moreover, we have naturally stumbled on the diffeomorphism 
\cite{39b} 
\be \label{2.7}
\Phi\; :\; T^\ast(G)\mapsto G^\Co;\; (p^j,h)\to g:=Hh=e^{-ip^j\tau_j/2}h\;.
\ee
The diffeomorphism (\ref{2.7}) has a further consequence : 
$(T^\ast(G),\omega)$ is a symplectic manifold while $G^\Co$ is a complex 
manifold. Thus, $T^\ast(G)$ is a symplectic manifold with a complex structure
which, as one can show (\cite{39b,41} and references therein), is 
$\omega$-compatible. In fact, $\omega$ is just given by (\ref{2.3}) 
with $P^e_i$ replaced by $p_i$ and the label $e=e'$ dropped.
Therefore, $T^\ast(G)$ is in fact a K\"ahler 
manifold and a Segal-Bargmann representation (wave functions depending on 
$g$) corresponds to a positive K\"ahler polarization \cite{50}.

Finally, let us compute the Segal-Bargmann transform corresponding to 
$C_\gamma$ (see \cite{38,41} for more details). As follows from the previous 
section, we have in 
the connection representation (wave functions depending on the $h_e$)
\be \label{2.8}
\hat{P}^e_j=\frac{i\hbar\kappa}{2} X^e_j \mbox{ where }
X^e_j=X_j(h_e),
\ee
and $X_j(h)$ denotes the right invariant vector fields on $G$ at $h$,
that is $X_j(h):=\mbox{tr}((\tau_j h)^T\partial/\partial h)$.
Thus, the coherent state transform is (following the notation of 
\cite{38})
\be \label{2.9}
\hat{W}_{\gamma t}:=e^{-\frac{\hat{C}_\gamma}{\hbar}}=
e^{\frac{t}{2}\Delta_\gamma} 
\ee
where we have defined the Laplacian on $\gamma$ by
\be \label{2.10}
\Delta_\gamma=\sum_{e\in E(\gamma)} \Delta_e,\; \Delta_e=\frac{1}{4}
\delta^{ij} X^e_i X^e_j 
\ee
and the heat kernel time parameter has the following interpretation in 
terms of the fundamental constants of the theory
\be \label{2.11}
t:=\frac{\hbar\kappa}{a^{n_D}}\;.
\ee
Notice that $a$ is just a parameter that we have put in by hand to make 
things dimensionless, for instance, it could be $1$cm in quantum general 
relativity in $D+1=4$ spacetime dimensions or $a=10^5$ for 
Yang-Mills in $D+1=4$ and thus is ``large".
The semiclassical limit $\hbar\to 0$ thus corresponds to $t\to 0$.
That $t$ is a tiny positive real number will be crucial in all the estimates 
that we are going to perform in this and the next paper of this series.

The factor of $1/4$ in the definition of $\Delta_e$ relative to 
$(X^e_j)^2$ is due to the factor of $1/2$ in the second Poisson bracket 
of (\ref{2.3}) and it is the same factor which gives $-\Delta_e$ 
the standard spectrum $j(j+1);\;j=0,\frac{1}{2},1,\frac{3}{2},..$ for the 
case of $G=SU(2)$.
 
We can also explicitly compute the quantum operator corresponding to 
$g_e$ in (\ref{2.6}) for arbitrary $G$. We have 
\ba \label{2.12}
\hat{g}_e &=& e^{t\Delta_\gamma/2}\hat{h}_e^{-t\Delta_\gamma/2}
=\sum_{n=0}^\infty\frac{(-t)^n}{2^n n!}[\hat{h}_e,\Delta_e]_{(n)}\nonumber\\
-[\hat{h}_e,\Delta_e] &=& \frac{1}{4}(X^i_e\tau_i\hat{h}_e+\tau_i\hat{h}_e 
X^i_e)=X^i_e\frac{\tau_i}{2}\hat{h}_e-\frac{(\tau_i)^2}{4}\hat{h}_e
\nonumber\\
&& (=(X^i_e\frac{\tau_i}{2}+\frac{3}{4})\hat{h}_e)
\ea
where the last line is the specialization to $G=SU(2)$.
Since $\Delta_\gamma$ commutes with $X^i_e$ we immediately find
\ba \label{2.13}
\hat{g}_e &=&
e^{t\hat{X}^i_e\frac{\tau_i}{4}-t\frac{\tau_i^2}{8}}\hat{h}_e
=e^{-i\hat{p}^j_e\frac{\tau_j}{2}-\frac{t\tau_j^2}{8}}\hat{h}_e
=e^{-i\hat{p}^j_e\frac{\tau_j}{2}} e^{-t\frac{\tau_j^2}{8}}\hat{h}_e
\nonumber\\
&& (=e^{\frac{3t}{8}}e^{-i\hat{p}^j_e\frac{\tau_j}{2}}\hat{h}_e)
\ea
since $itX^j_e/2=\hat{p}^e_j$
and in the third step we used that the matrix $\tau_j^2$ commutes with 
$\tau_i$. The last equality holds for $G=SU(2)$ only. Since the 
$\hat{p}_j$ are not mutually commuting the exponential in (\ref{2.13})
cannot be defined by the spectral theorem, however, we can define 
it through Nelson's analytic vector theorem.
Thus, we find precisely the quantization of 
the polar decomposition (\ref{2.6}) up to a factor of $e^{-\tau_j^2 t/8}$ 
which tends to unity linear in $t\to 0$. as to be expected.
In particular, for $G=U(1)$ we find with $\tau_j/2$ replaced by $i$
\be \label{2.15}
\hat{g}_e
=e^{\hat{p}_e+t/2}\hat{h}_e=e^{t/2} e^{\hat{p}_e}\hat{h}_e
\ee
Notice that one obtains
the first line of (\ref{2.12}) from (\ref{2.6}) if one replaces everywhere
$\{.,.\}$ by $[.,.]/(i\hbar)$ and phase space functions by 
operators which holds, of course, by the very construction of the map 
$\hat{W}_t$ \cite{38}.

\section{Peakedness Proofs for Gauge-Variant Coherent States}
\label{s3}

As outlined in \cite{32} the general form of the above transform guarantees
immediately that the {\it gauge-variant Coherent States}
\be \label{3.1}
\psi^t_{\gamma,\vec{g}}(\vec{h}):=
(\hat{W}_t\delta_{\gamma\mu_\gamma,\vec{h}'}(\vec{h}))_{|\vec{h}'\to\vec{g}}
\ee
obtained by heat kernel evolution followed by analytic continuation,
where $\vec{g}=\{g_e\}_{e\in E(\gamma)}$ and similarly for $\vec{h}$,
satisfy a number of desired properties. Here, for completeness we explicitly
recall that
\ba \label{3.2}
& & \delta_{\gamma\mu_\gamma,\vec{h'}}(\vec{h})=\prod_{e\in E(\gamma)}
\delta_{\mu_H,h'_e}(h_e),\nonumber\\
& & \delta_{\mu_H,h'}(h)=\sum_\pi d_\pi \chi_\pi(h' h^{-1})
\ea
where $d\mu_\gamma(\vec{h})=\otimes_{e\in E(\gamma)} d\mu_H(h_e)$ is simply
the Haar measure on $G^E$, the sum in (\ref{3.2}) runs over all distinct
irreducible representations $\pi$ of $G$ (pick once and for all a fixed 
representant from each equivalence class of those), $d_\pi=\dim(\pi)$
is the dimension of the representation space corresponding to $\pi$ and 
$\chi_\pi(.)=\mbox{tr}(\pi(.))$ is the character of $\pi$ which is a class
function and therefore depends only on the equivalence class of $\pi$.
It follows immediately that therefore the coherent states are explicitly 
given by
\ba \label{3.3}
& & \psi^t_{\gamma,\vec{g}}(\vec{h})=\prod_{e\in E(\gamma)}\psi^t_{g_e}(h_e)
\nonumber\\
& &\psi^t_g(h)=\sum_\pi d_\pi e^{-\frac{t}{2}\lambda_\pi}\chi_\pi(gh^{-1})
\ea
where $-\lambda_\pi\le0,\;=0$ only if $\pi$ is trivial, is the eigenvalue of 
the Laplacian in 
the representation $\pi$. For one copy of $G$, (\ref{3.3}) are precisely 
the states introduced by Hall \cite{39a} who proved various crucial 
functional analytic properties of these states, in particular that they
are entire analytic in $G^\Co$ and that heat kernel evolution is 
densely defined in the Hilbert space $L_2(G,d\mu_H)$. Moreover, he 
proved that the {\it Coherent State Transform} 
\be \label{3.4}
\hat{U}_t\; : \; L_2(G,d\mu_H)\mapsto {\cal H}L_2(G^\Co,d\nu_t);\;
f\mapsto (\hat{U}_t f)(g):=<\overline{\psi^t_g},f>
\ee
is a unitary transformation between two Hilbert spaces where $\nu_t$ is a 
certain measure to be defined later and ${\cal H}L_2$ means a space of 
square integrable holomorphic functions. This, of course, means that
the coherent states so defined satisfy the overcompleteness criterion 
already.

The product structure of 
the coherent states, that is, that the coherent state for a graph is just 
the product over its edges of the coherent states for the edges, is a huge 
simplification which basically will allow us to reduce all the estimates
to just estimates for one copy of $G$.

The properties mentioned above are :
\begin{itemize}
\item[(i)] {\it Eigenstates}\\
The coherent states labelled by $\vec{g}$ are simultanous eigenstates 
for each of the {\it annihilation operators} $\hat{g}_e^{AB},\;A,B=1,..,N$ 
constructed in the previous section. That is
\be \label{3.4a}
\hat{g}_e^{AB}\psi^t_{\gamma,\vec{g}}=g_e^{AB}\psi^t_{\gamma,\vec{g}}
\ee
\item[(ii)] {\it Expectation values}\\
From property (i) it immediately follows that the expectation value
of the sum of products of {\it normally ordered functions}, that is,
the product of any analytic function $f$ of the annihilation
operators $\hat{g}_e^{AB}$ and any
analytic function $f'$ of the {\it creation operators} 
$(\hat{g}_e^{AB})^\dagger$ in the state
$\psi^t_{\gamma,\vec{g}}$ is given by its classical value at $\vec{g},
\overline{\vec{g}}$. That is,
\be \label{3.5}
\frac{<\psi^t_{\gamma,\vec{g}},f'(\vec{\hat{g}}^\dagger) f(\vec{\hat{g}})
\psi^t_{\gamma,\vec{g}}>}{||\psi^t_{\gamma,\vec{g}}||^2} 
=f'(\overline{\vec{g}})f(\vec{g})
\ee
\item[(iii)] {\it Uncertainty bound}\\
The coherent states automatically saturate, {\it with equal weight (they 
are unquenched)}, the uncertainty bound for each pair of self-adjoint 
operators 
\be \label{3.6}
(\hat{x}_e^{AB},\hat{y}_e^{AB}):=
(\frac{1}{2}(\hat{g}_e^{AB}+(\hat{g}_e^{AB})^\dagger),
\frac{1}{2i}(\hat{g}_e^{AB}-(\hat{g}_e^{AB})^\dagger))
\ee
that is 
\be \label{3.7}
\frac{<\psi^t_{\gamma,\vec{g}},(\hat{x}_e^{AB}-x_e^{AB})^2
\psi^t_{\gamma,\vec{g}}>}{||\psi^t_{\gamma,\vec{g}}||^2} 
=\frac{<\psi^t_{\gamma,\vec{g}},(\hat{y}_e^{AB}-y_e^{AB})^2
\psi^t_{\gamma,\vec{g}}>}{||\psi^t_{\gamma,\vec{g}}||^2} 
=\frac{1}{2}
\frac{|<\psi^t_{\gamma,\vec{g}},[\hat{x}_e^{AB},\hat{y}_e^{AB}]
\psi^t_{\gamma,\vec{g}}>|}{||\psi^t_{\gamma,\vec{g}}||^2} 
\ee
where $x,y$ respectively are the expectation values of $\hat{x},\hat{y}$
respectively.
We will compute the actual value of the bound in a later subsection.
\end{itemize}
These properties are satisfied for any set of coherent states defined by
some complexifier $\hat{C}$ which satisfies certain sufficiently strong 
growth conditions on its eigenvalues (labelled by $\pi)$.
The peakedness properties that we are after are harder to prove. We will
do this in the next subsections for the gauge group $G=SU(2)$. The 
generalization to an arbirtrary compact gauge group is straightforward 
but technically difficult and will be displayed in a separate paper
\cite{42}. A sketch is contained in section \ref{s3.5}.
In appendix A we also consider the technically much 
simpler case of $G=U(1)$ and the interested reader is urged to consult that
appendix first before looking at the remainder of this section.
The graphical supplement to the remaining subsections can be found in 
Appendix B.

As is obvious from the tensor product structure of our states, it will be 
completely sufficient to establish the peakedness properties for one copy
of $G$ only and we can therefore drop the edge lable $e$ for the remainder 
of this section.

\subsection{Peakedness in the Connection Representation}
\label{s3.1}

The coherent states $\psi^t_g(h)$ are defined by the explicit series 
representation (\ref{3.3}) and we are interested in the limit $t\to 0$
of the probability distribution (with respect to Haar measure)
\be \label{3.8}
p^t_g(h):=\frac{|\psi^t_g(h)|^2}{||\psi^t_g||^2}
\ee
of which we would like to prove that it is concentrated at $h=u$ where
$g=Hu$ is the polar decomposition of $g\in SU(2)^\Co=SL(2,\Co)$. As the 
series in (\ref{3.3}) clearly converges worse and worse the smaller $t$
gets, the basic tool for all the estimates that follow is the elementary
Poisson Summation Formula\footnote{The authors are indebted to Brian
Hall for him pointing out the importance of this formula.}.
\begin{Theorem}[Poisson Summation Formula] \label{th3.1}~\\
Let $f$ be an $L_1(\Rl,dx)$ function such that the series
$$ \phi(y)=\sum_{n=-\infty}^\infty f(y+ns)$$ 
is absolutely and uniformly convergent for $y\in [0,s],s>0$.
Then
\be \label{3.9}
\sum_{n=-\infty}^\infty f(ns)=\frac{2\pi}{s}
\sum_{n=-\infty}^\infty \tilde{f}(\frac{2\pi n}{s})
\ee
where $\tilde{f}(k):=\int_{\Rl} \frac{dx}{2\pi} e^{-ikx} f(x)$
is the Fourier transform of $f$.
\end{Theorem}
The proof of this theorem can be found in any textbook on Fourier series,
see e.g. the classical book by Bochner \cite{51}. \\
The importance of this remarkable theorem for our purposes is that it 
converts a 
slowly converging series $\sum_n f(ns)$ as $s\to 0$ into a 
possibly rapidly converging series $\frac{1}{s}\sum_n \tilde{f}(2\pi 
n/s)$ of which in our case almost only the term with $n=0$ will be 
relevant. This is also crucial for numerical approximations as we will
see in appendix B.

The way the theorem is stated, it immediately applies to our problem 
only for the case $G=U(1)$ but one can actually generalize it to any
compact gauge group $G$ (see e.g. \cite{52}, \cite{39b} and 
references therein). Thus the method of proof displayed below for 
$G=SU(2)$ can be taken over to the general case. 

We begin with the following observation :
\be \label{3.10}
\psi^t_g(h)=\psi_H(hu^{-1})=\psi^t_{Huh^{-1}}(1)
\ee
if $g=Hu$ is the polar decomposition of $g$. Thus, we see that proving
that $p^t_g(h)$ is peaked at $h=u$ is equivalent to proving that 
$p^t_H(h)$ is peaked at $h=1$ independently of the positive definite,
Hermitean matrix $H$. By the same observation and the translation 
invariance of the Haar measure we see that $||\psi^t_g||=||\psi^t_H||$.  
In fact we find 
\be \label{3.11}
||\psi^t_g||^2=\psi^{2t}_{H^2}(1)
\ee
which one shows using the orthogonality relations 
\be \label{3.12}
\int_G d\mu_H(h) \overline{\pi(h)_{mn}}\pi'(h)_{m' n'}
=\frac{1}{d_\pi} \delta_{\pi\pi'} \delta_{mm'}\delta_{nn'}
\ee
(see e.g. \cite{53}, it is also one of the implications of the 
Peter\&Weyl theorem).

So far everything applies to any compact and connected $G$. We now 
specialize to $G=SU(2)$. In this case representations $\pi_j(g)_{mn}$
of dimension $d_j=2j+1$ are 
labelled by
half-integral spin quantum numbers $j=0,\frac{1}{2},1,..$ and magnetic 
quantum numbers $m,n\in\{-j,-j+1,..,j\}$, the eigenvalues of the 
Laplacian are $\lambda_j=j(j+1)$. In order to compute the character
$\chi_j(g)=\mbox{tr}(\pi_j(g)),\; g\in SL(2,\Co)$, we need the explicit 
form of the matrix elements. One finds (see, e.g. \cite{54})
\be \label{3.13}
\pi_j(g)_{mn}=\sum_l 
\frac{\sqrt{(j+m)! (j-m)! (j+n)! (j-n)!}}{(j-m-l)! (j+n-l)! (m-n+l)! l!}
a^{j+n-l} d^{j-m-l} b^{m-n+l} c^l
\ee
where the sum extends over all integers for which none of the factorials 
has negative arguments and 
\be \label{3.14}
g=\left( \begin{array}{cc} a & b\\ c & d \end{array} \right)\;
a,b,c,d\in \Co,\; ad-bc=1\;.
\ee
The eigenvalues $\lambda_1,\lambda_2$ of $g$ follow from the two equations
$\det(g)=\lambda_1\lambda_2=1,\;\mbox{tr}(g)=a+b=\lambda_1+\lambda_2$
which reveals 
\be \label{3.15}
\lambda_1=\lambda:=x+\sqrt{x^2-1},\;\lambda_2=\lambda_1^{-1}=x-\sqrt{x^2-1}
\mbox{ where }x=\frac{a+d}{2}\;.
\ee
Since both signs appear in (\ref{3.15}) there is no ambiguity in taking 
the square root of the complex number $x^2-1$.

Since the character is a class function invariant under conjugation we 
can assume $g$ to be diagonal in (\ref{3.13}) in which case the sum
over $l$ collapses to a single term $l=0$ and the sum over $m$ becomes 
a geometric series
\be \label{3.16}
\chi_j(g)=\sum_{m=-j}^j a^{j+m} d^{j-m}=\sum_{m=-j}^j \lambda^{2m} 
=\frac{\lambda^{2j+1}-\lambda^{-(2j+1)}}{\lambda-\lambda^{-1}}
\ee
which is invariant under $\lambda\leftrightarrow \lambda^{-1}$, the 
action of the Weyl subgroup. Formula
(\ref{3.16}) is a special case of the Weyl character formula \cite{53}.

We can now bring $\psi^t_g(1)$ into a form suitable for the Poisson 
summation formula
\ba \label{3.16a}
\psi^t_g(1)& = & \sum_j (2j+1)e^{-\frac{t}{2}j(j+1)}
\frac{\lambda^{2j+1}-\lambda^{-(2j+1)}}{\lambda-\lambda^{-1}}
\nonumber\\
&=& \frac{e^{t/8}}{\lambda-\lambda^{-1}}\sum_{n=1}^\infty
n e^{-t n^2/8} (\lambda^n-\lambda^{-n})\nonumber\\
&=& \frac{e^{t/8}}{\lambda-\lambda^{-1}}\sum_{n=-\infty}^\infty
n e^{-t n^2/8} \lambda^n
\ea
Next we notice that $\ln(\lambda)=\ach(x)$, where the choice of the branch 
cuts will be defined below, and define $s:=\sqrt{t}/2,z=\ach(x)/s$.
Then (\ref{3.16}) can be written as
\be \label{3.17}
\psi^t_g(1)= \frac{e^{t/8}}{2s\sqrt{x^2-1}}\sum_{n=-\infty}^\infty
(ns) e^{-(ns)^2/2} e^{(ns)z}=
\frac{e^{t/8}}{2s\sqrt{x^2-1}}\sum_{n=-\infty}^\infty f(ns)
\ee
where $f(x)=x\exp(-x^2/2+xz)$. This function certainly satisfies all the 
conditions for the application of the Poisson summation formula, its Fourier
transform is given by
\be \label{3.18}
\tilde{f}(k)=\frac{z-ik}{\sqrt{2\pi}}e^{-\frac{1}{2}(k+iz)^2} 
\ee
as can be shown by performing a contour integral. Thus we immediately find
the desired formula
\ba \label{3.19}
\psi^t_g(1) &=&
\frac{e^{t/8}}{2s\sqrt{x^2-1}}\frac{\sqrt{2\pi}}{s^2}
\sum_{n=-\infty}^\infty (\ach(x)-2\pi i n)
e^{-\frac{(2\pi n+i\sach(x))^2}{2s^2}} \nonumber\\
&=& \frac{4\sqrt{2\pi}e^{t/8}}{t^{3/2}}\frac{1}{\sqrt{x^2-1}}
\sum_{n=-\infty}^\infty (\ach(x)-2\pi i n)
e^{-2\frac{(2\pi n+i\sach(x))^2}{t}} 
\ea
Let likewise $y=\mbox{tr}(g g^\dagger)/2=\mbox{tr}(H^2)/2$ then
\be \label{3.20}
\psi^{2t}_{H^2}(1)=
\frac{2\sqrt{\pi}e^{t/4}}{t^{3/2}}\frac{1}{\sqrt{y^2-1}}
\sum_{n=-\infty}^\infty (\ach(y)-2\pi i n)
e^{-\frac{(2\pi n+i\sach(y))^2}{t}} 
\ee
thus 
\be \label{3.21}
p^t_H(h)=
\frac
{
\frac{16\sqrt{\pi}}{t^{3/2}}\frac{1}{|x^2-1|}
|\sum_{n=-\infty}^\infty (\ach(x)-2\pi i n)
e^{-2\frac{(2\pi n+i\sach(x))^2}{t}} |^2
}
{
\frac{1}{\sqrt{y^2-1}}
\sum_{n=-\infty}^\infty (\ach(y)-2\pi i n)
e^{-\frac{(2\pi n+i\sach(y))^2}{t}} 
}
\ee
Next we observe that upon writing $H=\exp(-i\tau_j p^j/2)$ we find 
$y=\ch(p),p=\sqrt{(p^j)^2}$ which allows us to write the probability
amplitude as
\be \label{3.22}
p^t_H(h)=
\frac
{
\frac{16\sqrt{\pi}}{t^{3/2}}\frac{1}{|x^2-1|}
|\sum_{n=-\infty}^\infty (\ach(x)-2\pi i n)
e^{-2\frac{(2\pi n+i\sach(x))^2+p^2/4}{t}} |^2
}
{
\frac{1}{\sh(p)}
\sum_{n=-\infty}^\infty (p-2\pi i n)
e^{-\frac{(2\pi n)^2+4i\pi n p}{t}} 
}
\ee
Let us first focus on the denominator $D^t_p$ in (\ref{3.22}) which can be 
written more explicitly as 
\be \label{3.23}
D^t_p=\frac{p}{\sh(p)}
[1+2\sum_{n=1}^\infty e^{-4\pi^2 n^2/t}\cos(\pi np/t)
+4\pi\sum_{n=1}^\infty
n e^{-4\pi^2 n^2/t}\frac{\sin(4\pi np/t)}{p}]
\ee
The term in the square brackets becomes at $p=0$ equal to
$$
1+2\sum_{n=1}^\infty e^{-4\pi^2 n^2/t}
+16\pi^2\sum_{n=1}^\infty \frac{n^2}{t}
e^{-4\pi^2 n^2/t} 
$$
which is still convergent and in fact for $t\to 0$ approaches the value $1$ 
exponentially fast with $t$. The same is true for $p\not=0$ as we show by 
means of the following lemma.
\begin{Lemma} \label{la3.0}
For any complex number $z$ we have $|\sin(z)/z|\le 
2\ch(\Im(z))<2\exp(|\Im(z)|)$. 
\end{Lemma}
Proof of Lemma \ref{la3.0} :\\
Let $z=x+iy$ then $\sin(z)=\sin(x)\ch(y)+i\cos(x)\sh(y)$. Using that
$|x/z|,|y/z|,|\cos(x)|\le 1$ we have 
\ba \label{3.23a}
|\frac{\sin(z)}{z}| &\le& |\frac{\sin(x)}{x}||\frac{x}{z}|\ch(y)
+|\cos(x)||\frac{\sh(y)}{y}||\frac{y}{z}| \nonumber\\
&\le&  \frac{\sin(|x|)}{|x|}\ch(y)
+\frac{\sh(|y|)}{|y|} 
\ea
Now $\sin(x)\le x$ for all $x\ge 0$ and employing the Taylor series
expansion of $\sh(y)$ we see that 
\be \label{3.23b}
|\frac{\sh(y)}{y}|\le 
\sum_{n=0}^\infty \frac{y^{2n}}{(2n+1)!}
\le \sum_{n=0}^\infty \frac{y^{2n}}{(2n)!}=\ch(y)
\ee
which concludes the proof.\\
$\Box$\\
With this information at our disposal we can estimate the absolute value 
of (\ref{3.23}) as follows
\ba \label{3.23c}
&& |D^t_p| \ge\frac{p}{\sh(p)}
[1+2\sum_{n=1}^\infty e^{-4\pi^2 n^2/t}\mbox{min}_p(\cos(4\pi np/t))
\nonumber\\ &&
+4\pi\sum_{n=1}^\infty
n e^{-4\pi^2 n^2/t}\mbox{min}_p(\frac{\sin(4\pi np/t)}{p})]
\nonumber\\
&=& \frac{p}{\sh(p)}
[1-2\sum_{n=1}^\infty e^{-4\pi^2 n^2/t}\mbox{max}_p(|\cos(4\pi np/t)|)
\nonumber\\ &&
-\frac{16\pi^2}{t}\sum_{n=1}^\infty
n^2 e^{-4\pi^2 n^2/t}\mbox{max}_p(|\frac{\sin(4\pi np/t)}{4\pi np/t}|)]
\nonumber\\ 
&\ge& \frac{p}{\sh(p)}
[1-2\sum_{n=1}^\infty e^{-4\pi^2 n^2/t}
-\frac{32\pi^2}{t}\sum_{n=1}^\infty n^2 e^{-4\pi^2 n^2/t}]
\nonumber\\ 
&=& \frac{p}{\sh(p)}
[1-e^{-4\pi^2/t}\sum_{n=1}^\infty e^{-4\pi^2 (n^2-1)/t}
(2+\frac{32\pi^2 n^2}{t})]
\nonumber\\ 
&\ge& \frac{p}{\sh(p)}
[1-e^{-4\pi^2/t}\sum_{n=0}^\infty e^{-4\pi^2 n^2/t}
(2+\frac{32\pi^2 (n+1)^2}{t})]
\ea
where in the last step we have used $(n-1)^2\le n^2-1$ valid for all integers
$n\ge 1$. The series in the last line of (\ref{3.23c}) is certainly 
still convergent for any $t>0$, the dominant term being the one at $n=0$
which at $t\to 0$ behaves as $1/t$. Since $\lim_{t\to 0} e^{-a/t}/t^n=0$
for all $a>0,n\in\Zl$ we find the first main result.
\begin{Lemma} \label{la3.1}
i) There exists a positive constant $K_t$ (independent of $p$), 
and exponentially vanishing with $t\to 0$ such that
$D^t_p\ge\frac{p}{\sh(p)}(1-K_t)$ for all $p\ge 0$.\\
ii) 
%There exists a positive constant $\tilde{K}_t$ (independent of $p$), 
%and exponentially vanishing with $t\to 0$ such that
For the same constant $K_t$ it holds that
$D^t_p\le\frac{p}{\sh(p)}(1+K_t)$ for all $p\ge 0$.
\end{Lemma}
The second part of this lemma is proved by similar methods, one just has 
to inverse signs in the estimates and replace $\min\leftrightarrow\max$
in the first line of (\ref{3.23c}).

The next step consists in the computation of $\ach(x)$. First of all 
we have with $h=\exp(\theta^j\tau_j),\;
\theta=\sqrt{(\theta^j)^2}\in [0,\pi],\tau_j=-i\sigma_j$ where $\sigma_j$
are the standard Pauli matrices, $\tau_i\tau_j=-\delta_{ij}+
\epsilon_{ijk}\tau_k$
\ba \label{3.24}
&& x=\frac{\mbox{tr}(Hh)}{2}=
\frac{\mbox{tr}([\ch(p/2)-i\frac{p^j}{p}\tau_j\sh(p/2)]
[\cos(\theta)+\frac{\theta^j}{\theta}\tau_j\sin(\theta)])}{2}
\nonumber\\
&=&\ch(p/2)\cos(\theta)+i\sh(p/2)\sin(\theta)\cos(\alpha)
\ea
where $\cos(\alpha)=(p^j\theta^j)/(p\theta)\in [-1,1]$. We wish to write
$x$ as $\ch(s+i\phi)$ for some $s\in\Rl, \phi\in[0,\pi]$ and it is a 
non-trivial question whether this is always possible.
\begin{Lemma} \label{la3.2}
For any complex number $z=R+iI$ there exist 
real numbers $s\in\Rl$ and $\phi\in [0,\pi]$ such that $\ch(s+i\phi)=z$.
These numbers are uniquely determined except in the case $I=0,|R|>1$
in which case the sign of $s$ is undetermined.
\end{Lemma}
Proof of Lemma \ref{la3.2} :\\
We will give a constructive proof as we will need the following formulae 
later on.\\
We have $\ch(s+i\phi)=\ch(s)\cos(\phi)+i\sh(s)\sin(\phi)$, thus if the 
statement of the lemma is true we must have 
\be \label{3.25}
\ch(s)\cos(\phi)=\Re(z)=:R\mbox{ and }
\sh(s)\sin(\phi)=\Im(z)=:I
\ee
The sign of $s$ coincides with that of $I$ while $\phi\ge\pi/2$ if $R\le 0$
and $\phi\le \pi/2$ if $R\ge 0$.
Using the trigonometric and hyperbolic relations 
$1=\cos^2(\phi)+\sin^2(\phi)=\ch^2(s)-\sh^2(s)$ we find after solving 
a system of quadratic equations unambiguously 
\ba \label{3.26}
\ch^2(s)&=& \frac{1}{2}(1+R^2+I^2+\sqrt{(1+R^2+I^2)^2-4R^2}) \nonumber\\
\cos^2(\phi)&=& \frac{1}{2}(1+R^2+I^2-\sqrt{(1+R^2+I^2)^2-4R^2}) 
\ea
Since $0\le\phi\le\pi$ we have $\sin(\phi)\ge 0$ and $\ch(s)\ge 1$ for 
either sign of $s$. Thus the only ambiguity in taking the square root of
(\ref{3.26}) appears in the definition of $\sh(s),\cos(\phi)$. However,
in the range of $s,\phi$ that we are considering we find uniquely
\ba \label{3.27}
\ch(s)&=& \frac{1}{\sqrt{2}}
\sqrt{1+R^2+I^2+\sqrt{(1+R^2+I^2)^2-4R^2}} \nonumber\\
\sh(s)&=& \frac{\mbox{sgn}(I)}{\sqrt{2}}
\sqrt{-1+R^2+I^2+\sqrt{(1+R^2+I^2)^2-4R^2}} \nonumber\\
\cos(\phi)&=& 
\frac{\mbox{sgn}(R)}{\sqrt{2}}\sqrt{1+R^2+I^2-\sqrt{(1+R^2+I^2)^2-4R^2}}
\nonumber\\
\sin(\phi)&=& 
\frac{1}{\sqrt{2}}\sqrt{1-R^2-I^2+\sqrt{(1+R^2+I^2)^2-4R^2}}
\ea
where sgn denotes the non-standard step function $\mbox{sgn}(x)=1$ if 
$x\ge 0$ and $\mbox{sgn(x)}=-1$ if $x<0$.
Since the functions $\cos$ and $\sh$ respectively are invertible on
$[0,\pi]$ and $\Rl$ respectively, above formulae define $\phi,s$ uniquely.
One can explicitly check that the squares of the first and third lines in
(\ref{3.27}) are always greater or smaller than one respectively for any
choice of $R,I$ and that the arguments of all square roots are 
non-negative.\\
We compute 
\be \label{3.28}
\ch(s)\cos(\phi)+i\sh(s)\sin(\phi)=\mbox{sgn}(R)|R|+i\mbox{sgn}(I)|I|
=z
\ee
since although sgn is a non-vanishing function, the function $\mbox{sgn}(x)
|x|$ vanishes anyway at $x=0$. This shows that the above choice for $s,\phi$
solves the task to reproduce $z$.
To see that $s,\phi$ are in fact uniquely determined unless $|R|>1,I=0$
we notice that an ambiguity can possibly arise only through the sign 
function, that is, if either $R$ or $I$ vanish.
\begin{itemize}
\item[i)] $R=0,I\not=0$\\
Then $\phi=\pi/2,\mbox{sgn}(s)=\mbox{sgn}(I)\not=0$ are uniquely determined.
\item[ii)] $R\not=0,I=0$\\
Then either $s=0$ or $\phi=0,\pi$.\\
Subcase a) : $|R|=1$.\\
Then $s=0$ and $\phi=0,\pi$ if $R=1,-1$ are uniquely determined.\\
Subcase b) : $|R|<1$.\\
If $s\not=0$ then necessarily $\phi=0,\pi$ so $|\ch(s)\cos(\phi)|>1>|R|$
which is not allowed. Thus $s$=0 and $\phi$ is uniquely determined.\\
Subcase c) : $|R|>1$.\\
If $s=0$ then $|\ch(s)\cos(\phi)|\le 1<|R|$ which is not allowed. Thus
$\phi=0,\pi$ according to the sign of $R$ but the sign of $s\not=0$ is 
ambiguous.
\item[iii)] $R=I=0$\\
Now necessarily $\phi=\pi/2,s=0$ are uniquely determined.
\end{itemize}
$\Box$\\
\\
We remark that the undeterminacy of the sign of $s$ in the case $|R|>1,
I=0$ does not affect us because applied to our situation we have 
$R=\ch(p/2)\cos(\theta), I=\sh(p/2)\sin(\theta)\cos(\alpha)$ and thus 
$|R|>1$ means that $p>0$ and either $\theta=0,\pi$ or $\alpha=\pi/2$. 
In the 
first case we simply have $h=\pm 1$ so $g=\pm H,x=\pm\ch(p/2), \lambda
=\pm\ch(p/2)+\sh(p/2)=\pm\exp(\pm p/2)$, thus we fix the signs by
$s:=p/2, \phi:=\theta=0,\pi$. In the second case, which is different from 
the first one only if $\theta\not=0,\pi$, we have $x=\ch(p/2)\cos(\theta),
\lambda=\ch(p/2)\cos(\theta)+\sqrt{\ch^2(p/2)\cos^2(\theta)-1}$ is 
real-valued because $|R|>1$ and either $\lambda$ or $-\lambda$ is positive if 
$\cos(\theta)$ is 
positive or negative respectively. In that case we define $\phi=0,\pi$ 
respectively and $s=\ach(|x|)=\ln(|x|+\sqrt{|x|^2-1})>0$ uniquely
so that $\lambda=e^{s+i\phi}=\pm e^s$ is uniquely determined.\\
\\
Consider now the exponent of the n-th term of the series in the numerator 
$N^t_H(h)$ of (\ref{3.22}) given by $-2[(2\pi n+i\ach(x))^2+p^2/4]/t$ and
whose real
part is given by $-2[(2\pi n-\phi)^2-s^2+p^2/4]/t$. 
We have the following elementary estimate 
\be \label{3.29}
(2\pi n-\phi)^2+p^2/4-s^2\ge 4\pi^2 (|n|-1)^2+[\phi^2+p^2/4-s^2]
\ee
for all $n\not=0$ which shows that it is important to know the sign
of the function $p^2/4-s^2+\phi^2$. In fact, we would like to show that it
is non-negative and vanishing if and only if $\theta=0$.
The following theorem is the first main theorem of this
subsection.
\begin{Theorem} \label{th3.2}
For all $p,\theta,\alpha$ it holds that the function
\be \label{3.30}
\delta^2(p,\theta,\alpha):=p^2/4-s^2(p,\theta,\alpha)
+\phi^2(p,\theta,\alpha)-\theta^2
\ee
is non-negative and zero if and only if either a) $\phi=\theta,|s|=p/2$ 
arbitrary and $|\cos(\alpha)|=1$ or b) $\alpha$ arbitrary and $s=p=0$ or
$\phi=\theta=0,\pi$.
\end{Theorem}
The proof of this theorem given below is elementary but lengthy,
therefore we will break it into several lemmas. 
\begin{Lemma} \label{la3.3}
The function $\delta^2$ in (\ref{3.30}) 
depends on $\alpha$ only through $r:=\cos^2(\alpha)\in [0,1]$ and is 
strictly monotonously decreasing as $r$ increases from $0$ to $1$ for all 
$p>0$ and all $\theta\in (0,\pi)$.
\end{Lemma}
Proof of Lemma \ref{la3.3} :\\
Since $f$ depends only on $|s|$ we can determine $|s|$ from 
$\ch(s)=\ch(|s|)$ and $\phi$ from $\ch(\phi)$. But both formulae in
(\ref{3.27}) depend on $\alpha$
only through $I^2=\sh^2(p)\sin^2(\theta)r$. Thus, in particular,
the sign ambiguity in $s$ is irrelevant as far as $f$ is concerned. 

Lets us define $\sigma=1+R^2+I^2$. Then $\sigma_{,r}=\sh^2(p)\sin^2(\theta)>0$
for $p>0,\theta\not=0,\pi$. We compute 
\ba \label{3.31}
\delta^2_{,r} &=& 2(\phi
[\acs(\frac{\mbox{sgn}(\cos(\theta))}{\sqrt{2}}
\sqrt{\sigma-\sqrt{\sigma^2-4R^2}})]_{,r}
\nonumber\\ &&
-|s| [\ach(\frac{1)}{\sqrt{2}}
\sqrt{\sigma+\sqrt{\sigma^2-4R^2}})]_{,r})\nonumber\\
&=& 
2\sigma_{,r}(-\phi\frac{\mbox{sgn}(\cos(\theta))}
{\sqrt{2}\sqrt{1-\frac{1}{2}(\sigma-\sqrt{\sigma^2-4 R^2})}}
\frac{1-\frac{\sigma}{\sqrt{\sigma^2-4 R^2}}}
{2\sqrt{\sigma-\sqrt{\sigma^2-4R^2}}}
\nonumber\\ && -|s|
\frac{1}
{\sqrt{2}\sqrt{\frac{1}{2}(\sigma+\sqrt{\sigma^2-4 R^2})-1}}
\frac{1+\frac{\sigma}{\sqrt{\sigma^2-4 R^2}}}
{2\sqrt{\sigma+\sqrt{\sigma^2-4R^2}}})
\nonumber\\
&=& 
\frac{\sigma_{,r}}{\sqrt{\sigma^2-4 R^2}}
(\phi\frac{1}{\sqrt{2-(\sigma-\sqrt{\sigma^2-4 R^2})}}
\mbox{sgn}(\cos(\theta))\sqrt{\sigma-\sqrt{\sigma^2-4R^2}}
\nonumber\\ && -|s|
\frac{1}
{\sqrt{\sigma+\sqrt{\sigma^2-4 R^2}-2}}
\sqrt{\sigma+\sqrt{\sigma^2-4R^2}})
\nonumber\\
&=& 
\frac{\sigma_{,r}}{\sqrt{\sigma^2-4 R^2}}
(\phi\frac{\sqrt{2}\cos(\phi)}{\sqrt{2}\sqrt{1-\cos^2(\phi)}}
-|s|\frac{\sqrt{2}\ch(|s|)}{\sqrt{2}\sqrt{\ch^2(|s|)-1}}
\nonumber\\
&=& 
\frac{\sigma_{,r}}{\sqrt{\sigma^2-4 R^2}}
(\phi\mbox{cot}(\phi)-|s|\mbox{coth}(|s|))
\ea
where in the last step we have observed that $|\sin(\phi)|=\sin(\phi)$
because $\phi\in [0,\pi]$ and that $|\sh(s)|=\sh(|s|)$. The last line 
in (\ref{3.31}) is evidently non-positive for $\theta\ge\pi/2$ (recall that
$\phi</=/>\pi/2$ iff $\theta</=/>\pi/2$). 
That this is also true for all of the range of $\theta$ follows from the 
following simple observation.
\begin{Lemma} \label{la3.4}
i) The function $x\mapsto x\mbox{cot}(x),\;x\in [0,\pi]$ is bounded 
from above by $1$ which is reached for $x=0$.\\
ii) The function $x\mapsto x\mbox{coth}(x),\;x\in [0,\infty)$ is bounded 
from below by $1$ which is reached for $x=0$.
\end{Lemma}
Proof of Lemma \ref{la3.4} :\\
i)\\
We simply compute
\be \label{3.32}
(x\mbox{cot}(x))'=\frac{\sin(2x)-2x}{2\sin^2(x)}\le 0
\ee
since $x\ge\sin(x)\forall x\ge 0$. Thus the function is monotonously
decreasing and therefore its maximum is attained at $x=0$ where its value is
$1$. Notice that the derivative exists even at $x=0$.\\
ii) Likewise we have
\be \label{3.33}
(x\mbox{coth}(x))'=\frac{\sh(2x)-2x}{2\sh^2(x)}\ge 0
\ee
since $x\le\sh(x)\forall x\ge 0$. Thus the function is monotonously
decreasing and therefore its minimum is attained at $x=0$ where its value 
is $1$. Notice that the derivative exists even at $x=0$.\\
$\Box$\\
\\
Using Lemma \ref{la3.4} we conclude that $\delta^2_{,r}\le 0$ and 
$\delta^2_{,r}=0$
only if either $\sigma_{,r}=\sh^2(p)\sin^2(\theta)=0$ or $\phi=s=0$.
If $\phi=s=0$ then $\ch(p/2)\cos(\theta)=1,\sh(p/2)\sin(\theta)
\cos(\alpha)=0$.
Let us exclude the values $p=0,\theta=0,\pi$ for the moment, then we find
that $f_{,r}<0$, except possibly at $r=0$ if also $\ch(p/2)\cos(\theta)=1$.  
Thus $\delta^2$ is strictly monotonously decreasing for all $r>0$ and 
since it is a continuous function of $r$ it is strictly monotonously
decreasing for all $r\in [0,1], p>0,\theta\in (0,\pi)$.\\
$\Box$\\
\\
Proof of Theorem \ref{th3.2} :\\
Using Lemma \ref{la3.3} we know that for $p>0,\theta\in (0,\pi)$ the 
function $\delta^2$ attains its minimum at $r=1$ for which we have 
$\ch(p/2)\cos(\theta)=\ch(s)\cos(\phi),\sh(s)\sin(\phi)=\pm\sh(p/2)
\sin(\theta)$
and thus $s=\pm p/2,\phi=\theta$. Therefore 
$\delta^2\ge 0$ for $p>0,\theta\not=0,
\pi$ and all $\alpha\in [0,\pi]$ and $f=0$ only at $\alpha=0,\pi$.

The remaining cases are $p=0$ or $\theta=0,\pi$. \\
Case $p=0$ :\\
Then $\ch(s)\cos(\phi)=\cos(\theta),\sh(s)\sin(\phi)=0$. Thus either 
$s=0,\phi=\theta$ or $\phi=0$ and then $\ch(s)=\cos(\theta)$ which is 
possible only if $\theta=0,s=0$. In both of these cases we find 
$\delta^2=0$.\\
Case $\theta=0$ :\\
Now $\ch(s)\cos(\phi)=\ch(p/2),\sh(s)\sin(\phi)=0$. Thus either
$\phi=0,|s|=p/2$ or $s=0$ and then $\cos(\phi)=\ch(p/2)$ which is possible
only if $p=0,\phi=0$. In both cases we find $\delta^2=0$.\\
Case $\theta=\pi$ :\\
Finally $\ch(s)\cos(\phi)=-\ch(p/2),\sh(s)\sin(\phi)=0$. Thus either
$\phi=\pi,|s|=p/2$ or $s=0$ and then $\cos(\phi)=-\ch(p/2)$ which is possible
only if $p=0,\phi=\pi$. In both cases we find $\delta^2=0$.\\

Collecting all the results we find $\delta^2\ge 0$ for all $p\ge 0,\theta\in 
[0,\pi],\alpha\in [0,\pi]$ and $\delta^2=0$ is possible only if either 
a) $\alpha=0,\pi$ while $\theta,p$ can be arbitrary or b) $p=0$ or
$\theta=0,\pi$ while $\alpha$ can be arbitrary.\\
$\Box$\\
\\
We now come back to the numerator $N^t_H(h)$ in (\ref{3.22}). The series
involved can be transformed into the following expression 
\ba \label{3.34}
&& \sum_{n=-\infty}^\infty (\ach(x)-2\pi i n)
e^{-2\frac{(2\pi n+i\sach(x))^2+p^2/4}{t}} \nonumber\\
&=& e^{-\frac{2}{t}(\phi^2+p^2/4-s^2-2is\phi)} 
\sum_{n=-\infty}^\infty (\ach(x)-2\pi i n)
e^{-\frac{8\pi^2 n^2}{t}} e^{-\frac{8i\pi n}{t}(s+i\phi)}
\nonumber\\
&=&
e^{-\frac{2}{t}(f+\theta^2-2is\phi)} 
[\ach(x)(1+2\sum_{n=1}^\infty e^{-\frac{8\pi^2 n^2}{t}}
\cos(\frac{8\pi n}{t}\ach(x))
\nonumber\\ &&
-4\pi\sum_{n=1}^\infty n e^{-\frac{8\pi^2 n^2}{t}}
\sin(\frac{8\pi n}{t}\ach(x))]
\nonumber\\
&=&
e^{-\frac{2}{t}(f+\theta^2-2is\phi)} 
\ach(x)[1+2\sum_{n=1}^\infty e^{-\frac{8\pi^2 n^2}{t}}
\cos(\frac{8\pi n}{t}\ach(x))
\nonumber\\ &&
-4\pi\sum_{n=1}^\infty n e^{-\frac{8\pi^2 n^2}{t}}
\frac{\sin(\frac{8\pi n}{t}\ach(x))}{\ach(x)}]
\ea
The square bracket expression is certainly regular at $x=1$, that 
is, $\ach(x)=0$ and still converges exponentially fast to $1$ similarly 
as for the denominator at $p=0$. The same holds at $\ach(x)\not=0$. This 
can be seen as follows : Taking the absolute value of (\ref{3.34}) 
we see that it can be estimated from above by (using Lemma \ref{la3.0}
and Theorem \ref{th3.2})
\ba \label{3.35}
|\ref{3.34}| &\le& 
e^{-\frac{2}{t}(\theta^2+\delta^2)} 
|\ach(x)|[1+2\sum_{n=1}^\infty e^{-\frac{8\pi^2 n^2}{t}}
|\cos(\frac{8\pi n}{t}\ach(x))|
\nonumber\\ &&
+32\pi^2\sum_{n=1}^\infty \frac{n^2}{t} e^{-\frac{8\pi^2 n^2}{t}}
|\frac{\sin(\frac{8\pi n}{t}\ach(x))}{8\pi n\ach(x)/t}|]
\nonumber\\
& \le & e^{-\frac{2}{t}(\theta^2+\delta^2)} 
|\ach(x)|[1+2\sum_{n=1}^\infty e^{-\frac{8\pi^2 n^2}{t}}
e^{\frac{8\pi n}{t}\phi}
\nonumber\\ &&
+64\pi^2\sum_{n=1}^\infty \frac{n^2}{t} e^{-\frac{8\pi^2 n^2}{t}}
e^{8\pi n\phi/t}]
\ea
We now consider two cases : \\
Case (A) : $0\le \phi\le (1-c)\pi$ where $0<c<1$ will be specified in the 
course of our derivation.\\
Case (B) : $(1-c)\pi\le \phi\le \pi$.\\

Turning to Case (A) we can further estimate (\ref{3.35}) by
\ba \label{3.36}
|\ref{3.34}| & \le & e^{-\frac{2}{t}(\theta^2+)\delta^2} 
|\ach(x)|[1+\sum_{n=1}^\infty e^{-\frac{8\pi^2 n^2}{t}}
e^{8\pi^2(1-c)n/t} (2+64\pi^2\frac{n^2}{t})]
\nonumber\\
& \le & e^{-\frac{2}{t}(\theta^2+\delta^2)} 
|\ach(x)|[1+e^{-8\pi^2c/t}\sum_{n=1}^\infty e^{-\frac{8\pi^2 (n^2-n)}{t}}
(2+64\pi^2\frac{n^2}{t})]
\nonumber\\
& \le & e^{-\frac{2}{t}(\theta^2+\delta^2)} 
|\ach(x)|[1+e^{-8\pi^2c/t}\sum_{n=1}^\infty e^{-\frac{8\pi^2 (n-1)^2}{t}}
(2+64\pi^2\frac{n^2}{t})]
\nonumber\\
& \le & e^{-\frac{2}{t}(\theta^2+\delta^2)} 
|\ach(x)|[1+e^{-8\pi^2c/t}\sum_{n=0}^\infty e^{-\frac{8\pi^2 n^2}{t}}
(2+64\pi^2\frac{(n+1)^2}{t})]
\ea 
where in the second step we have dropped the $n\ge 1$ multiplying $c$, 
in the third we used the estimate $(n-1)^2\le n^2-n$ valid for $n\ge 1$ and 
in the last we rewrote the series starting at $n=0$.
The term in the square bracket certainly converges to $1$ exponentially
fast with $t\to 0$ for any $c>0$ by an argument already mentioned.\\

Turning to Case (B) we notice that, as we have to divide the absolute 
value of the square of (\ref{3.35})
by $|x^2-1|$ we need to make sure that $\ach^2(x)/(x^2-1)$ is bounded at
$x=\pm 1$. At $x=1$ we find $\lim_{x\to 1} \ach^2(x)/(x^2-1)=
\lim_{x\to 1}\ach(x)/(x\sqrt{x^2-1})=\lim_{x\to 1}
(1/\sqrt{x^2-1})/(1/\sqrt{x^2-1})=1$ while at $x\to \infty$ we find 
$\lim_{x\to\infty}\ach^2(x)/(x^2-1)=\lim_{x\to\infty}\ln(x+\sqrt{x^2-1})/x
=0$. However at $x=-1$ we have $\ach(x)=\pm i\pi$ and so the expression 
is in danger to blow up. This is, however not the case. We simply have 
to write (\ref{3.34}) in the variable $\sigma:=\ach(x)-i\pi=s-i(\pi-\phi)$
which then becomes 
\ba \label{3.37}
&& \sum_{n=-\infty}^\infty (\ach(x)-2\pi i n)
e^{-2\frac{(2\pi n+i\sach(x))^2+p^2/4}{t}} \nonumber\\
&=&
\sum_{n=-\infty}^\infty (\sigma-i(2n-1)\pi)
e^{-2\frac{((2n-1)\pi+i\sigma)^2+p^2/4}{t}} \nonumber\\
&=& e^{-\frac{2}{t}(p^2/4-s^2+(\pi-\phi)^2+2is(\pi-\phi))}
\sum_{n=-\infty}^\infty 
(\sigma-i(2n-1)\pi) e^{-2\frac{[(2n-1)\pi]^2+2i(2n-1)\sigma\pi}{t}} 
\nonumber\\
&=& e^{-\frac{2}{t}(p^2/4-s^2+(\pi-\phi)^2+2is(\pi-\phi))}
[2\sigma\sum_{n=1,\mbox{odd}}^\infty e^{-2\frac{n^2 \pi^2}{t}}
\cos(4n\pi\sigma/t) 
\nonumber\\ &&
+2\pi\sum_{n=1,\mbox{odd}}^\infty 
n e^{-2\frac{n^2 \pi^2}{t}}
\sin(4n\pi\sigma/t)]
\nonumber\\
&=& e^{-\frac{2}{t}(p^2/4-s^2+(\pi-\phi)^2+2is(\pi-\phi)\sigma)} 
\sigma [2\sum_{n=1,\mbox{odd}}^\infty e^{-2\frac{n^2 \pi^2}{t}}
\cos(4n\pi\sigma/t)
\nonumber\\ && 
+8\pi^2\sum_{n=1,\mbox{odd}}^\infty 
\frac{n^2}{t} e^{-2\frac{n^2 \pi^2}{t}}
\frac{\sin(4n\pi\sigma/t)}{4n\pi\sigma/t}]
\ea
Using again Lemma \ref{la3.0}, Theorem \ref{th3.2} and the fact that
$(1-c)\pi\le\phi\le\pi$ we can 
estimate the absolute value of (\ref{3.34}) as 
\ba \label{3.38}
|\ref{3.34}| &\le& e^{-\frac{2}{t}(p^2/4-s^2+(\pi-\phi)^2)} |\sigma|
\sum_{n=1,\mbox{odd}}^\infty e^{-2\frac{n^2 \pi^2}{t}}
(2+16\pi^2 n^2/t) e^{4n\pi(\pi-\phi)/t} 
\nonumber\\
&\le& e^{-\frac{2}{t}(p^2/4-s^2+\phi^2+\pi^2-2\pi\phi)} |\sigma|
\sum_{n=1,\mbox{odd}}^\infty e^{-2\frac{n^2 \pi^2}{t}}
(2+16\pi^2 n^2/t) e^{4nc\pi^2/t} 
\nonumber\\
&\le& e^{-\frac{2}{t}(\theta^2+\delta^2-\pi^2)} |\sigma|
\sum_{n=1,\mbox{odd}}^\infty e^{-2\frac{n^2(1-2c)\pi^2}{t}}
(2+16\pi^2 n^2/t) e^{-4c(n^2-n)\pi^2/t} 
\nonumber\\
&\le& e^{-\frac{2}{t}(\theta^2+\delta^2-\pi^2+(1-2c)\pi^2)} |\sigma|
\sum_{n=1,\mbox{odd}}^\infty e^{-2\frac{(n^2-1)(1-2c)\pi^2}{t}}
(2+16\pi^2 n^2/t) 
\nonumber\\
&\le& e^{-\frac{2}{t}(\theta^2+\delta^2-2c\pi^2)} |\sigma|
\sum_{n=1,\mbox{odd}}^\infty e^{-2\frac{(n-1)^2(1-2c)\pi^2}{t}}
(2+16\pi^2 n^2/t) 
\ea
where in the last line we have assumed $c<1/2$ and used the estimate
$(n-1)^2\le n^2-1$ valid for $n\ge 1$.
At this point we choose some $c<1/2$ so that $\theta\ge \pi/2$. 
Then we have for any $0<d<1$, letting $n$ start at $0$ in the series,
\ba \label{3.39}
|\ref{3.34}| &\le& e^{-\frac{2}{t}((1-d)\theta^2+\delta^2+(d/4-2c)\pi^2)} 
|\sigma|
\sum_{n=0,\mbox{even}}^\infty e^{-2\frac{n^2(1-2c)\pi^2}{t}}
(2+16\pi^2(n+1)^2/t) 
\nonumber\\
&\le& |\sigma| e^{-\frac{2}{t}((1-d)\theta^2+\delta^2)}
[1+e^{-\frac{2}{t}(d/4-2c)\pi^2}
\sum_{n=0,\mbox{even}}^\infty e^{-2\frac{n^2(1-2c)\pi^2}{t}}
(2+16\pi^2(n+1)^2/t)]
\nonumber\\
\ea
where in the second step we have assumed $2c<d/4$ in order to isolate
the term with $n=0$.
We see that if we choose $c<d/8$ then the term in the square bracket 
converges exponentially fast to $1$ as $t\to 0$ and for $d<1$ the 
exponential prefactor decreases exponentially fast to zero as $t\to 0$
since $\theta\ge\pi/2$. 

Let us, for definiteness, choose $d=1/2,c=1/32$ which clearly also
satisfies $2c<1$. Then, putting
(\ref{3.36}) and (\ref{3.39}) together we
have shown :
\begin{Lemma} \label{la3.5}
i) For all $0\le\phi\le 31\pi/32$ there exists a positive constant
$K_t'$ (independent of $H,h$), decaying exponentially fast 
to $0$ as $t\to 0$ such that
\be \label{3.40}
|N^t_H(h)|
\le
\frac{16\sqrt{\pi}}{t^{3/2}}\frac{|\ach(x)|^2}{|x^2-1|}
e^{-\frac{4(\theta^2\delta^2)}{t}}(1+K_t')
\ee
where $x=\ch(p/2)\cos(\theta)+i\sh(p/2)\sin(\theta)\cos(\alpha)$.\\
ii) For all $31\pi/32\le\phi\le\pi$ there exists a positive constant
$K_t^{\prime\prime}$ (independent of $H,h$), decaying exponentially fast 
to $0$ as $t\to 0$ such that
\be \label{3.41}
|N^t_H(h)|
\le
\frac{16\sqrt{\pi}}{t^{3/2}}\frac{|\ach(x)-i\pi|^2}{|x^2-1|}
e^{-\frac{2(\theta^2+2\delta^2)}{t}}(1+K_t^{\prime\prime})
\ee
\end{Lemma}
Finally, combining Lemmata \ref{la3.1} and \ref{3.5} we find the following
uniform bound for the probability density in position space which is the
second main theorem of this subsection.
\begin{Theorem} \label{th3.3}
i) For all $0\le\phi\le 31\pi/32$ there exist positive constants
$K_t,K_t'$ (independent of $H,h$), decaying exponentially fast 
to $0$ as $t\to 0$ such that
\be \label{3.42}
p^t_H(h)\le
\frac{\frac{16\sqrt{\pi}}{t^{3/2}}\frac{|\ach(x)|^2}{|x^2-1|}
e^{-\frac{4(\theta^2+\delta^2)}{t}}(1+K_t')}
{\frac{p}{\sh(p)}(1-K_t)}
\ee
where $x=\ch(p/)\cos(\theta)+i\sh(p/)\sin(\theta)\cos(\alpha)$.\\
ii) For all $31\pi/32\le\phi\le\pi$ there exist positive constants
$K_t^{\prime\prime}$ (independent of $H,h$), decaying exponentially fast 
to $0$ as $t\to 0$ such that
\be \label{3.43}
p^t_H(h)\le
\frac{\frac{16\sqrt{\pi}}{t^{3/2}}\frac{|\ach(x)-i\pi|^2}{|x^2-1|}
e^{-\frac{2(\theta^2+2\delta^2)}{t}}(1+K_t^{\prime\prime})}
{\frac{p}{\sh(p)}(1-K_t)}
\ee
\end{Theorem}
Obviously, the bounds are not completely optimal but 
the remarkable and most important feature
is that the bound decays exponentially fast for $\theta\not=0$. At $\theta=0$
we have $\phi=0$ and $s=\pm p/2$ and the bound is given by the 
$p$-dependent value 
$$
\frac{(p/2)^2\sh(p)}{p\sh^2(p/2)}\frac{16\sqrt{\pi}}{t^{3/2}}
(1+K_t')/(1-K_t)
=\frac{p\sh(p)}{\ch(p)-1}\frac{8\sqrt{\pi}}{t^{3/2}}
(1+K_t')/(1-K_t)
$$
which defines a rather sharp peak as $t\to 0$ and that peak grows linearly 
with $p$. This is in contrast to the harmonic oscillator for which
the bound is also Gaussian suppressed in $x-q$ but it is also independent
of $t,p$. Of course, this is the effect of the non-Abelian nature of $G$
and due to the fact that $G$ is not a linear space.

Notice also, that the $e^{-4\delta^2/t}$ cannot be dispensed with :
For $\alpha=\theta=\pi/2$ it can happen that $s$ stays bounded while $p$ 
becomes large, in fact $s=0,\phi=\pi/2$ in this case.
Since $|\ach(x)|^2/|x^2-1|=[s^2+\phi^2]/[\sh^2(s)+\sin^2(\phi)]$ we obtain 
$|\ach(x)|^2/|x^2-1|=(\pi/2)^2$ 
and it seems that the peak grows exponentially
with $p$ in this case. However, this is not true : the function $\delta^2$
now takes the value $p^2/4$ and so the peak is in fact Gaussian damped 
with $p$ !

\subsection{Peakedness of the Overlap Function}
\label{s3.2}

We compute first the inner product between two coherent states and find
\be \label{3.44}
<\psi^t_g,\psi^t_{g'}>=\psi^{2t}_{HH'}(h)
\ee
where $g=Hu,g'=H'u'$ are the polar decompositions of $g,g'$ and
$h=u^{-1} u'$. Our objective is to show that the {\it Overlap Function}
for these coherent states given by
\be \label{3.45}
i^t(g,g'):=\frac{|<\psi^t_g,\psi^t_{g'}>|^2}
{||\psi^t_g||^2 ||\psi^t_{g'}||^2}=
\frac{[\psi^{2t}_{HH'}(h)]^2}{\psi^{2t}_{H^2}(1)\psi^{2t}_{(H')^2}(1)}
\ee
is peaked at $g=g'$ which in some sense would mean that the coherent state
labelled by $g$ represents a neighbourhood (whose size is controlled by
$t$) of the point $(p,u)$ defined by $g=Hu$ in the phase space $T^\ast G$.
The existence of a Segal-Bargmann Hilbert space in which wave functions
depend on phase space rather than momentum or configuration space will allow
us to specify the meaning of this statement precisely in a later subsection.

The idea of proof is to use Theorem \ref{th3.3} of the previous subsection.
However, in order to do that we must first compute the polar decomposition
of $HH'$ which is not necessarily a Hermitean, positive definite matrix any
longer. Using the parameterizations $H=\ch(p/2)-i\tau_j p^j\sh(p/2)/p,
H'=\ch(p'/2)-i\tau_j p'_j\sh(p'/2)/p'$ we write $HH'=\tilde{H}(H,H')
\tilde{u}(H,H')$ where $\tilde{H}$ and $\tilde{u}$ are uniquely determined
and then have $\psi^{2t}_{HH'}(u^{-1} u')=\psi^{2t}_{\tilde{H}}(\tilde{h})$
where $\tilde{h}=u^{-1} u'\tilde{u}^{-1}$. Suppose then that we can prove
that (\ref{3.45}) is peaked at $H=H'$ and $\tilde{h}=1$. Then, 
since $\tilde{u}=1$
and $\tilde{H}=H$ at $H=H'$, we have automatically shown that $i^t(g,g')$
is peaked at $g=g'$. This will be our strategy.

Let then $\tilde{H}=\exp(-i\tau_j \tilde{p}^j/2)$ and $\tilde{h}=
\exp(\tilde{\theta}^j\tau_j/2)$. We define as before 
$\tilde{p}=\sqrt{\tilde{p}^j
\tilde{p}^j}, \;\tilde{\theta}=\sqrt{\tilde{\theta}^j\tilde{\theta}^j}, \; 
\cos(\tilde{\alpha})=
\tilde{\theta}^j \tilde{p}^j/(\tilde{\theta}\tilde{p})$ and just have to compute $\tilde{p}$
in terms of $p_j,p'_j$. Defining also $p=\sqrt{p_j p_j},\;
p'=\sqrt{p_j' p_j'},\; \cos(\beta)=p_j p_j'/(p p')$ we have
\ba \label{3.46}
HH'&=&[\ch(p/2)\ch(p'/2)+\cos(\beta)\sh(p/2)\sh(p'/2)] 1_2
\nonumber\\ && 
-i\tau_j[\ch(p/2)\sh(p'/2)\frac{p'_j}{p'}+\ch(p'/2)\sh(p/2)\frac{p_j}{p}
\nonumber\\
&& -i\frac{\epsilon_{jkl} p_k p'_l}{pp'}\sh(p/2)\sh(p'/2)]
\nonumber\\
(HH')^\dagger&=&[\ch(p/2)\ch(p'/2)+\cos(\beta)\sh(p/2)\sh(p'/2)]1_2
\nonumber\\ &&
-i\tau_j[\ch(p/2)\sh(p'/2)\frac{p'_j}{p'}+\ch(p'/2)\sh(p/2)\frac{p_j}{p}
\nonumber\\
&& +i\frac{\epsilon_{jkl} p_k p'_l}{pp'}\sh(p/2)\sh(p'/2)]
\ea
Taking the product of these two matrices we find $\tilde{H}^2$ from which
we could compute $\tilde{p}_j$ but it turns out that we only need
$\tilde{p}$ which we get from the trace
\ba \label{3.46a}
&& \mbox{tr}(HH'(HH')^\dagger)=2[\ch^2(p)\ch^2(p')+\sh^2(p)\sh^2(p')
+\ch^2(p)\sh^2(p')\nonumber\\
&& +\ch^2(p')\sh^2(p)+4\ch(p)\ch(p')\sh(p)\sh(p')\cos(\beta)]
\ea
which equals $2\ch(\tilde{p})$. Using hyperbolic identities and addition
theorems it is possible to cast (\ref{3.46a}) into the following form
\be \label{3.47}
\tilde{p}=\ach((1+c)\ch^2(\frac{p+p'}{2})+(1-c)\ch^2(\frac{p-p'}{2})-1)
\ee
where we have used that the $\ch$ function is invertible on the positive
real line and $c=\cos(\beta)$ takes values in $[-1,1]$. The minimum
of the argument of (\ref{3.47}) with respect to $c$ at fixed $p,p'$
is given at $c=-1$ which is still positive.

Recalling (\ref{3.19}), (\ref{3.20}) we find
\ba \label{3.47a}
\psi^{2t}_{\tilde{H}}(\tilde{h}) &=&
\frac{2\sqrt{\pi}e^{t/4}}{t^{3/2}}\frac{1}{\sqrt{x^2-1}}
\sum_{n=-\infty}^\infty (\ach(x)-2\pi i n)
e^{-\frac{(2\pi n+i\sach(x))^2}{t}}
\nonumber\\
\psi^{2t}_{H^2}(1) &=&
\frac{2\sqrt{\pi}e^{t/4}}{t^{3/2}}\frac{1}{\sqrt{y^2-1}}
\sum_{n=-\infty}^\infty (\ach(y)-2\pi i n)
e^{-\frac{(2\pi n+i\sach(y))^2}{t}} 
\nonumber\\
\psi^{2t}_{(H')^2}(1) &=&
\frac{2\sqrt{\pi}e^{t/4}}{t^{3/2}}\frac{1}{\sqrt{(y')^2-1}}
\sum_{n=-\infty}^\infty (\ach(y')-2\pi i n)
e^{-\frac{(2\pi n+i\sach(y'))^2}{t}} 
\ea
where $x=\ch(s+i\phi)=
\ch(\tilde{p}/2)\cos(\tilde{\theta})+i\sh(\tilde{p}/2)\sin(\tilde{\theta})
\cos(\tilde{\alpha})$
and $y=\ch(p),y'=\ch(p')$. Therefore the overlap function is given
by
\ba \label{3.48}
i^t(g,g') &=& \frac{1}{|x^2-1|}
|\sum_{n=-\infty}^\infty (\ach(x)-2\pi i n)
e^{-\frac{(2\pi n+i\sach(x))^2}{t}}|^2 \times\\
&\times &
\frac{1}{\frac{1}{\sqrt{y^2-1}}
\sum_{n=-\infty}^\infty (\ach(y)-2\pi i n)
e^{-\frac{(2\pi n+i\sach(y))^2}{t}}} 
\times\nonumber\\
&\times&
\frac{1}{\frac{1}{\sqrt{(y')^2-1}}
\sum_{n=-\infty}^\infty (\ach(y')-2\pi i n)
e^{-\frac{(2\pi n+i\sach(y'))^2}{t}}}
\nonumber\\
&=& e^{-\frac{1}{t}(p^2+(p')^2-\tilde{p}^2/2)}
\frac{
\frac{1}{|x^2-1|}
|\sum_{n=-\infty}^\infty (\ach(x)-2\pi i n)
e^{-\frac{(2\pi n+i\sach(x))^2+\tilde{p}^2/4}{t}}|^2 
}
{ D^t_p D^t_{p'}} \nonumber
\ea
where $D^t_p$ was defined in (\ref{3.22}).
Consider now the exponential in front of the fraction in (\ref{3.48}).
\begin{Lemma} \label{la3.6}
The function
\be \label{3.49}
\Delta^2(p,p',c):=p^2+(p')^2-\tilde{p}^2/2
\ee
is positive definite, vanishing if and only if $p_j=p'_j$.
\end{Lemma}
Proof of Lemma \ref{la3.6} :\\
Showing that $f\ge 0$ is equivalent to proving that
$\tilde{p}\le\sqrt{2[p^2+(p')^2]}$ or (recall (\ref{3.47}))
\be \label{3.50}
\ch(\sqrt{2[p^2+(p')^2]})\ge(1+c)\ch^2(\frac{p+p'}{2})
+(1-c)\ch^2(\frac{p-p'}{2})-1
\ee
for any $p,p'\ge 0$ and $c\in [-1,1]$.
The derivative with respect to $c$ of the right hand side of (\ref{3.50})
is given by $\ch^2(\frac{p+p'}{2})-\ch^2(\frac{p-p'}{2})$ which is positive 
unless $p=p'=0$
in which case the derivative vanishes. However, at $p=p'=0$ both sides
of (\ref{3.50}) equal $1$ so that we are left with the remaining case
that not both of $p,p'$ vanish in which case the right hand side is
strictly monotonously increasing with $c$. Thus, the right hand side
takes its maximum at $c=1$. Thus, (\ref{3.50}) will be true for all
$c$ given $p,p'$ if and only if it is true at $c=1$ in which case it becomes
\ba \label{3.51}
&& \ch(\sqrt{2[p^2+(p')^2]})\ge 2\ch^2(\frac{p+p'}{2})-1=\ch((p+p'))
\nonumber\\
&\Leftrightarrow&
2[p^2+(p')^2]\ge (p+p')^2 \Leftrightarrow (p-p')^2\ge 0
\ea
Thus, in both cases the inequality is true and becomes an equality only
if $p_j=p'_j$.\\
$\Box$\\
Unfortunately it is not possible to prove the more intuitive result
$\Delta^2\ge (p_j-p'_j)^2$, in fact one can show that the opposite inequality
$\Delta^2\le (p_j-p'_j)^2$ holds. Therefore we must live with the function 
$\Delta$ as a replacement for $(p_j-p'_j)^2$.\\
\\
Now consider the remaining factor in (\ref{3.48}). We see that we can
apply Lemma \ref{la3.5} to its numerator and Lemma \ref{la3.1} to its
denominator, the only difference being that we have to replace
$t$ by $2t$ in the final estimate. Therefore we immediately find the main
theorem of this subsection.
\begin{Theorem} \label{th3.4}
i) For all $0\le\phi\le 31\pi/32$ there exist positive constants
$K_t,K_t'$ (independent of $g,g'$), decaying exponentially fast 
to $0$ as $t\to 0$ such that
\be \label{3.52}
i^t(g,g')\le
\frac{\frac{|\ach(x)|^2}{|x^2-1|}
e^{-\frac{\Delta^2+2\tilde{\theta}^2+2\delta^2}{t}}(1+K_t')}
{\frac{p}{\sh(p)}(1-K_t)\frac{p'}{\sh(p')}(1-K_t)}
\ee
where $x=\ch(\tilde{p}/2)\cos(\tilde{\theta})+i\sh(\tilde{p}/2)
\sin(\tilde{\theta})\cos(\tilde{\alpha})$ and 
$\Delta^2=p^2+(p')^2-\tilde{p}^2/2$.\\
ii) For all $31 \pi/32\le\phi\le\pi$ there exist positive constants
$K_t K_t^{\prime\prime}$ (independent of $g,g'$), decaying exponentially fast 
to $0$ as $t\to 0$ such that
\be \label{3.53}
i^t(g,g')\le
\frac{\frac{|\ach(x)-i\pi|^2}{|x^2-1|}
e^{-\frac{\Delta^2+\tilde{\theta}^2+2\delta^2}{t}}(1+K_t^{\prime\prime})}
{\frac{p}{\sh(p)}(1-K_t)\frac{p'}{\sh(p')}(1-K_t)}
\ee
\end{Theorem}
By its very definition, the overlap function is at most unity at
$g=g'$ by the Schwarz inequality and otherwise sharply damped at
$g\not=g'$ as the theorem reveals. In fact, as in the previous section,
either $\delta$ grows as $\tilde{p}^2/4$ as $p,p'\to\infty$ which leads to
Gaussian damping or $\delta$ stays bounded in which case $s$ grows as
$\tilde{p}/2$. In the latter case $|\ach(x)|^2/|x^2-1|$ behaves
as $\tilde{p}^2/(4\sh(\tilde{p}))\propto \tilde{p}^2 e^{\tilde{p}}$ 
while the denominator in
theorem \ref{th3.4} contributes a factor of $\sh(p)\sh(p')/(pp')$.
Now the overlap function is still Gaussian damped due to
$\Delta\not=0$ unless $\vec{p}=\vec{p}'$
in which case the two factors just discussed cancel each other as one
or both of $p\sim p'$ get large.

\subsection{Peakedness in the Electric Field Representation}
\label{s3.3}

We first need to define what we even mean by ``the electric field
representation".
\begin{Definition} \label{def3.1}
i) Let $|jmn>$ be the state defined by\\ 
$<h,jmn>:=\int_G d\mu_H(h')
\delta(h,h')|jmn>(h')=<\delta_h,jmn>=\pi_j(h)_{mn}$ and let
$\psi\in L_2(G,d\mu_H)$ be any state. Then we define the electric
field representation of $\psi$ by
\be \label{3.54}
\tilde{\psi}(jmn):=<jmn,\psi>
\ee
that is, the electric field representation of $\psi$ is nothing else than its
``Fourier coefficients" with respect to the complete orthogonal
system $|jmn>$ normalized by $||\;|jmn>||^2=1/d_j$.\\
ii) The Peter\&Weyl theorem guarantees that $\psi\mapsto\tilde{\psi}$
is a unitary transformation between $L_2(G,d\mu_H)$ and the Hilbert space
$\ell_2$ of sequences $(c_{jmn})$ of complex numbers equipped with 
the inner product $<c,c'>=\sum_{jmn} d_j \overline{c_{jmn}}c'_{jmn}$.
\end{Definition}
We have defined the electric field representation for a general compact group
$G$ where $j$ is some discrete label for a complete system of representants
from each equivalence class of irreducible representations and $m,n$ labels
its matrix elements. For $SU(2)$ $j$ is a half integral non-negative integer
and $m,n$ take the $d_j=2j+1$ values $-j,-j+1,..,j$.

We easily calculate that
\be \label{3.55}
\tilde{\psi}^t_g(jmn)=e^{-tj(j+1)/2}\pi_j(g)_{mn}
\ee
and are interested in the probability amplitude
\be \label{3.56}
p^t_g(jmn):=\frac{|\tilde{\psi}^t_g(jmn)|^2}{||\psi^t_g||^2}
\ee
for the momentum of the particle to be in the configuration $jmn$ in the
state $\psi^t_g$. The precise relation between the classical numbers $p^j$
and the quantum numbers $jmn$ will become clear shortly.

We notice the following elementary estimates : Let $g=Hu=u(u^{-1}Hu)=uH'$
be the unique left and right polar decompositions of $g$. Define
$X_{m'}:=\pi_j(H)_{mm'},\overline{Y_{m'}}:=\pi_j(u)_{m' n},
X'_{m'}:=\pi_j(H')_{m'n},\overline{Y'_{m'}}:=\pi_j(u)_{mm'}$, then
by the Schwarz inequality 
\ba \label{3.57}
|\pi_j(g)_{mn}|&=&|\sum_{m'} \overline{Y_{m'}} X_{m'}|=|<Y,X>|\le
||Y||\; ||X|| \nonumber\\
|\pi_j(g)_{mn}|&=&|\sum_{m'} \overline{Y'_{m'}} X'_{m'}|=|<Y',X'>|\le
||Y'||\; ||X'||
\ea
where the inner product is the Hermitean inner product of the $d_j$
dimensional representation space corresponding to $\pi_j$. 
But $||Y||^2=||Y'||^2=1$
by the unitarity of $u$ while $||X||^2=\pi_j(H^2)_{mm}$ and
$||X'||^2=\pi_j((H')^2)_{nn}$ by the hermiticity of $H,H'$. We summarize
this observation in the following Lemma.
\begin{Lemma} \label{la3.7}
The matrix elements of $\pi_j(g)_{mn}$ have the factorizing bound
\be \label{3.58}
|\pi_j(g)_{mn}|^2\le \sqrt{\pi_j(H^2)_{mm}}\sqrt{\pi_j((H')^2)_{nn}}
\ee
for all $-j\le m,n\le j$ where $H,H'$ are the left and right polar
decompositions of $g=Hu=uH'$.
\end{Lemma}
This factorization property will be crucial later on when we project
the gauge-variant coherent states on a general graph to the gauge invariant
subspace of the Hilbert space.

In the considerations that follow we will again specialize to $G=SU(2)$.
The following Lemma, recalling (\ref{2.15}), justifies the name ``electric
field representation".
\begin{Lemma} \label{la3.8}
Let $^R p_e^j:=p_e^j$ as in section \ref{s2} and
$^L p_e^j:=-\frac{1}{2}\mbox{tr}(h_e\tau_j h_e^{-1} \tau_k)\;^R p_e^j$
(recall (\ref{2.4})). Then, dropping the label $e$, 
\ba \label{3.59}
^R\hat{p}_3 |jmn>&=&-it m |jmn> \nonumber\\
^L\hat{p}_3 |jmn>&=&-it n |jmn> \nonumber\\
(^R\hat{p}_j)^2 |jmn>&=&(^L\hat{p}_j)^2 |jmn>=+t^2 j(j+1) |jmn>
\ea
that is, the three operators $^R\hat{p}_3,\;^L\hat{p}_3,\;
(^R\hat{p}_j)^2$ are simultanuously diagonizable with $|jmn>$ as eigenstates.
Moreover, the magnetic quantum numbers $mt,nt$ have the interpretation of
the 3-component of $^R p_j$ and $^L p_j$ respectively while for large
$p$ the quantum number $jt$ has the interpretation of the norm of
$^R p_j$ which equals the norm of $^L p_j$.
\end{Lemma}
Proof of Lemma \ref{la3.8} :\\
The proof follows almost immediately from the fact that
$^R\hat{p}_j=-it\;^R X_j/2,\;^L\hat{p}_j=-it\;^L X_j/2$ where $^R X,\;^L X$
denotes the right or left invariant vector field on $G$ which certainly
commute with each other and their square gives four times the Laplacian.
The eigenvalues displayed can be easily computed from the fact
that $^R X_j=\frac{d}{ds}_{|s=0}L_{\exp(s\tau_j)}$ and
$^L X_j=\frac{d}{ds}_{|s=0}R_{\exp(s\tau_j)}$ where $R_h,L_h$ denote
right and left translation on $G$ and from the explicit matrix
element formula (\ref{3.13}) by expanding 
$\pi_j(\exp(s\tau_j)g)_{mn},\;\pi_j(g\exp(s\tau_j))_{mn}$ around
$s=0$.\\
$\Box$\\
We need the following lemma.
\begin{Lemma} \label{la3.9}
The functions $p^t_g(jmn)$ are bounded as $j,m,n\to\infty$
with a peak at $(j+1/2)t=p$ for all $m,n$.
\end{Lemma}
Proof of Lemma \ref{la3.9} :\\
Recalling Lemma \ref{la3.1} we have first of all
\be \label{3.60}
p^t_g(jmn)\le \frac{\sh(p)}{p}\frac{t^{3/2}e^{-t/4}}{2\sqrt{\pi}}
\frac{e^{-p^2/t}}{1-K_t}e^{-tj(j+1)}|\pi_j(g)_{mn}|^2
\ee
for some positive constant $K_t$ decaying exponentially to zero as
$t\to 0$.
We have the elementary estimate
\be \label{3.61}
|\pi_j(g)_{mn}|^2\le\sum_{mn} |\pi_j(g)_{mn}|^2 
=\chi_j(g g^\dagger)=\chi_j(H^2)=\frac{\sh((2j+1)p)}{\sh(p)}
\ee
and therefore after simple algebraic manipulations
\be \label{3.62}
p^t_g(jmn)\le \frac{1}{p}\frac{t^{3/2}}{4\sqrt{\pi}}
\frac{1}{1-K_t}e^{-\frac{[(j+1/2)t-p]^2}{t}}
\ee
for all $p\ge 1$, say, and, using again Lemma \ref{la3.0}, we find
\be \label{3.63}
p^t_g(jmn)\le (2j+1)\frac{t^{3/2}}{2\sqrt{\pi}}
\frac{1}{1-K_t}e^{-\frac{[(j+1/2)t-p]^2}{t}}
\ee
for all $0\le p\le 1$. From these estimates peakedness is obvious at the
value claimed.\\
$\Box$\\
\\
Up to now all estimates were for general $p$. From now on we restrict
attention to large $p$ (that is, of order unity or larger) as it is of
interest in applications to semi-classical approximations. As the
probability amplitude is then small, according to the previous lemma,
unless $jt\approx p$, we can restrict attention to the case that
$j$ is large in what follows.

The next theorem is the main result of this subsection.
\begin{Theorem} \label{th3.5}
The diagonal matrix elements $\pi_j(H)_{mm},\pi_j(H')_{nn}$ are
for large $p,p',j$ peaked at $m/j=p_3/p$ and $n/j=p'_3/p'$ where
$H=\exp(-ip_j\tau_j/2),\;H'=\exp(-ip'_j\tau_j/2)$. The maximal value
of $\pi_j(H)_{mm}$ at $m\approx [p_3/p] j$ is given by $\approx e^{pj}$.
\end{Theorem}
Proof of Theorem \ref{th3.5} :\\
We display the proof for $H$, the one for $H'$ is identical.\\
We will discuss separately the following two cases :\\
\\
Case I) $|p_3/p|<1$ : \\
Employing the explicit formula (\ref{3.13}) at $m=n$ we find, using
$ad-bc=1$
\be \label{3.64}
\pi_j(H)_{mm}=(ad)^j (\frac{a}{d})^m\sum_l
\left( \begin{array}{c} j+m\\ l \end{array} \right)
\left( \begin{array}{c} j-m\\ l \end{array} \right) [1-\frac{1}{ad}]^l
\ee
where, as usual, the sum over $l$ is over all integers such that no 
factorials
have negative arguments. Since $a=\ch(p/2)+\sh(p/2)p_3/p,
d=\ch(p/2)-\sh(p/2)p_3/p$
we have $ad=\ch^2(p/2)-\sh^2(p/2)(p_3/p)^2$ which is large if $p$ is large
unless $p_3\approx \pm p$ which we excluded.

For large $p$ we can therefore replace $1-1/(ad)$ by $1$ and can
use the addition theorem for binomial coefficients
$$
\sum_l \left( \begin{array}{c} \alpha\\ l \end{array} \right)
\left( \begin{array}{c} \beta\\  \gamma-l \end{array} \right)
=\left( \begin{array}{c} \alpha+\beta\\ \gamma \end{array} \right)
$$
to arrive at
\be \label{3.65}
\pi_j(H)_{mm}\approx (ad)^j (\frac{a}{d})^m
\left( \begin{array}{c} 2j \\ j-m \end{array} \right)
\ee
For large $j$ to which we have focussed attention to, and if also
$j\pm m$ are large, more precisely, if $|m/j|<1$, we can apply the crudest
version of Stirling's formula $n!\approx (n/e)^n$ to estimate the
factorials. Introducing the abbreviations $s=p_3/p, t=m/j$ we have
$ad=\ch^2(p/2)-s^2\sh^2(p/2)\approx \exp(p)(1-s^2)/4, 
a/d\approx (1+s)/(1-s)$,
thus
\ba \label{3.66}
\pi_j(H)_{mm}&\approx&
(ad)^j (\frac{2j}{e})^{2j} (a/d)^m
\frac{e^{2j}}{(j+m)^{j+m} (j-m)^{j-m}}
\nonumber\\
&\approx& \frac{e^{pj}}{4^j} (1-s^2)^j (2j)^{2j} (\frac{1+s}{1-s})^m
\frac{1}{(1+t)^{j+m} (1-t)^{j-m} j^{2j}}
\nonumber\\
&=& e^{pj} (1-s^2)^j  (\frac{1+s}{1-s})^{jt}
\frac{1}{(1+t)^{j(1+t)} (1-t)^{j(1-t)}}
\nonumber\\
&=& e^{pj}(\frac{1-s^2}{1-t^2})^j (\frac{(1+s)(1-t)}{(1-s)(1+t)})^{jt}
\nonumber\\
&=:& e^{pj} (1-s^2)^j e^{jf(t)}
\ea
Let us compute the extrema of the function $f(t)$. We have
\be \label{3.67}
\dot{f}=\frac{2t}{1-t^2}+\ln(\frac{(1+s)(1-t)}{(1-s)(1+t)})
-t(\frac{1}{1-t}+\frac{1}{1+t})=
\ln(\frac{(1+s)(1-t)}{(1-s)(1+t)})
\ee
which vanishes precisely at $t=s$. Moreover,
$\frac{d^2 f}{dt^2}=-\frac{2}{1-t^2}<0$ for all $t\in [0,1]$, thus $t=s$
is the only local and therefore the global maximum. We conclude that
$f(t)\le f(s)$ and thus $\pi_j(H)_{mm}\le e^{jp}$ where the maximum
is taken at $m/j=p_3/p$. Notice that our intermediate assumption that 
$|m/j|<1$ is justified in retrospect as well. Expanding $f$ around $t=s$
we get $f(t)=f(s)+f^{\prime\prime}(s)(t-s)^2+o((t-s)^3)=
-\ln(1-s^2)-\frac{(t-s)^2}{1-s^2}$ so that 
\be \label{3.67a}
\pi_j(H)_{mm}\approx e^{2jp} e^{-j\frac{(m/j-p_3/p)^2}{1-(p_3/p)^2}}
\ee
Case II) $|p_3/p|\approx 1$ :\\
In this case $p_1/p=p_2/p\approx 0$ and the sum over $l$ in (\ref{3.64})
collapses to a single term $l=0$ and we find
\be \label{3.68}
\pi_j(H)_{mm}\approx (ad)^j (\frac{a}{d})^m
\ee
Since $a/d=(\ch(p/2)+s\sh(p/2))/(\ch(p/2)-s\sh(p/2))=\exp(sp)$ for 
$s\approx \pm 1$
while $ad=\ch^2(p/2)-s^2\sh^2(p/2)\approx 1$ we get
\be \label{3.69}
\pi_j(H)_{mm}\approx  e^{smp}
\ee
which obviously takes its maximum at $m=s j$, that is, $t=m/j=s$ again.
The maximum value is given by $\pi_j(H)_{mm}=e^{jp}$. Thus
\be \label{3.69a}
\pi_j(H)_{mm}\approx e^{jp} e^{jp(sm/j-1)}\approx e^{jp} 
e^{-jp|m/j-p_3/p|}
\ee
$\Box$\\
\\
Notice that at $m=p_3/p j$ we have $\pi_j(H)_{mm}\approx e^{2pj}$ while 
we have shown already that $|\pi_j(g)_{mn}|\le \sqrt{\chi_j(H^2)}\approx
e^{(j+1/2)p}$. This means that for large $p,j$ the function 
$|\pi_j(H)_{mn}|\le\sqrt{\pi_j(H^2)_{mm}}\le e^{pj}$ is indeed 
concentrated at $m=n\approx j p_3/p$. This can be shown explicitly 
by repeating the above analysis and varying besides $m$ also $n$.

We summarize the results of this subsection in the following theorem.
\begin{Theorem} \label{th3.6}
The probability amplitude $p^t_g(jmn)$ is, for large $p$, peaked at 
$jt\approx p$ and $mt\approx p^R_3,nt\approx p^L_3$. More precisely,
there exists a constant $K_t$ exponentially vanishing as $t\to 0$ 
and independent of $g$ such that
\ba \label{3.70}
p^t_g(jmn) &\stackrel{<}{\sim}& \frac{1}{p}\frac{t^{3/2}}{4\sqrt{\pi}}
%\lesssim,\le\approx
\frac{1}{1-K_t}
e^{-j/2\frac{(m/j-(^R p_3)/p)^2}{1-(^R p_3/p)^2}}
e^{-j/2\frac{(n/j-(^L p_3)/p)^2}{1-(^L p_3/p)^2}}
e^{-\frac{[(j+1/2)t-p]^2}{t}}\nonumber\\
&& \mbox{ if } |^{R/L}p_3/p|<1
\nonumber\\
p^t_g(jmn) &\stackrel{<}{\sim}& \frac{1}{p}\frac{t^{3/2}}{4\sqrt{\pi}}
\frac{1}{1-K_t}
e^{-jp|m/j-(^R p)_3/p|}e^{-jp|n/j-(^L p)_3/p|}
e^{-\frac{[(j+1/2)t-p]^2}{t}}\nonumber\\
&& \mbox{ if } 
|^{R/L}p_3/p| \stackrel{<}{\sim} 1 \ea
\end{Theorem}
The careful reader will notice a seemingly crucial difference between the 
configuration and momentum
representation : While the peak in the configuration representation
grows with $t\to 0$, in the momentum representation it sinks with $t\to 0$.
However, this is only apparently so : notice that the configuration Hilbert 
space is an $L_2$ space since the operators $\hat{h}_{AB}$ have continuous
spectrum while the momentum Hilbert space is an $\ell_2$ space since the 
operators $^R\hat{p}_j,\;^L\hat{p}_j,\hat{p}_j^2$ have discrete spectrum.
Now let $\xi^t_g=\psi^t_g/||\psi^t_g||$ then 
\be \label{3.70a}
1=||\xi^t_g||^2=\sum_{jmn} p^t_g(jmn)
=\sum_{jmn} \Delta p_j\; \Delta^R p_m \;\Delta^L p_n \frac{p^t_g(jmn)}{t^3/2}
\ee 
where $p_j=jt,\;^R p_m=mt,\;^L p_n=nt$ with $2j\in\Nl,\;j-m,j-n\in \Zl,\;
|m|,|n|\le j$ and therefore $2\Delta p_j=\Delta^R p_m=\Delta^L p_n=t$.
It follows from (\ref{3.70}) that $p^t_g(jmn)=
\tilde{p}^t_g(p_j,^R p_m, ^L p_n)$ evidently depends only on
$p_j,^R p_m,^L p_n$ and thus (\ref{3.70a}) is a Riemann sum, as $t\to 0$,
approximating the integral
\be \label{3.70b}
1=\int_0^\infty dp_j 
\int_{-p_j}^{p_j} d^R p_m \int_{-p_j}^{p_j} d^L p_m 
\frac{\tilde{p}^t_g(p_j,^R p_m, ^L p_n)}{2 t^3}
\ee
In other words, as $t\to 0$ the momentum spectrum approaches a continuum one 
and the corresponding propability amplitude is up to a constant factor given 
by (\ref{3.70}) divided by $t^3$ whose peak evidently also grows with
$t\to 0$ just as in the configuration representation. Thus, the apparent 
difference of the peak behaviour for the two
representations is absent in the limit $t\to 0$ if we use a contiuum 
spectrum approximation.

The trick to use an approximate continuum momentum representation as in
(\ref{3.70a}), (\ref{3.70b}) will be used in the proof of Ehrenfest theorems 
in \cite{33}. In particular, we see from the explicit expression
(\ref{3.70}) that $\tilde{p}^t_g(p_j,^R p_m,^L p_n)\to
\delta(p(g),jt)\delta(^R p(g),mt)\delta(^L p(g),nt)$ approaches a 
$\delta$ distribution with respect to the measure (\ref{3.70b}).

\subsection{Uncertainty Relation and Phase Space Bounds}
\label{s3.4}

In this subsection we will compute explictly the Heisenberg uncertainty
bound for the operators $\hat{g}_{AB}$, verify that it corresponds to the 
bound to be expected from the Poisson bracket $\{g_{AB},\overline{g_{AB}}\}$
and finally will see explicitly that the overlap function $i^t(g,g')$ times
$1/t^3$ can be interpreted as the probability density to find the system
at the phase space point $g'$ in the state $\hat{U}_t\psi^t_g$ with respect
to the Liouville measure on phase space. 

We will first need the so-called averaged heat kernel measure $\nu_t$ on 
$G^\Co$ which one can obtain either by the methods derived in 
\cite{55} (and advertized in \cite{39a}) which are specific 
for the heat kernel coherent states or by the more general method derived in
\cite{38} for an arbitrary family of coherent states. We will give a 
direct derivation below for $SU(2)$ as we wish to be as explicit as possible.
\begin{Lemma} \label{la3.10}
The measure $\nu_t$ underlying the map defined in (\ref{3.4}) is given 
for $G=SU(2)$ by
\be \label{3.71}
d\nu_t(g):=d\mu_H(u) d\sigma_t(H):=d\mu_H(u)
[\frac{2\sqrt{2} e^{-t/4}}{(2\pi t)^{3/2}} 
\frac{\sh(p)}{p} e^{-p^2/t} d^3p]=\nu_t(g)d\Omega
\ee
where $g=Hu$ is the polar decomposition of $g$, $d^3p$ is the 
standard Lebesgue measure on $\Rl^3$ and $d\Omega=d\mu_H d^3p$ is the 
Liouville measure on $T^\ast G$.
\end{Lemma}
Proof of Lemma \ref{la3.10} :\\
First of all, to see that $d\Omega(g):=d\mu_H(u)d^3p$
with $g=Hu,\;H=\exp(-i\tau_g p_j/2)$ is the Liouville measure
on $G^\Co\cong T^\ast G$ for the case $G=SU(2)$ (up to normalization) 
it is heplful to think of $SU(2)$ as the sphere $S^3$. The phase space 
$\hat{N}= T^\ast G$ can then be thought of as the symplectic reduction of the  
phase space $N=T^\ast \Rl^4$ under the co-isotropic constraint 
$C:=(x^1)^2+(x^2)^2+(x^3)^2+(x^4)^2-1$. Writing the symplectic structure
$\omega=\sum_{I=1}^4 dP_I\wedge dx^I$ on $N$ in terms of radial and polar
coordinates defined by 
$$
\vec{x}=:r\vec{n}:=r(\sin(\theta)\sin(\phi)\cos(\varphi),
\sin(\theta)\sin(\phi)\sin(\varphi),
\sin(\theta)\cos(\phi),
\cos(\theta))
$$ 
with $r\in [0,\infty);\; \theta,\phi\in [0,\pi];\;
\varphi\in [-\pi,\pi]$ as well as adapted normal and tangential (to  
$S^3$) momenta defined by $P^\perp_I:=(P_J n^J) n^I,
P^\parallel_I:=P_I-P^\perp_I$ repectively (the latter of which
are Dirac observables) and then pulling it back to the constraint
surface $C=0$ gives the Liouville measure on $T^\ast S^3\times \Rl^3$
which is a product measure on $S^3$ times the Lebesgue measure on $\Rl^3$.
The same measure on $S^3$ can be obtained as the effective measure induced 
by $\delta(C)d^4x$ which is obviously proportional to the Haar measure  
on $SU(2)$ as it is invariant under $SO(4)\cong SU(2)\times SU(2)$
(i.e. left and right translations).\\
\\
We now must verify the isometry relation 
\be \label{3.72}
<\hat{U}_t\psi,\hat{U}_t\psi'>_{\nu_t}=<\psi,\psi'>_{\mu_H}
\ee
for any two $\psi,\psi'\in L_2(G,d\mu_H)$. It will be sufficient to check 
this on a basis, say the basis $|jmn>$ introduced in the previous 
subsection for which $(\hat{U}_t|jmn>)(g)=e^{-tj(j+1)/2}\pi_j(g)_{mn}$. 
Using the polar decomposition and writing $\pi_j(g)=\pi_j(H)\pi_j(u)$
we see that we can take advantage of the orthogonality relations of the
$\pi_j(u)_{mn}$ under Haar measure if we make the ansatz
$d\nu_t(g)=d\mu_H(u) d\sigma_t(H)$. Thus, 
choosing $\psi=|jmn>,\psi'=|j'm'n'>$, we immediately find the condition  
\be \label{3.73}
\delta_{mm'}e^{tj(j+1)}=\int d\sigma_t(H) \pi_j(H^2)_{m'm}
\ee
for all $j,m,m'$. We can produce the required Kronecker $\delta$ on 
the right hand side of (\ref{3.73}) if we choose the measure $d\sigma_t(H)$
to be invariant under $SO(3)$, the homomorphic image of $SU(2)$ under the  
vector (or spin $1$) representation because in that case 
\ba
&& \pi_j(u)_{nm'}
[\int d\sigma_t(H)\pi_j(H^2)_{m'm}]
=[\int d\sigma_t(H)\pi_j((uHu^{-1})^2)_{nm'}]\pi_j(u)_{m'm}
\nonumber\\
&=& [\int d\sigma_t(H)\pi_j(H^2)_{nm'}]\pi_j(u)_{m'm}
\nonumber
\ea
that is, the matrix $A_{m'm}=\int d\sigma_t(H)\pi_j(H^2)_{m'm}$
commutes with the irreducible representation $\pi_j$ of $SU(2)$ and 
is therefore proportional to the unit matrix by one of Schur's lemmata.
We therefore are led to the ansatz $d\sigma_t(H)=d^3p f_t(p)$ where
the positive function $f_t(p)$ only depends on $p=\sqrt{p_j p_j}$ and 
$H=\exp(-ip_j\tau_j/2)$ as before. Then 
$A_{m'm}=\mbox{tr}(A)\delta_{mm'}/d_j$ and we are left with the condition 
that
\be \label{3.74}
(2j+1)e^{tj(j+1)}=\int_{\Rl^3} d\sigma_t(H) \chi_j(H^2)
=4\pi\int_0^\infty p^2 dp f_t(p) \frac{\sh((2j+1)p)}{\sh(p)}
\ee
for all $j$. We see that we can produce the $(2j+1)$ factor on the right
hand side of (\ref{3.74}) if we can do an integration by parts. We therefore
write $f_t(p)=g'_t(p)\sh(p)/p^2$ and find the condition
\be \label{3.75}
e^{tj(j+1)}=-4\pi\int_0^\infty dp g_t(p) \ch((2j+1)p)
\ee
provided that $g_t(p)$ is finite at $p=0$ where $\sh((2j+1)p)$ vanishes 
and that $g_t(p)$ decays faster at infinity than $\sh((2j+1)p)$. Finally,
assuming that $g_t(p)=g_t(-p)$ is invariant under reflection we find the 
condition
\be \label{3.76}
e^{\frac{t}{4}(2j+1)^2-t/4}=-2\pi\int_\Rl dp g_t(p) \exp((2j+1)p)
\ee
which we recognize as the moment problem for a Gaussian. Thus we define
$g_t(p)=-k_t\exp(-s p^2/2)$ and find
\be \label{3.77}
e^{\frac{t}{4}(2j+1)^2-t/4}
=2\pi k_t/\sqrt{s} \int_\Rl dx e^{-x^2/2} \exp((2j+1)x/\sqrt{s}) 
=\sqrt{2\pi}^3 k_t/\sqrt{s} e^{\frac{(2j+1)^2}{2s}}
\ee
from which we read off $s=2/t,\; 
k_t=\frac{e^{-t/4}\sqrt{2/t}}{\sqrt{2\pi}^3}$. Notice that $g_t$
is indeed finite at $p=0$, decays faster than any exponential of $p$ at
$\infty$ and is reflection invariant.
Therefore
\ba \label{3.78}
d\sigma_t(H)&=& f_t(p) d^3p=\sh(p)/p^2 g_t'(p) d^3p
=-k_t\sh(p)/p^2(-2p e^{-p^2/t}/t) d^3p \nonumber\\
&=& \frac{\sh(p)}{p}\frac{2\sqrt{2} e^{-t/4}}{(2\pi t)^{3/2}}e^{- p^2/t}
d^3p
\ea
$\Box$\\
\\
The next Lemma is sometimes called the reproducing kernel property and holds
completely generally for any system of coherent states defined by a  
complexifier \cite{32}.
We will state and prove it only for the group case for general $G$ (see
\cite{39a} for more details).
\begin{Lemma} \label{la3.11}
The coherent state transform of a coherent state at the value $g'$
is the same as taking inner products in the $L_2$ Hilbert
space with the coherent state with label $(g')^\star$ where
$g^\star=(g^{-1})^\dagger$ 
%if $G$ is unitary) 
is the unique involution on $G^\Co$ 
with the property that $g^\star=g$ if and only if $g\in G$. That is,
\be \label{3.79}
(\hat{U}_t\psi^t_g)(g')=\psi^{2t}_g(g')=
<\psi^t_{(g')^\star},\psi^t_g>_{\mu_H}
\ee
\end{Lemma}
Proof of Lemma \ref{la3.11} :\\
The proof is trivial. We have 
\be \label{3.80}
<\psi^t_{g'},\psi^t_g>=\sum_\pi d_\pi e^{-t\lambda_\pi}
\chi_\pi(g (g')^\dagger)
\ee
while 
\be \label{3.81}
(\hat{U}_t\psi^t_g)(g')=\sum_\pi d_\pi e^{-t\lambda_\pi}\chi_\pi(g (g')^{-1})
=\psi^{2t}_g(g')
\ee
$\Box$\\
Notice that in the polar decomposition of $g=Hu$ we have $g^\star=H^{-1}u$
which corresponds to $p_j\to-p_j,\; u\to u$ as it should be.

The following theorem clarifies the meaning of the overlap function of 
subsection \ref{s3.2}. We will do this here only for $SU(2)$. The statement
for general $G$ can be found in \cite{39b}.
\begin{Theorem} \label{th3.7}
The overlap function $i^t(g,g')$ approaches exponentially fast with $t\to 0$ 
the function $p^t(g,(g')^\star)\frac{\pi t^3}{2}$ where 
$p^t(g,g')$ denotes the probability density to find the system at the 
phase space point $g'$ in the state $\hat{U}_t\psi^t_g$ with respect to
the Liouville measure on the phase space $T^\ast G$.
\end{Theorem} 
Proof of Theorem \ref{th3.7} :\\
The probability density of the image of the normalized coherent state 
$\psi^t_g$
under the coherent state transform at the phase space point $g'$ in the 
Bargmann-Segal Hilbert space with respect to the Liouville measure
$d\Omega=d\mu_H(u)d^3 p$ on $T^\ast G$ is given by
\be \label{3.82}
p^t(g,g')=\nu_t(g')\frac{|(\hat{U}_t\psi^t_g)(g')|^2}{||\psi^t_{g}||^2}
\ee
where we have used the isometry property of $\hat{U}_t$, that is, the norm
in the denominator of (\ref{3.82}) can be computed in either Hilbert space.
Using Lemma \ref{la3.11} and the definition of the overlap function
we have 
\ba \label{3.83}
p^t(g,g') &=&[\nu_t(g') \; ||\psi^t_{(g')^\star}||^2]
\;\frac{|<\psi^t_{(g')^\star},\psi^t_g>|^2}
{||\psi^t_{g}||^2 ||\psi^t_{(g')^\star}||^2}
\nonumber\\
&=&[\nu_t(g') \; ||\psi^t_{g'}||^2] i^t(g,(g')^\star)
\ea
where we have used the fact that $||\psi^t_g||$ depends on $p$ only.
Now using Lemma \ref{la3.1} and the explicit expression for $\nu_t(g')$
given in (\ref{3.71}) we find for the factor multiplying $i^t(g,(g')^\star)$
in (\ref{3.83}) the estimate
\be \label{3.84}
%[\frac{8\sqrt{2} e^{-t/4}}{(2\pi t)^{3/2}} 
%\frac{\sh(2p)}{p} e^{-4p^2/t}] from \nu_t
%[\frac{2\sqrt{\pi}e^{t/4}}{t^{3/2}}\frac{2p}{\sh(2p)}e^{4p^2/t}]
% from ||\psi^t_g||^2
\frac{2}{\pi t^3}(1-K_t)
\le \frac{p^t(g,g')}{i^t(g,(g')^\star)}
\le \frac{2}{\pi t^3}(1+\tilde{K}_t)
\ee
for some constants $K_t,\tilde{K}_t$, independent of $g,g'$, exponentially 
vanishing with $t\to 0$. \\
$\Box$\\
Since $i^t(g,g')$
is peaked at $g=g'$ where it equals unity and has a decay width of order 
$\sqrt{t}$ we conclude that like for a particle moving in $\Rl^3$ the phase 
space volume occupied by a coherent state with respect to the measure
$d\mu_H(h) d^3p$ is given by $\propto (2\pi t)^3\propto \hbar^3$. 
In particular, we obtain the interpretation that the normalized coherent 
states with label $g$ in the Bargmann-Segal Hilbert space are concentrated 
at the phase space point $g^\star$ with respect to the Liouville measure.
If we would have defined the map $\hat{U}_t$ through heat kernel evolution
followed by {\it antianalytical extension}, i.e. 
$(\hat{U}_t\psi)(g):=(\hat{W}_t\psi)(h)_{h\to g^\star}$ (which for 
$g\in G$ does not make any difference) then the coherent state labelled 
by $g$ in the Bargmann-Segal Hilbert space is concentrated at $g$ since 
the measure $d\nu_t(g)=d\nu_t(g^\star)$ is involution invariant and so
the unitarity and peakedness properties are preserved. With this 
definition of $\hat{U}_t$ the strange asymmetry $g\leftrightarrow g^\star$ is 
removed and we assume this to have been done from now on.\\
\\
Let us now compute the actual uncertainty bound. By the Heisenberg
uncertainty relation for the self-adjoint operators 
\be \label{3.85}
\hat{x}_{AB}:=\frac{1}{2}(\hat{g}_{AB}+(\hat{g}_{AB})^\dagger),\;
\hat{y}_{AB}:=\frac{1}{2i}(\hat{g}_{AB}-(\hat{g}_{AB})^\dagger)
\ee
where $(.)^\dagger$ means the adjoint with respect to $L_2(G,d\mu_H)$
we have for any $A,B\in\{-1/2,1/2\}$ and for any state
\be \label{3.86}
<(\Delta\hat{x}_{AB})^2>^{1/2}<(\Delta\hat{y}_{AB})^2>^{1/2}
\ge \frac{|<[\hat{g}_{AB},\hat{g}_{AB}^\dagger]>|}{4}
\ee
and for coherent states the bound is saturated with equal contributions from
$\hat{x},\hat{y}$. From this we conclude easily that
\be \label{3.87}
\frac{|\sum_{A,B} <[\hat{g}_{AB},\hat{g}_{AB}^\dagger]>|}{4}
\le \sum_{A,B}
\frac{|<[\hat{g}_{AB},\hat{g}_{AB}^\dagger]>|}{4}
\le 4\mbox{max}_{A,B}
<(\Delta\hat{x}_{AB})^2>^{1/2}<(\Delta\hat{y}_{AB})^2>^{1/2}
\ee
We will compute the quantity on the left hand side of (\ref{3.87}) 
instead of the individual bounds as this is easier and because it 
gives a uniform bound. It also gives an idea of the individual bounds
because they are all of the same order as one can easily check.

We begin with the computation of the Poisson brackets (remember that
$g=Hh$)
\ba \label{3.88}
\{g_{AB},\overline{g_{AB}}\}&=&
\frac{\partial H_{AC}}{\partial p_j}\{p_j,h^{-1}_{BD}\} h_{CB} H_{DA}
-\frac{\partial H_{DA}}{\partial p_j}\{p_j,h_{CB}\} h^{-1}_{BD} H_{AC}
\nonumber\\
&& +\{p_j,p_k\}\frac{\partial H_{AC}}{\partial p_j} h_{CB}
\frac{\partial H_{DA}}{\partial p_k} h^{-1}_{BD}
\ea
and using the symplectic structure given in (\ref{2.7}) we find
\ba \label{3.89}
\sum_{A,B} \{g_{AB},\overline{g_{AB}}\}&=&
-\frac{\kappa}{a}\frac{\partial}{\partial p_j}\mbox{tr}(H^2\tau_j)
\nonumber\\
&=&2i\frac{\kappa}{a}(\ch(p)+2\frac{\sh(p)}{p})
\ea
where we have made use of $H^2=\ch(p)-i\tau_j \frac{p_j}{p}\sh(p)$. 
Notice that
the right hand side of (\ref{3.89}) depends on the phase space point which
is different from the situation with $T^\ast \Rl$.\\

We now compare this with the expectation value of the sum of commutators
with respect to the normalized coherent state $\psi^t_g$. In order to do
this we need the following Lemma about the Clebsch-Gordan decomposition.
\begin{Lemma} \label{la3.12}
For any $g\in SL(2,\Co)$ we have 
\ba \label{3.90}
g_{A_0 B_0} \pi_j(g)_{A_1..A_{2j},B_1..B_{2j}}&=&
\pi_{j+1/2}(g)_{A_0..A_{2j},B_0..B_{2j}}\nonumber\\
&&
-\frac{d_{j-1/2}}{d_j} \epsilon_{A_0 (A_1}
\pi_{j-1/2}(g)_{A_2..A_{2j}),(B_2..B_{2j}}\epsilon_{B_1)B_0}
\ea
where all A's and B's take the values $\pm 1/2$, $\epsilon_{AB}$ is the 
totally skew tensor density of weight minus one in two dimensions and 
$(.)$ denotes 
total symmetrization of indices to be taken as an idempotent operation.
\end{Lemma}
The proof requires elementary linear algebra and is left to the reader.
One uses the fact that the space of totally symmetric spinors of rank $2j$
provide the representation space of the irreducible representation of
spin $j$ of $SU(2)$, that is, 
$\pi_j(g)_{A_1..A_{2j},B_1..B_{2j}}
=\pi_j(g)_{A_1+..+A_{2j},B_1+..+B_{2j}}$ in terms of the former notation 
with magnetic quantum numbers.

We now use the fact that 
\be \label{3.91}
\hat{g}_{AB}=e^{t\Delta/2}\hat{h}_{AB} e^{-t\Delta/2},\;
(\hat{g}_{AB})^\dagger=e^{-t\Delta/2}(\hat{h}^{-1})_{BA} e^{t\Delta/2}
\ee
and that $g^{-1}=\epsilon g^T\epsilon^{-1}$ for any $g\in SL(2,\Co)$.
The computations are rather tedious and lengthy. We will not display all 
the steps but only the main stations of the calculation which 
require frequent use of Lemma \ref{la3.12} and relabelling of indices.
One first checks that indeed 
\be \label{3.92}
(\hat{g}_{AB}\psi_g^t)(h)=g_{AB} \psi^t_g(h)
\ee
so that we have easily
\be \label{3.93}
<\psi^t_g,\sum_{A,B} (\hat{g}_{AB})^\dagger\hat{g}_{AB}\psi^t_g>
=\mbox{tr}(g^\dagger g)||\psi^t_g||^2
\ee
where the dagger in the last line denotes the matrix adjoint. Using the 
$SL(2,\Co)$ Mandelstam identity $\mbox{tr}(g)\chi_j(g)=
\chi_{j+1/2}(g)+\chi_{j-1/2}(g)$ which one derives from Lemma \ref{la3.12} 
one can write (\ref{3.93}) in the equivalent form 
\be \label{3.94}
<\psi^t_g,\sum_{A,B} (\hat{g}_{AB})^\dagger\hat{g}_{AB}\psi^t_g>
=\sum_j d_j e^{-t\lambda_j}(\chi_{j+1/2}(H^2)+\chi_{j-1/2}(H^2))
\ee
On the other hand, tedious calculations reveal that
\be \label{3.95}
<\psi^t_g,\sum_{A,B} \hat{g}_{AB}(\hat{g}_{AB})^\dagger\psi^t_g>
=\sum_j d_j e^{-t\lambda_j} \chi_j(H^2)
[\frac{d_{j+1/2}}{d_j}e^{-t(\lambda_j-\lambda_{j+1/2})}
-\frac{d_{j-1/2}}{d_j}e^{-t(\lambda_j-\lambda_{j-1/2})}]
\ee
Taking the difference of (\ref{3.95}) and (\ref{3.94}) and relabelling
summation indices one arrives at
\ba \label{3.96}
&& <\psi^t_g,\sum_{A,B}[\hat{g}_{AB},(\hat{g}_{AB})^\dagger]\psi^t_g>
\nonumber\\
&=&2\sum_j e^{-t\lambda_j} \chi_j(H^2)
[d_{j-1/2}\sh(t(j+1/4))-d_{j+1/2}\sh(t(j+3/4))]
\ea
Introducing the parameter $T=\sqrt{t}/2$ and the function\\
$f(x)=\exp(-x^2)(x-T)\sh(px/T)\sh(2Tx-T^2)$ one can cast (\ref{3.96})
in a form suitable for an appeal to the Poisson summation formula
\be \label{3.96a}
<\psi^t_g,\sum_{A,B}[\hat{g}_{AB},(\hat{g}_{AB})^\dagger]\psi^t_g>
=2\frac{e^{t/4}}{T\sh(p)}\sum_{n=-\infty}^\infty f(nT) 
\ee
Computing the Fourier transform of $f$ and applying the Poisson summation
formula we end up with
\ba \label{3.96b}
&& <\psi^t_g,\sum_{A,B}[\hat{g}_{AB},(\hat{g}_{AB})^\dagger]\psi^t_g>
=\frac{4\sqrt{\pi}e^{t/4}}{t^{3/2}\sh(p)}\sum_{n=-\infty}^\infty
\times\nonumber\\
&\times& \{(p/2-i\pi n)e^{-t/4} e^{-4(\pi n +i(p/2+T^2))^2/t}
-(p/2+i\pi n+2T^2)e^{t/4} e^{-4(\pi n -i(p/2+T^2))^2/t}
\nonumber\\
&& +(-p/2+i\pi n+2T^2)e^{t/4} e^{-4(\pi n +i(p/2-T^2))^2/t}
+(p/2+i\pi n)e^{-t/4} e^{-4(\pi n -i(p/2-T^2))^2/t}\}
\nonumber\\ &&
\ea
Recalling (\ref{3.20}) that
\be \label{3.97}
||\psi^t_g||^2=
\frac{4\sqrt{\pi}e^{t/4}}{t^{3/2}}\frac{1}{\sh(p)}
\sum_{n=-\infty}^\infty (p/2-i\pi n)
e^{-4\frac{(\pi n+ip/2)^2}{t}} 
\ee
we see that the prefactors in front of the sums in (\ref{3.96}) and 
(\ref{3.97}) equal each other and an analysis similar to that which has 
led to Lemma \ref{la3.1} reveals that there exist positive constants 
$K_t,K'_t,\tilde{K}_t,\tilde{K}'_t$ exponentially vanishing with $t\to 0$ 
such that 
\ba \label{3.98}
&& 2\frac{\ch(p)(1-e^{t/2})-t\frac{\sh(p)}{p}-\tilde{K}'_t}{1+K'_t}
\nonumber\\ &\le&
\frac{<\psi^t_g,\sum_{A,B}[\hat{g}_{AB},(\hat{g}_{AB})^\dagger]\psi^t_g>}
{||\psi^t_g||^2}
\nonumber\\ &\le&
2\frac{\ch(p)(1-e^{t/2})-t\frac{\sh(p)}{p}+\tilde{K}_t}{1-K_t}
\ea
We conclude that (recall (\ref{3.89}))
\ba \label{3.99}
\frac{<\psi^t_g,\sum_{A,B}[\hat{g}_{AB},(\hat{g}_{AB})^\dagger]\psi^t_g>}
{||\psi^t_g||^2}
&=&-t(\ch(p)+2\sh(p)/p)[1+O(t^2)] \nonumber\\
&=&i\hbar\sum_{A,B} \{g_{AB},\overline{g_{AB}}\}[1+O(t)]
\ea
that is, the uncertainty bound in terms of commutators in the coherent state 
$\psi^t_g$ is given precisely by the value of the associated Poisson bracket
at the phase space point $g$ up to corrections quadratic in $t$.

The fact that the bound depends on the label of the coherent state is due 
to the fact that we use the operators $\hat{x}_{AB},\hat{y}_{AB}$ rather than
the operators 
$\hat{p}_j,\hat{q}_{AB}=(\hat{h}_{AB}+(\hat{h}_{AB})^\dagger)/2$, say
for which we get the uncertainty bound
\be \label{3.100}
\frac{|<\psi^t_g,[\hat{p}_j,\hat{q}_{AB}]\psi^t_g>|}{2||\psi^t_g||^2}
=t\frac{|<\psi^t_g,[(\tau_j\hat{h})_{AB}-(\hat{h}^{-1}\tau_j)_{BA}]\psi^t_g>|}
{4||\psi^t_g||^2}
\le t/2
\ee
since $\hat{h}_{AB}$ is a bounded operator on $L_2(G,d\mu_H)$ with bound $1$.

Summarizing, our coherent states saturate the uncertainty bound in 
precisely the way as they should and occupy a phase space volume (with 
respect to Liouville measure) of order $t^3$ exactly as the harmonic 
oscillator coherent states.

\subsection{Extension to Groups of Higher Rank} 
\label{s3.5}

Looking at the method of proof for all the theorems proved in the present
section for $G=SU(2)$ we realize three basic steps : \\
I) The determination of the exact complexification $g=g(p,h)$ of the 
configuration space of the phase space
$T^\ast G$ induced by the complexifier $p_j^2/2$ in order to determine what
quantity precisely should be peaked in either representation.\\
II) The use of the Poisson summation formula which transforms a slowly 
converging 
series into a rapidly converging one, allowing us to essentially drop all but
one term in estimates.\\
III) The separate estimate of the series for disjoint ranges of the group angles due
to the singular nature of functions that multiply the series, in our case 
$0\le \theta\le \frac{31}{32}\pi$ and $\frac{31}{32}\pi\le\theta\le\pi$ 
respectively, by rewriting the series in terms of parameters, here $\delta$,
which cancel the singularities and allow to obtain uniform bounds.\\
\\
How does this generalize to arbitrary compact gauge groups ? \\
I) The analysis in section \ref{s2} has revealed that for a general compact,
semi-simple gauge 
group the complexification is simply polar decomposition. So this generalizes 
immediately to any compact, semi-simple gauge group. \\
II) In \cite{52} a Poisson summation formula is derived for any compact gauge group
(see also \cite{39b} and references therein). Basically, one uses that the 
coherent states depend only on the characters in the various representations.
The characters in turn can be reduced to the a maximal torus $T^r$ (maximal 
Abelian subgroup) in a rank $r$ gauge group (generated by a $r$-dimensional Cartan 
subalgebra) moded by the action of the Weyl group. For instance, in the case 
$G=SU(2)$ there is the maximal torus $e^{\theta\tau_3},\theta\in[0,\pi]$ and the 
action of the Weyl group is given by $\tau_3\to -\tau_3$ and we have seen that
indeed our characters were invariant under the inversion 
$\lambda\leftrightarrow \lambda^{-1}$ with $\lambda=\ch(\theta)+\sh(\theta)=
e^\theta,
\lambda^{-1}=\ch(-\theta)+\sh(-\theta)=e^{-\theta}=\lambda{-1}$.
Then one can in fact carry out the Poisson resummation which is again of the 
form of a series times a product of $r$ singular factors of the type
$1/\sh(\theta)$ that we have seen in the case of $SU(2)$ which simply come out
of Weyl's character formula \cite{53}. The series part of that formula again 
obviously has the typical Gaussian damping factor that we have seen in the 
case of $G=SU(2)$.\\ 
III) For each of these singular prefactors we must make a separate estimate as 
outlined in this paper which is no problem in principle although the number of 
cases to be discussed grows as $2^{N-1}$ !\\
\\
Concluding, while possibly technically quite difficult, the methodof proof 
displayed in this paper for $G=SU(2)$ can be taken over, 
without principal changes,
to arbitrary compact, semi-simple $G$. We will come back to this in 
\cite{42}.

\section{Peakedness Proofs for Gauge-Invariant Coherent States}
\label{s4}

We first have to compute gauge-invariant coherent states from 
non-gauge-invariant ones. We will do this by the group averaging
procedure (\cite{7} and references therein). The idea is then to use the 
peakedness proofs
of the previous section exploiting that the gauge group to be averaged
over has unit volume. As will become obvious in this section, the peakedness
proofs for the gauge-invariant case are under much less control than for the 
gauge-variant case. Fortunately, as already mentioned in \cite{32}
for most of the applications of coherent states we can stick to the 
gauge-variant ones so that the lack of completeness in this section is not
very serious. We leave the improvement of the estimates of the present 
section for future research.
\begin{Definition} \label{def4.1}
Let $\gamma$ be a graph and $E(\gamma)$ the set of its oriented edges and 
$V(\gamma)$ the set of its vertices. Let 
\be \label{4.1}
\psi^t_{\gamma,\vec{g}}(\vec{h})=\prod_{e\in E(\gamma)} \psi^t_{g_e}(h_e)
\ee
be the family of gauge-variant coherent states on $\gamma$. Then we 
define a family of gauge-invariant coherent states on $\gamma$ by
\be \label{4.2}
\Psi^t_{\gamma,\vec{g}}(\vec{h}):=\eta_\gamma\cdot
\psi^t_{\gamma,\vec{g}}(\vec{h}):=
\prod_{v\in V(\gamma)}\int_G d\mu_H(h_v)\prod_{e\in E(\gamma)}
\psi^t_{g_e}(h_{e(0)}h_e(h_{e(1)})^{-1})
\ee  
\end{Definition}
The operation $\eta_\gamma$ can actually be applied to any gauge-variant
state, the result is obviously a gauge-invariant state. The following 
Lemma is elementary.
\begin{Lemma} \label{la4.1}
Let, as in section \ref{s1}, $T_{\gamma,\vec{j},\vec{J}}$ be a complete
orthonormal basis of spin-network states. Then
\be \label{4.3}
\Psi^t_{\gamma,\vec{g}}(\vec{h})
=\sum_{\vec{j},\vec{J}}e^{-\frac{t}{2}\sum_{e\in E(\gamma)} \lambda_{j_e}}
T_{\gamma,\vec{j},\vec{J}}(\vec{g}) 
\overline{T_{\gamma,\vec{j},\vec{J}}(\vec{h})} 
\ee
\end{Lemma}
Proof of Lemma \ref{la4.1} :\\
Let $\Delta_\gamma=\sum_{e\in E(\gamma)}\Delta_e$. By its very definition
we have 
\be \label{4.4}
\psi^t_{\gamma,\vec{g}}=
(e^{t\Delta_\gamma/2}\delta^{non-inv}_{\gamma,\vec{h}'})_{|\vec{h'}\to\vec{g}}
\ee
where $\delta^{non-inv}_{\gamma,\vec{h}'}$ is the distribution defined by
$\delta^{non-inv}_{\gamma,\vec{h}}(f_\gamma)=f_\gamma(\vec{h})$ for any
smooth function $f_\gamma$ cylindrical with respect to $\gamma$. On
the other hand 
\be \label{4.5}
\sum_{\vec{j},\vec{J}}e^{-\frac{t}{2}\sum_{e\in E(\gamma)} \lambda_{j_e}}
T_{\gamma,\vec{j},\vec{J}}(\vec{g}) 
\overline{T_{\gamma,\vec{j},\vec{J}}(\vec{h})}
=(e^{t\Delta_\gamma/2}\delta^{inv}_{\gamma,\vec{h}'})_{|\vec{h}'\to\vec{g}}
\ee
since 
$$
\sum_{\vec{j},\vec{J}}
T_{\gamma,\vec{j},\vec{J}}(\vec{h}') 
\overline{T_{\gamma,\vec{j},\vec{J}}(\vec{h})}
$$
is indeed a representation of the $\delta-$distribution on smooth 
gauge invariant functions cylindrical with respect to $\gamma$ as one can 
check on a spin-network basis.

Since $\Delta_\gamma$ is gauge-invariant it commutes with the operation
$\eta_\gamma$ and it remains to show that 
$\eta_\gamma\cdot \delta^{non-inv}_{\gamma,\vec{h}}$ is another 
representation of $\delta^{inv}_{\gamma,\vec{h}}$. But if $f_\gamma$
is gauge invariant, smooth and cylindrical with respect to $\gamma$
we have (we use the notation $\vec{h}^{\vec{h}'}:=
\{h'_{e(0)}h_e h_{e(1)}^{\prime\;-1}\}_{e\in E(\gamma)}$ for the gauge 
transformed vector of holonomies) 
\ba \label{4.6}
&& [\eta_\gamma\cdot \delta^{non-inv}_{\gamma,\vec{h}}](f_\gamma)
\nonumber\\
&=& \prod_{e\in E(\gamma)} \int_G d\mu_H(h^{\prime\prime}_e)
[\prod_{v\in V(\gamma)} \int_G d\mu_H(h'_v)\prod_{e\in E(\gamma)}
\delta(h_e^{\prime\prime},h'_{e(0)}h_e 
h_{e(1)}^{\prime\; -1})]f_\gamma(\vec{h}^{\prime\prime}) 
\nonumber\\
&=& \prod_{e\in E(\gamma)} \int_G d\mu_H(h^{\prime\prime}_e)
[\prod_{v\in V(\gamma)} \int_G d\mu_H(h'_v)\prod_{e\in E(\gamma)}
\delta(h_{e(0)}^{\prime\;-1}h_e^{\prime\prime}h'_{e(1)},h_e)]
f_\gamma(\vec{h}^{\prime\prime}) 
\nonumber\\
&=& \prod_{e\in E(\gamma)} \int_G d\mu_H(h^{\prime\prime}_e)
\prod_{e\in E(\gamma)}\delta(h_e^{\prime\prime},h_e)
[\prod_{v\in V(\gamma)} \int_G d\mu_H(h'_v)
f_\gamma(\vec{h}^{\prime\prime\; \vec{h}'})]
\nonumber\\
&=& \prod_{e\in E(\gamma)} \int_G d\mu_H(h^{\prime\prime}_e)
\prod_{e\in E(\gamma)}\delta(h_e^{\prime\prime},h_e)
f_\gamma(\vec{h}^{\prime\prime})]
\nonumber\\
&=& f_\gamma(\vec{h})
\ea
by the gauge invariance of $f_\gamma$ and the normalization of
the Haar measure. 

Thus we have shown that the vectors in ${\cal H}_\gamma$ on the left hand 
side and right hand side of (\ref{4.3}) have equal inner product with a 
dense set of vectors. Thus they must be the same in the $L_2$ sense.\\
$\Box$\\
\\
Notice that the properties (i), (ii) and (iii) mentioned at the beginning of
section \ref{s3} automatically hold also for gauge-invariant coherent states
provided that we use only analytically continued entire gauge invariant 
functions
in the connection representation and their complex conjugates as the
classical counterparts of operators to be measured by them. The point of
working with the integral representation (\ref{4.2}) rather than with
the explicit formula (\ref{4.3}) is two-fold : First of all, we have
established the peakedness proofs for the gauge-variant states (\ref{4.1})
already and wish to combine those with the integral formula (\ref{4.2})
while with (\ref{4.3}) we would need to start from scratch. Secondly,
the spin network states are not as explicitly known as one might think,
the complication coming from the space of vertex contractions which involves
the difficult calculus of the $3nj$ symbol \cite{16} for vertices of 
valence $n+2$.

In the next subsections we will need the following Lemma.
\begin{Lemma} \label{la4.2}
The relation between gauge-invariant and non-gauge-invariant inner products
is given by
\be \label{4.7}
<\Psi^t_{\gamma,\vec{g}},\Psi^t_{\gamma,\vec{g}'}>
=\prod_{v\in V(\gamma)} \int_G d\mu_H(h_v)
<\psi^t_{\gamma,\vec{g}},\psi^t_{\gamma,\vec{g}^{\prime\;\vec{h}}}>
\ee
\end{Lemma}
Proof of Lemma \ref{la4.2} :\\
The proof follows easily by the invariance properties of the Haar measure.
Notice that $\psi^t_g(h_1 h h_2^{-1})=\psi^t_{h_1^{-1}g h_2}(h)$ for one 
copy of the group, therefore 
\be \label{4.8}
\psi^t_{\gamma,\vec{g}}(\vec{h}^{\vec{h}'})
=\psi^t_{\gamma,\vec{g}^{(\vec{h}^{\prime\;-1})}}(\vec{h})
\ee
Then we have
\ba \label{4.9}
&& <\Psi^t_{\gamma,\vec{g}},\Psi^t_{\gamma,\vec{g}'}>
\nonumber\\
&=&\prod_{e\in E(\gamma)} \int_G d\mu_H(\tilde{h}_e)
\prod_{v\in V(\gamma)} \int_G d\mu_H(h_v)\int_G d\mu_H(h'_v)
\overline{\psi^t_{\gamma,\vec{g}}(\vec{\tilde{h}}^{\vec{h}})}
\psi^t_{\gamma,\vec{g}^{\prime\; (\vec{h}^{\prime\;-1})}}(\vec{\tilde{h}})
\nonumber\\
&=&\prod_{e\in E(\gamma)} \int_G d\mu_H(\tilde{h}_e)
\prod_{v\in V(\gamma)} \int_G d\mu_H(h_v)\int_G d\mu_H(h'_v)
\overline{\psi^t_{\gamma,\vec{g}}(\vec{\tilde{h}})}
\psi^t_{\gamma,\vec{g}^{\prime\; (\vec{h}^{\prime\; -1})}}
(\vec{\tilde{h}}^{(\vec{h}^{-1})})
\nonumber\\
&=&\prod_{e\in E(\gamma)} \int_G d\mu_H(\tilde{h}_e)
\prod_{v\in V(\gamma)} \int_G d\mu_H(h_v)\int_G d\mu_H(h'_v)
\overline{\psi^t_{\gamma,\vec{g}}(\vec{\tilde{h}})}
\psi^t_{\gamma,\vec{g}^{\prime\; (\vec{h}(\vec{h}^{\prime\; -1}))}}
(\vec{\tilde{h}})
\nonumber\\
&=&\prod_{e\in E(\gamma)} \int_G d\mu_H(\tilde{h}_e)
\prod_{v\in V(\gamma)} \int_G d\mu_H(h_v)
\overline{\psi^t_{\gamma,\vec{g}}(\vec{\tilde{h}})}
\psi^t_{\gamma,\vec{g}^{\prime\; \vec{h}}}(\vec{\tilde{h}})
\nonumber\\
&=&\prod_{v\in V(\gamma)} \int_G d\mu_H(h_v)
<\psi^t_{\gamma,\vec{g}},\psi^t_{\gamma,\vec{g}^{\prime\;\vec{h}}}>
\ea
$\Box$\\
So far we have defined everything for a general gauge group. We now 
specialize again to $SU(2)$ since we have proved peakedness theorems
only for $SU(2)$ in section \ref{s3}. Notice, however, that the proofs 
generalize to any $G$ once peakedness is established for the non-gauge
invariant states.

\subsection{Peakedness of the Overlap Function}
\label{s4.1}

In section (\ref{s3.2}) we showed that the overlap function for 
two coherent states with labels $g,g'$ is strongly peaked at $g=g'$
for one copy of the gauge group. This immediately implies that the 
gauge-non-invariant overlap function on $\gamma$ defined by
\be \label{4.10}
i^t_\gamma(\vec{g},\vec{g}')=\prod_{e\in E(\gamma)} i^t(g_e,g_e')
\ee
is strongly peaked at $\vec{g}=\vec{g}'$. We define the gauge invariant
overlap function on $\gamma$ by
\be \label{4.11}
I^t_\gamma(\vec{g},\vec{g}'):=
\frac{|<\Psi^t_{\gamma,\vec{g}},\Psi^t_{\gamma,\vec{g}'}>|^2}
{||\Psi^t_{\gamma,\vec{g}}||^2 ||\Psi^t_{\gamma,\vec{g}'}||^2}
\ee
Then the following theorem holds.
\begin{Theorem} \label{th4.1}
The peakedness of $I^t_\gamma(\vec{g},\vec{g}')$ at 
$[\vec{g}]=[\vec{g}']$ is implied by the peakedness of 
$i^t_\gamma(\vec{g},\vec{g}')$ at 
$\vec{g}=\vec{g}'$. Here $\vec{g}=\vec{H}\vec{u}$ denotes the polar 
decomposition of $\vec{g}$ and $[\vec{g}]:=\{\vec{g}^{\vec{h}'};\;
\vec{h}'\in {\cal G}_\gamma\}$ the gauge equivalence class of $\vec{g}$.
\end{Theorem}
Proof of Theorem \ref{th4.1} :\\
Defining
\be \label{4.12}
j^t_\gamma(\vec{g},\vec{g}'):=
\frac{<\psi^t_{\gamma,\vec{g}},\psi^t_{\gamma,\vec{g}'}>}
{||\psi^t_{\gamma,\vec{g}}||\; ||\psi^t_{\gamma,\vec{g}'}||}
\ee
so that $|j^t_\gamma|^2=i^t_\gamma$ we have, using Lemma \ref{la4.2}
\ba \label{4.13}
&& I^t_\gamma(\vec{g},\vec{g}')\nonumber\\
&=&\frac
{
\prod_{v\in V(\gamma)}\int d\mu_H(h_v) d\mu_H(h'_v)
<\psi^t_{\gamma,\vec{g}},\psi^t_{\gamma,\vec{g}^{\prime\;\vec{h}}}>
\overline{<\psi^t_{\gamma,\vec{g}},\psi^t_{\gamma,\vec{g}^{\prime\;\vec{h}'}}>}
}
{
\prod_{v\in V(\gamma)}\int d\mu_H(h_v) d\mu_H(h'_v)
<\psi^t_{\gamma,\vec{g}},\psi^t_{\gamma,\vec{g}^{\vec{h}}}>
<\psi^t_{\gamma,\vec{g}'},\psi^t_{\gamma,\vec{g}^{\prime\;\vec{h}'}}>
}
\nonumber\\
&=&\frac
{
\prod_{v\in V(\gamma)}\int d\mu_H(h_v) d\mu_H(h'_v)
j^t_\gamma(\vec{g},\vec{g}^{\prime\;\vec{h}}) \;
||\psi^t_{\gamma,\vec{g}}||\; 
||\psi^t_{\gamma,\vec{g}^{\prime\;\vec{h}}}|| \;
\overline{j^t_\gamma(\vec{g},\vec{g}^{\prime\;\vec{h}'})} \;
||\psi^t_{\gamma,\vec{g}}||\; 
||\psi^t_{\gamma,\vec{g}^{\prime\;\vec{h}'}}|| 
}
{
\prod_{v\in V(\gamma)}\int d\mu_H(h_v) d\mu_H(h'_v)
j^t_\gamma(\vec{g},\vec{g}^{\vec{h}})\;
||\psi^t_{\gamma,\vec{g}}||\; 
||\psi^t_{\gamma,\vec{g}^{\vec{h}}}|| \;
j^t_\gamma(\vec{g}',\vec{g}^{\prime\;\vec{h}'})\;
||\psi^t_{\gamma,\vec{g}'}|| \; 
||\psi^t_{\gamma,\vec{g}^{\prime\;\vec{h}'}}|| 
}
\nonumber\\ &&
\ea
Notice that 
$||\psi^t_{\gamma,\vec{g}^{\vec{h}}}||=||\psi^t_{\gamma,\vec{g}}||$.
The group integrals are very difficult to perform exactly in the case of a 
general graph (in fact, even for $G=U(1)$ the problem can be mapped to an
Ising model !) and we will
confine ourselves to an exact computation in appendix \ref{sa} for a simple 
graph and only for $G=U(1)$. However, peakedness can still be established 
heuristically as follows :

We know from section \ref{s3.2} that $j^t_\gamma(\vec{g},\vec{g}')$
is peaked at $\vec{g}=\vec{g}'$ with decay width of order $t$.
Thus the integral over $\vec{h}$ of 
$j^t_\gamma(\vec{g},\vec{g}^{\prime\;\vec{h}})$ will be large only
if there exists $\vec{h}$ such that $\vec{g},\vec{g}^{\prime\;\vec{h}}$ 
are lying in the same phase cell in which case we say that
$[\vec{g}]\approx[\vec{g}']$. Let $V_\gamma$ be the volume with respect
to $\prod_v d\mu_H(h_v)$ of the region $R_\gamma(\vec{g},\vec{g}')$ of
those $\vec{h}$ such 
that $\vec{g},\vec{g}^{\prime\;\vec{h}}$ are lying in the same phase cell.
By translation invariance of the Haar measure this volume is independent
of $\vec{g},\vec{g}'$ once it is true that $[\vec{g}]\approx[\vec{g}']$. 
Therefore 
$j^t_\gamma(\vec{g},\vec{g}^{\prime\;\vec{h}})\approx 1$ 
if $\vec{h}\in R_\gamma(\vec{g},\vec{g}')$ and $[\vec{g}]\approx[\vec{g}']$
and  
$j^t_\gamma(\vec{g},\vec{g}^{\prime\;\vec{h}})\approx 0$ otherwise. In other
words,  
\be \label{4.14}
j^t_\gamma(\vec{g},\vec{g}^{\prime\;\vec{h}})\approx 
\chi_{R_\gamma(\vec{g},\vec{g}')}(\vec{h})j^t_\gamma([\vec{g}]_0,[\vec{g}']_0)
\ee
where $\chi$ denotes the set-theoretic characteristic function. Here it 
is understood that we choose from each gauge equivalence class $[\vec{g}]$ 
once and for all a representant $[\vec{g}]_0$.
Since $i^t_\gamma$ almost takes only the values $0$ or $1$ as $t\to 0$
we see that the choice of the representant is irrelevant. 

Thus, the numerator in (\ref{4.13}) is given approximately by
$$
V_\gamma^2 
|j^t_\gamma([\vec{g}]_0,[\vec{g}']_0)|^2\;
||\psi^t_{\gamma,\vec{g}}||^2 \;||\psi^t_{\gamma,\vec{g}'}||^2
=V_\gamma^2 \; i^t_\gamma([\vec{g}]_0,[\vec{g}']_0)\;
||\psi^t_{\gamma,\vec{g}}||^2 \; ||\psi^t_{\gamma,\vec{g}'}||^2
$$
while the denominator is approximately given by
$$
V_\gamma^2 \;
||\psi^t_{\gamma,\vec{g}}||^2 \; ||\psi^t_{\gamma,\vec{g}'}||^2
$$
Summarizing, we find 
\be \label{4.15}
I^t_\gamma(\vec{g},\vec{g}')\approx i^t_\gamma([\vec{g}]_0,[\vec{g}']_0)
\ee
meaning that there exists a gauge $\vec{h}$ such that
$I^t_\gamma(\vec{g},\vec{g}')\approx 
i^t_\gamma(\vec{g},\vec{g}^{\prime\;\vec{h}})$.\\
$\Box$\\
Another argument proceeds as follows : it may be difficult to do in practice 
but
it is possible in principle to separate $\vec{g}$ or $\vec{p},\vec{h})$
into 1) gauge invariant quantities that are non-vanishing on the constraint 
surface of the phase space on the one hand and 2) pure gauge quantities and 
those that vanish on constraint surface on the other hand. The gauge-variant
overlap function is Gaussian peaked with respect to both sets of quantities
and doing the integrals in (\ref{4.13}) on the constraint surface does not 
change this behaviour with respect to the first set of quantities, in other
words, if not the gauge invariant data of $\vec{g},\vec{g}'$ are close to 
each other then $I^t(\vec{g},\vec{g}')$ is still small.\\
Remark :\\
Given a generic graph $\gamma$ with $|E(\gamma)|>2$ edges and $|V(\gamma)|>1$
vertices the number of configuration degrees of freedom before taking the 
Gauss constraint into account is $|E(\gamma)|\dim(G)$ and after
$(|E(\gamma)|-|V(\gamma)|)\dim(G)$ if $G$ is non-Abelian and 
$(|E(\gamma)|-|V(\gamma)|+1)\dim(G)$ if $G$ is Abelian since in that
case the gauge transformations at one of the vertices can be absorbed into
those of another. Therefore, the volume $V_\gamma$ 
of the pure gauge degrees of freedom that contribute to 
$I^t(\vec{g},\vec{g}')$ in (\ref{4.13}) should be of the order
$V(\gamma)=\sqrt{t}^{|V(\gamma)|\dim(G)}$ and 
$V(\gamma)=\sqrt{t}^{(|V(\gamma)|-1)\dim(G)}$ respectively since the decay 
width of our coherent states is $\sqrt{t}$ for all degrees of freedom 
(unquenched). This is confirmed in our example calculation in appendix 
\ref{sa}.

\subsection{Peakedness in the Connection Representation}
\label{s4.2}

In section \ref{s3.1} we showed that the non-gauge-invariant probability 
density in the configuration representation is peaked at $h_e=u_e$
for all $e\in E(\gamma)$ in the state $\psi^t_{g_e}$ where $g_e=H_e u_e$
is the polar decomposition of $g_e$. 
Thus we have also shown that 
the non-gauge-invariant probability density on the whole graph $\gamma$
\be \label{4.16}
p^t_{\gamma,\vec{g}}(\vec{h}):=
\frac{|\psi^t_{\gamma,\vec{g}}(\vec{h})|^2}{||\psi^t_{\gamma,\vec{g}}||^2}
\ee
is peaked at $\vec{h}=\vec{u}$. We define the gauge-invariant probability 
density by
\be \label{4.17}
P^t_{\gamma,\vec{g}}(\vec{h}):=
\frac{|\Psi^t_{\gamma,\vec{g}}(\vec{h})|^2}{||\Psi^t_{\gamma,\vec{g}}||^2}
\ee
Then the following theorem is easy to prove.
\begin{Theorem} \label{th4.2}
The peakedness of $P^t_{\gamma,\vec{g}}(\vec{h})$ at 
$[\vec{h}]=[\vec{u}]$ is implied by the peakedness of 
$p^t_{\gamma,\vec{g}}(\vec{h})$ at 
$\vec{h}=\vec{u}$. Here $\vec{g}=\vec{H}\vec{u}$ denotes the polar 
decomposition of $\vec{g}$ and $[\vec{h}]:=\{\vec{h}^{\vec{h}'};\;
\vec{h}'\in {\cal G}_\gamma\}$ the gauge equivalence class of $\vec{h}$.
\end{Theorem}
Proof of Theorem \ref{th4.2} :\\
Let us define the quantity
\be \label{4.18}
b^t_{\gamma,\vec{g}}(\vec{h}):=
\frac{\psi^t_{\gamma,\vec{g}}(\vec{h})}{||\psi^t_{\gamma,\vec{g}}||}
\ee
so that $|b^t_{\gamma,\vec{g}}(\vec{h})|^2=p^t_{\gamma,\vec{g}}(\vec{h})$.
Then we have by Lemma \ref{la4.2} and Definition \ref{def4.1}
\ba \label{4.19}
&& P^t_{\gamma,\vec{g}}(\vec{h})\nonumber\\
&=& \frac
{
\prod_{v\in V(\gamma)}\int_G d\mu_H(u_v)d\mu_H(u'_v)
\psi^t_{\gamma,\vec{g}}(\vec{h}^{\vec{u}})
\overline{\psi^t_{\gamma,\vec{g}}(\vec{h}^{\vec{u}'})}
}
{||\Psi^t_{\gamma,\vec{g}}||^2}
\nonumber\\
&=& \frac
{
||\psi^t_{\gamma,\vec{g}}||^2
\prod_{v\in V(\gamma)}\int_G d\mu_H(u_v)d\mu_H(u'_v)
b^t_{\gamma,\vec{g}}(\vec{h}^{\vec{u}})
\overline{b^t_{\gamma,\vec{g}}(\vec{h}^{\vec{u}'})}
}
{||\Psi^t_{\gamma,\vec{g}}||^2}
\nonumber\\
&\approx& \frac
{
\prod_{v\in V(\gamma)}\int_G d\mu_H(u_v)d\mu_H(u'_v)
b^t_{\gamma,\vec{g}}(\vec{h}^{\vec{u}})
\overline{b^t_{\gamma,\vec{g}}(\vec{h}^{\vec{u}'})}
}
{V_\gamma}
\ea
where in the last line we have used a result established in the course of 
the proof of Theorem \ref{th4.1}. Now from section \ref{3.1} we know that
$b^t_{\gamma,\vec{g}}(\vec{h}^{\vec{u}})$ is not small only if
there exists $\vec{u}$ such that $\vec{h}^{\vec{u}},\vec{U}$ 
lie in the same configuration cell in which case we say that
$[\vec{U}]\approx[\vec{h}]$. 
Here,$\vec{g}=\vec{H}\vec{U}$ is the polar decomposition of $\vec{g}$. 
The volume of the region $R_\gamma(\vec{U},\vec{h})$ of $\vec{u}$'s such 
that this condition is satisfied is $V_\gamma$ again. 
Using the same notation as in Theorem \ref{th4.1} and choosing
from each class $[\vec{h}]$ a representant $[\vec{h}]_0$ such
that $[\vec{h}]_0=[\vec{U}]_0$ if $[\vec{U}]=[\vec{h}]$ where 
$[\vec{U}]_0$ is determined by $[\vec{g}]_0$ for $g=HU$ we see that
\be \label{4.20}
b^t_{\gamma,\vec{g}}(\vec{h}^{\vec{u}})
\approx
b^t_{\gamma,[\vec{g}]_0}([\vec{h}]_0)\chi_{R_\gamma(\vec{g},\vec{h})}(\vec{u})
\ee
Therefore (\ref{4.19}) becomes
\be \label{4.21}
P^t_{\gamma,\vec{g}}(\vec{h})
\approx V_\gamma p^t_{\gamma,[\vec{g}]_0}([\vec{h}]_0)
\ee
$\Box$

\subsection{Peakedness in the Electric Field Representation}
\label{s4.3}

The gauge-non-invariant coherent states for a graph in the electric field 
representation are simply given by the product of the ones for each edge 
\be \label{4.22}
\tilde{\psi}^t_{\gamma,\vec{g}}(\vec{j},\vec{m},\vec{n}):=
\prod_{e\in E(\gamma)}\tilde{\psi}^t_{g_e}(j_e m_e n_e)
\ee
Alternatively they can be defined as the inner product of the state
$\psi^t_{\gamma,\vec{g}}$ defined above with the 
state $|\vec{j}\vec{m}\vec{n}>$ given by
\be \label{4.23}
<\vec{h},\vec{j}\vec{m}\vec{n}>
=\prod_{e\in E(\gamma)}\pi_{j_e}(h_e)_{m_e n_e}
\ee
Similarly, we define the gauge-invariant coherent states in the electric 
field representation by
\be \label{4.24}
\tilde{\Psi}^t_{\gamma,\vec{g}}(\vec{j},\vec{J}):=
<\vec{j}\vec{J},\Psi^t_{\gamma,\vec{g}}>
\ee
where 
\be \label{4.25}
<\vec{h},\vec{j}\vec{J}>=T_{\gamma,\vec{j},\vec{J}}(\vec{h})
\ee
is a spin-network state. Clearly, these gauge invariant Fourier 
coefficients belong to an $\ell_2$ Hilbert space of sequences 
equipped with an inner product isometric to the one on the $L_2$ space
and it is given by $\sum_{\vec{j}\vec{J}} 
\overline{a_{\vec{j}\vec{J}}}b_{\vec{j}\vec{J}}$. Thanks to 
Lemma (\ref{la4.1}) we can explicitly compute (\ref{4.24}) to be 
\be \label{4.26}
\tilde{\Psi}^t_{\gamma,\vec{g}}(\vec{j},\vec{J})=
e^{-\frac{t}{2}\sum_{e\in E(\gamma)}j_e(j_e+1)}T_{\gamma,\vec{j},\vec{J}}(\vec{g})
\ee
Finally we define the gauge-invariant probability amplitude in the 
electric field representation by
\be \label{4.26a}
P^t_{\gamma,\vec{g}}(\vec{j}\vec{J}):=
\frac{|\tilde{\Psi}^t_{\gamma,\vec{g}}(\vec{j}\vec{J})|^2}
{||\Psi^t_{\gamma,\vec{g}}||^2}
\ee

In order to exploit the peakedness properties established in
subsection \ref{s3.3} we must know the explicit definition of 
$T_{\gamma,\vec{j},\vec{J}}(\vec{g})$.
\begin{Lemma} \label{la4.3}
Denote by $N(v)$ the valence of a vertex $v$ of a graph $\gamma$ and split
each edge $e$ of $\gamma$ into two halves with outgoing orientations from
those endpoints that are vertices of $\gamma$. 
For each vertex $v$ of $\gamma$, choose a labelling of the split
edges $f^v_k,\; k=1,.., N(v)$ incident at it. Given an unsplit
edge $e$, let natural numbers $k(e),l(e)$ be defined by 
$e=f^{e(0)}_{k(e)}\circ (f^{e(1)}_{l(e)})^{-1}$ and define 
$j^v_k=j_e$ if $k=k(e),v=e(0)$ or $k=l(e),v=e(1)$. Also, for each
vertex $v$ choose a recoupling scheme $(J^v_{k-1} j^v_{k+1})\to J^v_k,\;
k=1,..,N(v)-1,J^v_0=j^v_1,\;J^v_{N(v)-2}=j^v_{N(v)},\;J^v_{N(v)-1}=0$.
Finally, let $\vec{j}^v=\{j^v_1,..,j^v_{N(v)}\},\;
\vec{m}^v=\{m^v_1,..,m^v_{N(v)}\},\;
\vec{J}^v=\{J^v_1,..,J^v_{N(v)-3}\}$.\\
A spin-network basis is then given by 
\be \label{4.27}
T_{\gamma,\vec{j},\vec{J}}(\vec{h})
=\prod_{v\in V(\gamma)}  
{[}\prod_{k=1}^{N(v)} \pi_{j^v_k}(h_{f^v_k})_{m^v_k n^v_k}
c^v_{\vec{j}^v\vec{m}^v;\vec{J}^v}]
{[}\prod_{e\in E(\gamma)}\pi_{j_e}(\epsilon)_{n^{e(0)}_{k(e)} n^{e(1)}_{l(e)}}]
\ee
where $\epsilon$ is the totally skew tensor density of weight one in two 
dimensions and 
\be \label{4.28}
c^v_{\vec{j}^v\vec{m}^v;\vec{J}^v}
=<j^v_1 m^v_1.. j^v_{N(v)} m^v_{N(v)}|j^v_1..j^v_{N(v)};J^v_1..J^v_{N(v)-3}>
\ee
is the Clebsch-Gordan-coefficient for recoupling of $N(v)$ angular momenta.
\end{Lemma}
Proof of Lemma \ref{la4.3} :\\
We simply have to compute the inner products of two of the states 
in (\ref{4.27}) with labels $\vec{j},\vec{J}$ and $\vec{j}'\vec{J}'$
respectively. 
Clearly we get the non-vanishing result if and only if 
$\vec{j}^v=\vec{j}^{\prime\; v},\vec{m}^v=\vec{m}^{\prime\; v},
\vec{n}^v=\vec{n}^{\prime\; v}$ for all $v$
in which case (\ref{4.27}) gets mutiplied by
$\prod_{e\in E(\gamma)} 1/d_{j_e}$. Thus we find for the inner product
\be \label{4.29}
=\delta_{\vec{j}\vec{j}'}
\prod_{k=1}^{N(v)} 
{[}c^v_{\vec{j}^v\vec{m}^v;\vec{J}^v}
c^v_{\vec{j}^v\vec{m}^v;\vec{J}^{\prime v}}]
{[}\prod_{e\in E(\gamma)}\frac{1}{d_{j_e}}
\pi_{j_e}(\epsilon)_{n^{e(0)}_{k(e)} n^{e(1)}_{l(e)}}
\pi_{j_e}(\epsilon)_{n^{e(0)}_{k(e)} n^{e(1)}_{l(e)}}]
\ee
Performing the sum over $\vec{m}^v$ produces a 
$\delta_{\vec{J}^v,\vec{J}^{\prime v}}$
due to the completeness relations for the CG-coefficients. Performing the 
sum over $\vec{n}$ produces a $\prod_e \chi_{j_e}(\epsilon\epsilon^T)=
\prod_e d_{j_e}$. Thus altogether
\be \label{4.30}
<\vec{j}\vec{J},\vec{j}'\vec{J}'>=
\delta_{\vec{j}\vec{j}'}\delta_{\vec{J}\vec{J}'}
\ee
$\Box$\\
We also must compute the multiple CG-coefficients in terms of the 
elementary $3j$-symbols
$<j_1 m_1 j_2 m_2|j_1 j_2;j m>=\delta_{m,m_1+m_2}
<j_1 m_1 j_2 m_2|j_1 j_2;J m_1+m_2>,\;
\mbox{max}(|m_1+m_2|,|j_1-j_2|)\le J\le j_1+j_2$ for which approximation 
formulae for large $j$'s exist.
\begin{Lemma} \label{la4.4}
\ba \label{4.31}
&& <j_1 m_1..j_N m_N|J_1.. J_{N-3} J_{N-2}=j_N J_{N-1}=0 M=0>
=\delta_{m_1+..+m_N,0}\times\nonumber\\
&&\times \prod_{k=1}^{N-1}
<J_{k-1} n_{k-1} j_{k+1} m_{k+1}|J_{k-1} j_{k+1}; J_k n_k>
\ea
where $n_k=\sum_{l=1}^{k+1} m_l, J_0=j_1,J_{N-2}=j_N,J_{N_1}=0$.
\end{Lemma}
Proof of Lemma \ref{la4.4} :\\
This follows from iterating the definition of the CG-coefficients
as unitary transformation coefficients between the two orthonormal
bases $|j_1 m_1 j_2 m_2>:=|j_1 m_1>\otimes |j_2 m_2>$ and
$|j_1 j_2;jm>$. See also the standard literature on angular momentum, e.g.
\cite{56}.\\
$\Box$\\
\\
The idea is now the following : We have shown in section \ref{s3.3}
that $p^t_{\gamma,\vec{g}}(\vec{j}\vec{m}\vec{n})$ is peaked 
at the values $tj_e=p_e,tm_e=^R p_e^3,tn_e=^L p_e^3$ where
the momenta displayed correspond to the polar decompositions 
$g_e=(^R H_e) u_e=u_e(^L H_e)$ and $^{R/L} H
=\exp(-i\tau_j (^{R/L} p_e^j/2))$.
Let us denote these values as 
$\vec{j}(\vec{g}),\vec{m}(\vec{g}),\vec{n}(\vec{g})$.
Using again Theorem \ref{th4.1} and the explicit formula (\ref{4.27})
we find that
\be \label{4.32}
P^t_{\gamma,\vec{g}}(\vec{j},\vec{J})
\approx V_\gamma [\prod_{v\in V(\gamma)} c^v(\vec{j}^v\vec{J}^v\vec{m}^v)]
p^t_{\gamma,\vec{g}}(\vec{j},\vec{m},\vec{n})
\ee
where now the $m^v_k$ contain both $m_e$ and $n_e$ according to formula
(\ref{4.27}).
That is, the gauge invariant states  
$\tilde{\Psi}^t_{\gamma,\vec{g}}(\vec{j},\vec{J})$ are nothing else than 
the non-gauge-invariant states 
$\tilde{\psi}^t_{\gamma,\vec{g}}(\vec{j},\vec{m},\vec{n})$ contracted with 
Clebsch-Gordan coefficients. 
Since $p^t_{\gamma,\vec{g}}(\vec{j},\vec{m},\vec{n})$ is small unless
$j_e,m_e,n_e$ take the values determined by $\vec{g}$ above,
a crude estimate of the value of (\ref{4.32}) is that 
\be \label{4.32a}
P^t_{\gamma,\vec{g}}(\vec{j},\vec{J})
\approx V_\gamma [\prod_{v\in V(\gamma)} 
c^v(\vec{j}^v(\vec{g})\vec{J}^v\vec{m}^v(\vec{g}))] 
p^t_{\gamma,\vec{g}}(\vec{j}\vec{m}\vec{n}) 
\ee
Combining this with Theorem \ref{th3.6} we find 
\begin{Theorem} \label{th4.3}
For large $p_e$ there exists a constant $K_t$ exponentially vanishing as 
$t\to 0$ and independent of $\vec{g}$ such that 
\ba \label{4.33}
P^t_{\gamma,\vec{g}}(\vec{j}\vec{J})
&\stackrel{<}{\sim}& V_\gamma [\prod_{v\in V(\gamma)} 
c^v(\vec{j}^v(\vec{g})\vec{J}^v\vec{m}^v(\vec{g}))^2] \times
\\
&& \times 
[\prod_{e\in E(\gamma)}
\frac{1}{2p_e}\frac{t^{3/2}}{4\sqrt{\pi}}\frac{1}{1-K_t}\times
\nonumber\\
&& \times 
e^{-j/2\frac{(m_e/j_e-(^R p^3_e)/p_e)^2}{1-(^R p^3_e/p_e)^2}}
e^{-j/2\frac{(n_e/j_e-(^L p^3_e)/p_e)^2}{1-(^L p^3_e/p_e)^2}}
e^{-\frac{(\frac{(2j_e+1)t}{2}-p_e)^2}{t}}\mbox{ if } 
|^{R/L}p^3_e/p_e|<1 \nonumber\\
P^t_{\gamma,\vec{g}}(\vec{j},\vec{J})
&\stackrel{<}{\sim}& V_\gamma [\prod_{v\in V(\gamma)} 
c^v(\vec{j}^v(\vec{g})\vec{J}^v\vec{m}^v(\vec{g}))^2] \times
\nonumber\\
&& \times 
[\prod_{e\in E(\gamma)}
\frac{1}{2p_e}\frac{t^{3/2}}{4\sqrt{\pi}}\frac{1}{1-K_t}\times
\nonumber\\
&& \times 
e^{-j_e p_e|m_e/j_e-(^R p)^3_e/p_e|}
e^{-j_e p_e|n_e/j_e-(^L p)^3_e/p_e|}
e^{-\frac{(\frac{(2j_e+1)t}{2}-p_e)^2}{t}}\mbox{ if } 
|^{R/L}p^3_e/p_e| \stackrel{<}{\sim} 1 \nonumber
\ea
other mixed cases being treated similarly.
\end{Theorem}
Theorem \ref{th4.3} is not entirely satisfactory since one would prefer
to know at which values of $\vec{J}$ the probability amplitude is peaked.
One might hope that the Clebsh Gordan coefficient itself is peaked
at certain values of $j$ if the values of $j_1 j_2 m_1 m_2$ are given which
is the case if we perform the approximation (\ref{4.32a}). 

To investigate this question we review pieces of the beautiful paper 
\cite{57} which rigorizes the classical work of Ponzano and Regge
\cite{58}.\\
Given the values $j_1,j_2,m_1,m_2,j$ we can construct the following 
quantities : Let $j_3:=j,m_3:=m_1+m_2$. Then we define
\ba \label{4.34}
\lambda_i &:=&\sqrt{j_i^2-m_i^2} \nonumber\\
\beta^2&:=&
[(\lambda_1+\lambda_2)^2-\lambda_3^2]
[\lambda_3^2-(\lambda_1-\lambda_2)^2]
\ea
The interpretation of the $\lambda_i$ is clear : if we interprete the
$j_i$ as the length of vectors $\vec{p}_i$ in $\Rl$ satisfying
$\vec{p}_1+\vec{p}_2=\vec{p}_3$ and $m_i$ as 
their 3-components
then the $\lambda_i$ are the lengths of the projections of these vectors 
into the 1-2 plane. Furthermore, it is easy to see by methods of 
two-dimensional Euclidean geometry that $\beta^2$ is proportional to 
the square of a triangle with side lengths $\lambda_1,\lambda_2,\lambda_3$
provided $\beta^2\ge 0$ : This defines the (classically) allowed region.
Namely, it is easy to see that $\beta^2\ge 0$ is equivalent to
$\lambda_1+\lambda_2\ge \lambda_3 \ge |\lambda_1-\lambda_2|$.
However, there are quantum mechanically allowed ranges of the $j_i,m_i$
which satisfy $\beta^2<0$ which defines the (classically) forbidden region.
A nice graphical illustration of these regions in parameter space can be 
found in \cite{57}. The asymptotic behaviour of the CG-coefficients as 
the $j_i$ get large can be obtained by casting the Racah formula
\cite{56} for the CG-coefficients into an integral formula and 
performing a steepest descent contour deformation and a saddle point 
approximation. These deformations need to be 
discussed separately for the allowed and the forbidden region.\\
I) Allowed region :\\
We define five angles : Consider a triangle in two-dimensional
Euclidean space with side lengths $\lambda_i$. 
Let $0\le \gamma_1,\gamma_2\le \pi$ respectively be the angle between
the sides of a triangle of length $\lambda_1,\lambda_3$ and
$\lambda_2,\lambda_3$ respectively. Furthermore, consider a tetrahedron
spanned by the vectors $\vec{p}_1,\vec{p}_2$ and an additional vector 
$\vec{p}$ which has large and positive 3-component and small 
1,2-components. Let $0\le \chi_i\le \pi$ be the angle
between the outward unit vectors of those faces of the tetrahedron 
intersecting in the edge which corresponds to the vector $\vec{p}_i$.
Finally, take the limit of $R\to \infty$ of $\vec{p}\to R\vec{e}_3$
where $\vec{e}_3$ is the standard unit vector of Euclidean space in the 
3-direction. Then \cite{57}
\ba \label{4.35}
|<j_1 m_1 j_2 m_2|j_3 m_3>|^2 &\approx& \frac{4j_3}{\pi|\beta|}
\cos^2(\chi-\pi[j+\frac{3}{4}]) \mbox{ where }\nonumber\\
\chi &=& m_2 \gamma_2-m_1\gamma_1+\sum_{i=1}^3(j_i+\frac{1}{2})\chi_i
\ea
II) Forbidden region :\\
The way one obtains the $\gamma_i,\chi_i$ in the allowed region is actually
by first computing $\cos(\gamma_i),\cos(\chi_i)$ by analytical 
formulae. The corresponding expressions take in the 
allowed region values in $[-1,1]$. In the forbidden region these values 
become positive and of modulus greater than one. Thus, the angles 
become imaginary or the cosines turn into hyperbolic cosines. 
Furthermore, in the allowed region there are two saddle points which
give rise to the cosine in (\ref{4.35}) upon adding their contribution 
while in the forbidden region there is only one saddle point so that 
one ends up only with one exponential function of real argument.
Continuing to call the ``angles" $\gamma_i,\chi_i$ which now take range 
in the positive real axis one finds that
\ba \label{4.36}
|<j_1 m_1 j_2 m_2|j_3 m_3>|^2 &\approx& \frac{4j_3}{\pi|\beta|}
\exp(-2\chi) \mbox{ where }\nonumber\\
\chi &=& m_2 \gamma_2+m_1\gamma_1+\sum_{i=1}^3(j_i+\frac{1}{2})\chi_i
\ea
Strictly speaking, the forbidden region subdivides into six subregions 
and the ``angles" are a bit differently defined in each subregion but
the essential behaviour of (\ref{4.36}) stays the same.\\
\\
Let us now analyze (\ref{4.35}), (\ref{4.36}) which we interprete 
as the probability amplitude $p(j):=p_{j_1 j_2 m_1 m_2}(j)$ for the 
system of two angular momenta
of modulus $j_1,j_2$ and 3-components $m_1,m_2$ to couple to resulting 
angular momentum $j_3=j$ with 3-component $m_3=m_1+m_2$. We are 
interested in the maximum of that function as $j$ varies in its quantum 
mechanically allowed range max$(|m_1+m_2|,|j_1-j_2|)\le j\le j_1+j_2$. The 
explicit formula for $p(j)$ in terms of $j_i,m_i$ is very complicated and 
the attempt to find the exact critical point leads to unfeasable 
transcendent equations so that we stick here to a qualitative analysis. 

In the allowed region the amplitude of the CG-coefficient is governed by 
the relatively simple function $j/|\beta|$ while it oscillates rapidly as 
we change $j$ due to the $j\pi$ term in the argument of the cosine. The 
$\chi_i,\gamma_i$ on the other hand are slowly varying. Thus, the 
critical point can be analyzed by studying the function 
\be \label{4.36a}
f(x):=j^2/|\beta|^2=
\frac{j^2}
{[(\lambda_1+\lambda_2)^2-\lambda_3^2][\lambda_3^2-(\lambda_1-\lambda_2)^2]}
=:
\frac{x+m^2}
{[\lambda_+^2-x^2][x-\lambda_-^2]}
\ee
where we have defined $\lambda_\pm=\lambda_1\pm \lambda_2,x=\lambda^2$.
One finds that 
\be \label{4.37}
f'(x)=\frac{1}{\beta^4}[j^4-(m^2+\lambda_+^2)(m^2+\lambda_-^2)]
\ee
Thus the critical value is at
\be \label{4.38}
j_0:=\root 4 \of{(m^2+\lambda_+^2)(m^2+\lambda_-^2)}
\ee
and has the following interpretation : Suppose $j^2$ is the square of the 
vector $\vec{p}_1+\vec{p}_2$ then
$$
\lambda_-^2+m^2\le 
j^2=j_1^2+j_2^2+2m_1 m_2+2\vec{p}^\perp_1\vec{p}^\perp_2\le 
\lambda_+^2+m^2
$$
where $\vec{p}_i^\perp$ are the projections into the 1-2 plane. Thus,
$j_0$ is {\it the geometric mean of the classical extremal values of $j^2$}.
On the other hand, the expectation value of the operator $\hat{j}^2$
is approximately given by $j_1^2+j_2^2+2m_1 m_2$ which is the algebraic
mean of the two extremal values of $j^2$. The geometric mean is never bigger 
than the algebraic mean. Furthermore we see that
$\sqrt{m^2+\lambda_-^2}\le j_0\le\sqrt{m^2+\lambda_+^2}$ which means that
(\ref{4.37}) is less/bigger than zero for $j</> j_0$ which means that
$j=j_0$ is the only minimum. Clearly, the formula (\ref{4.35}) must
break down at $\lambda=\lambda_\pm$ where it diverges while $0\le p(j)\le 1$.

In the forbidden region (\ref{3.36}) is exponentially damped. The function
in front of the exponential factor is given by $-f(x)$ and so the critical
point $j_0$ is now a maximum which however has to compete with the 
exponential dampedness.\\
\\
The qualitative behaviour of $p(j)$ can therfore be summarized as follows :\\
If there is an allowed region then $p(j)$ is rapidly oscillating with $j$
in that region where the envelope is given by a function which has a minimum
at $j=j_0$ and is increasing towards the values of $j$ corresponding to
$\lambda=\lambda_\pm$. In the forbidden region $p(j)$ is exponentially 
damped where the decay width depends on $j_1 j_2 m_1 m_2$.
In the transition region between allowed and forbidden region we have to
join these curves smoothly. If there is no allowed region (e.g. in the case
$m_2=\pm j_2$) there is only exponential dampedness and the peak is
at the transition point $\lambda=\lambda_\pm$.

In conclusion, $p(j)$ generically does not display any peakedness 
properties, 
the best that one can say is that the expectation value of $j$ is $j_0$
given above. This agrees qualitatively by fitting the values of 
$<\hat{j}^2>,<(\Delta\hat{j}^2)^2>$ into a Gaussian distribution.
Of course, it is not surprising that the values of the recoupling momenta
$J^v_k,k=1,..,N(v)-3$ are not so sharply peaked as not even classically
any value of $j$ in the range allowed by $j_1,j_2,m_1,m_2$ is distinguished.\\
\\
Thus, in order to make progress in that direction one must go back to 
(\ref{4.32})
and repeat the analysis by first summing over all $m_e,n_e$ and then determine
the peakedness properties with respect to $\vec{j},\vec{J}$. This, however, is 
beyond
the scope of the present paper.

\subsection{Phase Space Bounds and Heisenberg Uncertainty Relation}
\label{s4.4}

These follow essentially from the non-gauge-invariant ones by straightforward 
but
tedious calculations and will be left to the ambitious reader.\\
\\
\\
\\
\\
{\large Acknowledgements}\\
\\
We are very grateful to Brian Hall for extensive discussions about 
the coherent states introduced by himself and in particular for pointing 
out the importance of the Poisson summation formula in estimates. T.T.
also thanks The University of California at San Diego for hospitality and 
financial support. We also thank Laurent Freidel for 
bringing reference \cite{57} to our attention. O.W. thanks the 
Studienstiftung des Deutschen Volkes for financial support.

\begin{appendix}

\def\hg{\hat{g}}
\def\hgd{\hat{g}^{\dagger}}
\def\vg{\vec{g}}
\def\vh{\vec{h}}
\def\vu{\vec{u}}
\def\vgp{\vec{g}'}
\def\vhp{\vec{h}'}
\def\dl{\Delta}

\section{The $U(1)$ case}
\label{sa}

In this appendix we will 
apply the results of this paper to the case of $U(1)$ as the 
gauge group. 
As will become clear, the much simpler structure of $U(1)$ leads to a 
considerable 
simplification of the derivation of all the results. The main reason for 
this is, 
of course, the fact that $U(1)$ is Abelian and as a consequence of this that 
all its irreducible 
representations are one-dimensional. This means that one has to 
deal with numbers only instead of matrices.

\subsection{Peakedness Proofs for Gauge-Variant Coherent states}
\label{sa.1}

We recall from (\ref{3.3}) the general form of a coherent state:
\be
\psi^t_g(h)=\sum_\pi d_\pi e^{-\frac{t}{2}\lambda_{\pi}} \chi_{\pi}(gh^{-1})
\ee

For the case of $U(1)$, $d_\pi =1$, 
the set of all irreducible representations can be 
parameterized by the set of 
integers which we denote by $n$, and the eigenvalue of the 
Laplacian is $-n^2$. 
Furthermore, we parametrize the $U(1)$ element $h$ by the angle
 $\theta ,\theta\in [0,2\pi]$ and $g$ by $\phi\in [0,2\pi]$ and $p\in\Rl $.
Summarizing, we have
\ba \label{a1.1}
& & h=e^{i\theta } \nonumber\\
& & g=e^{i(\phi -ip)} \nonumber\\
& & \chi_\pi (gh^{-1}) = e^{in(\phi - \theta )} e^{np} \mbox{ for } \pi=n
\ea
and thus
\be \label{a1.2}
\psi^t_g(h)= 
\sum_{n=-\infty}^{\infty} e^{-\frac{t}{2} n^2} e^{in(\phi - \theta )} 
e^{np}.
\ee

\subsubsection{Peakedness in the Connection Representation}
\label{sa.1.1}

As in the main text we define
\be \label{a2.1}
p^t_g(h)=\frac{|\psi^t_g(h)|^2}{||\psi^t_g||^2},
\ee
for which we would 
like to prove peakedness at $\theta=\phi$, or, equivalently, at 
$\phi=0$ for $\psi^t_g(1)$. For the norm of $\psi$ we immediately get
\be \label{a2.2}
||\psi^t_g||^2=\psi^{2t}_{H^2}(1)=\sum_n e^{-tn^2} e^{2np}.
\ee

Now we have 
to write the formula for $\psi^t_g(1)$ in a form suitable for applying the 
Poisson formula:
\ba \label{a2.3}
\psi^t_g(1)&=&\sum_n e^{-\frac{t}{2}n^2} e^{in\phi} e^{np} \nonumber \\
        &= &\sum_n e^{-(ns)^2/2} e^{i(ns)\frac{\phi}{s}} e^{(ns) \frac{p}{s}} 
\nonumber \\
           &=& \sum_n f(ns),
\ea
where we introduced $s=\sqrt{t}$. Thus we have the following function
\be \label{a2.4}
f(x)=e^{-x^2/2} e^{ix \frac{\phi}{s}} e^{x\frac{p}{s}}
\ee
which satisfies all conditions for Poisson summation formula. We obtain for the 
Fourier transform
\be \label{a2.5}
\tilde{f}(k)= \frac{1}{\sqrt{2\pi}} e^{-\frac{1}{2}(k^2+\frac{\phi^2}{s^2} 
+2ik\frac{p}{s} -2k\frac{\phi}{s} - 2i\frac{\phi p}{s^2} -\frac{p^2}{s^2})}.
\ee
Applying Poisson's formula then leads to
\be \label{a2.6}
\psi^t_g(1)=\sqrt{\frac{2\pi}{t}} \sum_n e^{-\frac{2\pi^2 n^2 
+ \frac{1}{2} \phi^2 
+ 2i\pi np -i\phi p -2\pi n\phi -\frac{1}{2} p^2}{t}}.
\ee
From this we can immediately read off the 
Poisson transformed form for the norm of 
$\psi^t_g$ as well, for which we obtain
\be \label{a2.7}
\psi^{2t}_{H^2}(1)=\sqrt{\frac{\pi}{t}} 
\sum_n e^{-\frac{\pi^2 n^2 +2i\pi np -p^2}{t}}.
\ee
Now we are ready to calculate the probability amplitude. 
Inserting (\ref{a2.6}) and 
(\ref{a2.7}) into (\ref{a2.1}) we find
\ba \label{a2.8}
p^t_g(h)&=& 2\sqrt{\frac{\pi}{t}} \frac{|\sum_n e^{-\frac{2\pi^2 n^2 
+ \frac{1}{2} \phi^2 + 2i\pi np -i\phi\pi -2\pi n\phi 
-\frac{1}{2} p^2}{t}}|^2}{\sum_n e^{-\frac{\pi^2 n^2 +2i\pi np -p^2}{t}}} 
\nonumber \\
        &=& 2\sqrt{\frac{\pi}{t}} 
\frac{e^{-\frac{\phi^2}{t}} |\sum_n e^{-\frac{2\pi^2 n^2 
-2\pi n(\phi -ip)}{t}} |^2}{\sum_n e^{-\frac{\pi n^2+2i\pi np}{t}}}.
\ea
Our next step is to 
determine bounds for the denominator which we denote by $D^t_p$:
\ba \label{a2.9}
D^t_p &=& \sum_{n=-\infty}^{\infty} e^{-\frac{\pi n^2}{t}} (\cos(2\pi np/t)
-i\sin(2\pi np/t)) \nonumber \\
      &=& 1+ 2 \sum_{n=1}^{\infty}  e^{-\frac{\pi n^2}{t}} \cos(2\pi np/t)
\ea
So a lower bound is given by
\ba \label{a2.10}
|D^t_p|&\geq& 1+2 \sum_{n=1}^{\infty}  e^{-\frac{\pi n^2}{t}} 
\min_p (\cos(2\pi np/t)) \nonumber \\
       &=& 1-2\sum_{n=1}^{\infty} e^{-\frac{\pi n^2}{t}} \nonumber \\
       &=:& 1-K_t 
\ea
where $K_t$ goes to zero exponentially fast when $t$ goes to zero. 
By an equivalent estimate with signs reversed we get the following upper bound: 
$|D^t_p| \leq 1+K_t$. \\
For the numerator we obtain the following estimate:
\ba \label{a2.11}
|N^t_p| &\leq &  2\sqrt{\frac{\pi}{t}} e^{-\frac{\phi^2}{t}} 
\sum_n| e^{-\frac{2\pi^2 n^2 -2\pi n(\phi -ip)}{t} |^2} \nonumber \\
&=& 
2\sqrt{\frac{\pi}{t}} e^{-\frac{\phi^2}{t}} \sum_n  e^{-\frac{4\pi^2 n^2}{t}} 
e^{\frac{4\pi n\phi}{t}}  \nonumber \\ 
&=&  2\sqrt{\frac{\pi}{t}} e^{-\frac{\phi^2}{t}} (1+ \sum_{n=1}^{\infty} 
e^{-\frac{4\pi^2 n^2}{t}} \ch(\frac{4\pi n\phi }{t}) ) \nonumber \\
&\leq & 2\sqrt{\frac{\pi}{t}} e^{-\frac{\phi^2}{t}} (1+ \sum_{n=1}^{\infty} 
e^{-\frac{4\pi^2 n^2}{t}} \ch(\frac{8\pi n }{t}) ) \nonumber \\ 
         &=&  2\sqrt{\frac{\pi}{t}} e^{-\frac{\phi^2}{t}} (1+ \tilde{K_t} )
\ea
with $\tilde{K_t} \rightarrow 0$ exponentially fast for $t\rightarrow 0$. \\

We summarize the results of this subsection in the following theorem :
\begin{Theorem} \label{tha2.1}
There exist 
positive constants $K_t,\tilde{K_t} $ (independent of $p$ and $\theta$ ), 
decaying exponentially fast to $0$ as $t\to 0$ such that
\be \label{a2.12}
p^t_g(1) \leq \frac{ 2\sqrt{\frac{\pi}{t}} e^{-\frac{\phi^2}{t}} 
(1+ \tilde{K_t} )}{1-K_t}
\ee
\end{Theorem}

\subsubsection{Peakedness of the Overlap Function}
\label{sa.1.2}

We recall from (\ref{3.45}) the expression for the overlap function:
\be \label{a3.1}
i^t(g,g'):= 
\frac{|<\psi^t_g, \psi^t_{g'} >|^2}{||\psi^t_g||^2 ||\psi^t_{g'} ||^2} 
= \frac{|\psi^{2t}_{HH'} (h)|^2}{\psi^{2t}_{H^2} (1) \psi^{2t}_{(H')^2} (1)}
\ee
In our case $H=e^p,H'=e^{p'}$ and therefore $HH'=e^{p+p'}=\tilde{H}$, while 
$h=u' u^{-1}=:\tilde{h}=e^{i(\phi'-\phi)}$. 
We would like to show that this overlap 
function is sharply 
peaked at $g=g'$, that is at $\phi =\phi'$ and $p=p'$. To make 
conclusions 
about the convergence 
behaviour for $t\to 0$, we again need the Poisson transformed 
expressions. These do not 
have to be calculated anew again, but can mutatis mutandis 
simply be taken over from the last subsection. We obtain :
\be \label{a3.2}
\psi^{2t}_{\tilde{H}} (\tilde{h}) = 
\sqrt{\frac{\pi}{t}} \sum_n e^{-\frac{2\pi^2 n^2 
+ \frac{1}{2} (\phi -\phi')^2 + 2i\pi n(p+p') -i(\phi -\phi')(p+p') 
-2\pi n(\phi-\phi') -\frac{1}{2} (p+p')^2}{t}}
\ee
\be \label{a3.3}
\psi^{2t}_{H^2} (1)=\sqrt{\frac{\pi}{t}} 
\sum_n e^{-\frac{\pi^2 n^2 +2i\pi np -p^2}{t}}
\ee
\be \label{a3.4}
\psi^{2t}_{(H')^2}(1)=\sqrt{\frac{\pi}{t}} 
\sum_n e^{-\frac{\pi^2 n^2 +2i\pi np' -(p')^2}{t}}
\ee
Inserting these results into (\ref{a3.1}) leads to
\be \label{a3.5}
i^t(g,g') = \frac{e^{-\frac{\frac{1}{2} (\phi -\phi')^2}{t}} 
e^{-\frac{\frac{1}{2} (p-p')^2}{t}} | \sum_n e^{-\frac{2\pi^2 n^2  
+ 2i\pi n(p+p') -i(\phi -\phi')(p+p') 
-2\pi n(\phi-\phi') }{t}} |^2} {D^t_p D^t_{p'}}
\ee
where $D^t_p$ was defined 
above. Now the argument goes completely analogously to the 
last subsection, that is, one 
calculates bounds for the (square of the modulus of 
the) series and for $D^t_p$, for the latter one can actually just take over the 
previous results.

The final result is :
\begin{Theorem}
There exist constants $K_t,\tilde{K_t'} $ (independent of $g,g'$), decaying 
exponentially fast to $0$ as $t\to 0$ such that
\be \label{a3.6}
i^t(g,g') \leq \frac{e^{-\frac{\frac{1}{2} (\phi -\phi')^2}{t}} 
e^{-\frac{\frac{1}{2} (p-p')^2}{t}} (1+\tilde{K_t'} )} {(1-K_t)(1-K_t)}
\ee
\end{Theorem}

\subsubsection{Peakedness in the Electric Field Representation}
\label{sa.1.3}

In section (\ref{s3.3}) this representation was essentially defined as the 
``Fourier coefficients" 
of $\psi^t_g $ with respect to the orthonormal system $|jmn>$, 
given by
\be \label{a4.1}
\tilde{\psi}^t_g (jmn)=e^{-tj(j+1)/2} \pi_j (g)_{mn}
\ee
In the case of $U(1)$, where all 
irreducible representations are one-dimensional, 
the corresponding orthonormal 
system is labelled only by the set of integers $|n>$, 
with $n$ corresponding to the $j$ above. 
So for the state $\psi^t_g $ in the electric 
field representation we have
\be \label{a4.2}
\tilde{\psi}^t_g (n)=e^{-tn^2/2} e^{in\phi } e^{np}.
\ee
The aim of this section is to show that
\be \label{a4.3}
p^t_g(n):=\frac{|\tilde{\psi}^t_g (n)|^2}{||\psi^t_g ||^2} = \frac{e^{-tn^2} 
e^{2np}}{||\psi^t_g ||^2}
\ee
is peaked at $tn=p$. 
This would complement the peakedness property in the connection 
representation, leading to 
the conclusion that the coherent states used here have the 
desirable property to be 
"localised" at the phase space point given by $g$. The proof 
is straightforward given the results of the previous sections. Recalling that
\be \label{a4.4}
||\psi^t_g ||^2 = \psi^{2t}_{H^2} (1) 
= \sqrt{\frac{\pi}{t}} e^{\frac{p^2}{t} } D^t_p
\ee
and the estimate for $D^t_p$ we can conclude that
\ba \label{a4.5}
p^t_g(n) & \leq & 
\frac{\sqrt{\frac{t}{\pi}} e^{-tn^2} e^{2np} e^{-p^2/t}}{(1-K_t)} 
\nonumber \\
&=& \frac{\sqrt{\frac{t}{\pi}} e^{-\frac{(tn-p)^2}{t}}}{(1-K_t)}
\ea

From this we 
immediately see that $p^t_g$ is bounded as $n\to \infty$ and that it is 
peaked sharply at $tn=p$ as desired. Other than in the $SU(2)$ case there is no 
qualitatively different behaviour according to whether $p$ is large or not. 
The reason is simply that $p$ shows up in the exponent only in this case.
Notice again that 
in order to get an approximately continuum  momentum distribution
we should introduce $p_n=nt$ 
as a summation variable and have to divide (\ref{a4.5})
by $\Delta p_n=t$ which then approaches 
indeed a $\delta$-distribution as $t\to 0$.

\subsubsection{Uncertainty Relation and Phase Space Bounds}
\label{sa.1.4}

There are two 
things we would like to show in this section. First we want to verify 
for the case 
of $U(1)$ that the overlap function $i^t_(g,g')$ can be interpreted as 
the probability 
density $p^t(g,g') $ to find the system at the phase space point $g'$ 
in the state 
$\hat{U}_t \psi^t_g $ (see section \ref{s3} for the definition of 
$\hat{U}_t $) times the volume of a phase space cell. 
Second we will 
calculate the commutator between the operators $\hg$ and $\hgd $ to 
verify that it 
has the correct semiclasssical limt, that is, the Poisson bracket 
between $g$ and $\bar{g}$, 
thus ensuring the validity of the Heisenberg uncertainty 
bound.

Our first task is to determine the measure $d\nu_t$ on the target space of 
the coherent state transform. We recall its definition from (\ref{3.4}):
\be \label{a5.1}
\hat{U}_t\; : \; L_2(G,d\mu_H)\mapsto {\cal H}L_2(G^\Co,d\nu_t);\;f\mapsto 
(\hat{U}_t f)(g):=<\overline{\psi^t_g},f>
\ee
where in our case $G$ is $U(1)$, $d\mu_H=\frac{d\phi}{2\pi}$ and 
$G^{\Co} =\Co-\{0\}$. 
The measure $d\nu_t$ is to be determined from the unitarity 
requirement of the transform. It is easiest to check that requirement given by
\be \label{a5.2}
<\hat{U}_t f,\hat{U}_t f'>_{\nu_t} = <f,f'>_{\mu_{H}} 
\ee
for any $f,f'$ $ \in L_2 (U(1),d\theta)$ 
on the basis of the electric field representation. 
There we have $(\hat{U}_t |n>)(g)
= e^{-n^2t/2} e^{in\phi} e^{np}$ so the unitarity 
condition reads
\be \label{a5.3}
\int \frac{d\phi}{2\pi} 
d\sigma (H) e^{-n^2t/2} e^{-in\phi} e^{np} e^{-(n')^2 t/2} 
e^{in'\phi } e^{n' p} = \int \frac{d\phi}{2\pi} e^{i\phi (n-n')}
\ee
where we made the product ansatz 
$d\nu_t = d\mu_{H} d\sigma_t$. This simplifies to
\be \label{a5.4}
\int d\sigma_t (H) e^{-tn^2} e^{2np} = 1
\ee
which by inspection is solved by a 
Gaussian measure in $p$. The precise result is 
given in the following Lemma: 
\begin{Lemma}
The measure $\nu_t$ on the target space of the 
coherent state transform is given by
\be \label{a5.5}
d\nu_t(g) = d\mu_{H}(u) d\sigma_t(H) = d\mu_{H}(u) \sqrt{\frac{1}{\pi t}} 
e^{-\frac{p^2}{t}} dp=:\nu_t(g) d\Omega 
\ee
where $g=Hu $ is the polar decomposition of $g$ and $d\Omega=d\mu_H(u) dp$ is 
the 
Liouville measure on $T^\ast U(1)$. 
\end{Lemma}

Now we are ready to address our first problem. 
We take over from section (\ref{s3.4}) 
the general expression for $p^t(g,g')$:
\be \label{a5.6}
p^t(g,g') = \nu_t (g') \frac{| (\hat{U}_t \psi^t_g)(g')|^2}{||\psi^t_g||^2} = 
\nu_t (g') ||\psi^t_{g'} ||^2 i^t(g,(g')^{\star}) 
\ee
where $\star $ 
in the $U(1)$ case is just complex conjugation. Then using the previous
results (\ref{a2.7}), (\ref{a2.9}), (\ref{a2.10}) that
\be \label{a5.7}
\sqrt{\frac{\pi}{t}} e^{\frac{p^2}{t}} (1-K_t) \leq ||\psi^t_{g'} ||^2 \leq 
\sqrt{\frac{\pi}{t}} e^{\frac{p^2}{t}} (1+ K_t) \ee
and the expression for $\nu_t $ we find:
\be \label{a5.8}
\frac{2\pi}{(2\pi t)} (1-K_t) \leq \frac{p^t (g,g')}{i^t(g,(g')^{\star})} 
\leq \frac{2\pi}{(2\pi t)} (1+ K_t)
\ee
for some constant $K_t$ exponentially vanishing for $t\to 0$. 
We summarize :
\begin{Theorem}
The overlap function 
$i^t(g,g')$ approaches exponentially fast with $t\to 0$ the 
function $p^t (g,(g')^\star)t$ where $p^t (g,g')$ denotes the 
probability density 
to find the system at the phase space point $g'$ in the state 
$\hat{U}_t \psi^t_g $ with respect to the measure $d\Omega$ in the 
phase space $T^{\ast} U(1)$. 
\end{Theorem}

Thus the phase space 
volume occupied by a coherent state with respect to the measure 
$d\mu_{H} dp$ is given by $\propto t \propto \hbar $.

We now come to the commutator calculation. The classical variables are $g$ and 
$\bar{g}$ where $g=Hu = e^p e^{i\phi }$. For their Poisson bracket we get
\ba \label{a5.9}
\{ g,\bar{g} \} &=& \{ e^p e^{i\phi },e^p e^{-i\phi} \} \nonumber \\
        &=& \{e^p,e^{-i\phi } \} e^{i\phi } e^p + \{ e^{i\phi }, e^p\} e^p 
e^{i\phi} \nonumber \\
     &=& -i e^p e^{-i\phi} e^{i\phi} e^p - i e^{i\phi} e^p e^p e^{-i\phi} 
\nonumber \\
                &=& -2i e^p
\ea
This Poisson bracket 
should be proportional to the first order term (in t) of the 
expression
\be \label{a5.10}
\frac{<\psi^t_g, [\hg,\hgd ]\psi^t_g >}{||\psi^t_g ||^2}
\ee
One of the characteristic properties of 
coherent states is that $\hg \psi^t_g (h) = 
g\psi^t_g (h)$, so one term of the commutator is easy :
\be \label{a5.11}
< \psi^t_g , \hgd \hg \psi^t_g > 
= (\bar{g} g) ||\psi^t_g ||^2 = e^{2p} ||\psi^t_g ||^2
\ee
The second term requires a bit more work. We recall the following expressions
\ba \label{a5.12}
&& \psi^t_g (h) = \sum_n e^{-n^2t/2} (gh^{-1} )^n \nonumber \\
&& \hgd = e^{-t \Delta /2} (\hat{h}^{-1} )  e^{t \Delta /2}
\ea
from which we calculate
\ba \label{a5.13}
(\hgd \psi^t_g )(h) & = &  e^{-t \Delta /2} (\hat{h}^{-1}) \sum_n  e^{-n^2t/2}  
e^{-n^2t/2} (gh^{-1} )^n \nonumber \\
      &=&  e^{-t \Delta /2} \sum_n  e^{-n^2t/2} g^n h^{-n -1} \nonumber \\
              &=& \sum_n  e^{-t n^2}  e^{\frac{t}{2} (n+1)^2} g^n h^{-n-1}
\ea  
Applying $\hg $ in a similar way we obtain
\be \label{a5.14}
(\hg \hgd \psi^t_g )(h) = \sum_n  e^{-tn^2}  e^{\frac{t}{2} (n+1)^2}  
e^{\frac{t}{2} (n+1)^2}  e^{-\frac{t}{2} n^2} g^n h^{-n}
\ee
and thus after some simple algebra
\ba \label{a5.15}
<\psi^t_g , \hg \hgd \psi^t_g > &=& \sum_n  
e^{2nt+t} e^{-tn^2} e^{2np} \nonumber \\
               &=&   \sum_n  e^{-t(n-1)^2} e^{2t} e^{2np} \nonumber \\
               &=& e^{2t} e^{2p} ||\psi^t_g ||^2
\ea
where we relabelled the summation index in the last line. Combining all results 
we find
\ba \label{a5.16}
\frac{<\psi^t_g, [\hg,\hgd ]\psi^t_g >}{||\psi^t_g ||^2} &=& (e^{2t} -1) e^{2p} 
\nonumber \\
             &=& 2te^{2p} + O(t^2) \nonumber \\
             &=& it \{ g, \bar{g} \} + O(t^2)
\ea
which is the desired result.

\subsection{Peakedness Proofs for Gauge-Invariant Coherent States}
\label{sa.2}

In section 
\ref{s4} of the main text first the notion of gauge-invariant coherent 
states was introduced and several properties were proved. This was done for a 
general compact Lie group so there is no need to repeat that part here for 
$U(1)$. Then the discussion was specialized to the case of $SU(2)$ where the 
peakedness in 
the connection representation, the electric field representation and for the 
overlap 
function was proved. We will not repeat those proofs here for the case of 
$U(1)$, rather we will illustrate them by means of a concrete example. To avoid 
tedious book-keeping 
problems we take a very simple graph $\gamma_0$ for the coherent 
state to be 
considered. It consists of two vertices $v_1,v_2$ which are connected by 
three edges $e_1,e_2,e_3$. 
Without loss of generality we can assume that the edges are outgoing from 
the same vertex. 
This example will be underlying all the discussion in the following 
subsections.

\subsubsection{Peakedness of the Overlap function}
\label{sa.2.1}

In this subsection we want to illustrate theorem \ref{th4.1} by explicitly 
calculating 
the last line of (\ref{4.13}) for our example. First we determine the form of 
$j^t_{\g}(\vg,\vgp)$ and then perform the integrations over the gauge 
group of 
which there 
are two in our case. One remark about the notation: "$\approx $" will stand for 
equality in the limit $t\to 0$, that is, terms which vanish to first 
order in this 
limit are omitted. A general $J^t_{\gamma}(\vg,\vgp)$ is given in terms of
\be \label{a6.1}
j^t_{\g}(\vg,\vgp^{\vh}) = 
\frac{<\psi^t_{\g,\vg},\psi^t_{\g,\vgp^{\vh}}>}{||\psi^t_{\g,\vg}||\;
||\psi^t_{\g,\vgp^{\vh}}||}
\ee
The expression for the norms for our special graph $\g$ can be taken over 
from 
the previous sections: 
\be \label{a6.2}
||\psi^t_{\g,\vg}|| = (\frac{\pi}{t})^{3/4} 
e^{\frac{\sum_{i=1}^{3} p_i^2}{2t}} 
(1-K_t)
\ee
\be \label{a6.3}
||\psi^t_{\g,\vgp^{\vh}}|| = ||\psi^t_{\g,\vgp}|| = 
(\frac{\pi}{t})^{3/4} e^{\frac{\sum_{i=1}^{3} (p_i')^2 }{2t}} (1-K_t')
\ee
To calculate the numerator we introduce the two gauge angles $\theta_1,
\theta_2$, 
associated with the gauge transformations at $v_1,v_2$ and their difference 
$\Delta \theta = \theta_2 - \theta_1 $. From (\ref{a3.2}) we find by inserting 
$\Delta \theta$ in the right place
\ba \label{a6.4}
<\psi^t_{\g,\vg},\psi^t_{\g,\vgp^{\vh}}> & =& (\frac{\pi}{t})^{3/2} 
e^{\frac{-3(\Delta \theta )^2 }{4t}} 
e^{\frac{-\Delta \theta \sum_{i=1}^{3} (\phi_i -\phi_i ') }{2t}} 
e^{\frac{i \Delta \theta \sum_{i=1}{3} (p_i + p_i') }{2t}} \times \nonumber \\
& & \times  e^{\frac{-\frac{1}{2} \sum_{i=1}^{3} (\phi_i - \phi_i')^2 }{2t}} 
e^{\frac{i \sum_{i=1}^{3} (\phi_i -\phi_i') (p_i +p_i') }{2t}} 
e^{\frac{\frac{1}{2} \sum_{i=1}^{3} (p_i +p_i')^2 }{2t}} (1+\tilde{K}_t )
\ea
Only the first three exponents are relevant for the integration so we will 
separate 
them out. Also, due to the special dependence of the expression on the gauge 
angles we 
perform a change of integration variables. Instead of integrating over 
$\theta_1$ an
d $\theta_2$ we will integrate over $\theta_{-} = (\theta_2 -\theta_1)/2 = 
\Delta \theta /2$ and  $\theta_{+} = (\theta_2 +\theta_1)/2 $ where the 
latter integration is trivial. We obtain for the integral
\ba \label{a6.5}
Int   &: = & \frac{1}{4\pi^2} \int_{-\pi}^{\pi} d\theta_{-} 
\int_{|\theta_{-}|}^{2\pi -|\theta_{-} |} d\theta_{+} 
2 e^{\frac{-3\theta_{-} ^2 }{t}} 
e^{\frac{- \theta_{-} (\sum_{i=1}^{3} (\phi_i -\phi_i ') -i (p_i + p_i'))}{t}} 
\nonumber \\
&=& \frac{1}{4\pi^2} 4 \int_{-\pi}^{\pi} d\theta_{-} (2\pi - |\theta_{-} |) 
e^{\frac{-(3\theta_{-}^2+\theta_{-}(\sum_{i=1}^{3} (\phi_i-\phi_i ')
-i(p_i+p_i')))}{t}}
\nonumber \\
&=&  \frac{1}{4\pi^2} 4 \sqrt{t} \int_{-\pi /\sqrt{t}}^{\pi /\sqrt{t}} dx 
(\pi - \sqrt{t}|x |) e^{-3(x^2 + \frac{x ( \sum_{i=1}^{3} (\phi_i -\phi_i ') 
+i (r_i + r_i'))}{3\sqrt{t}})} \nonumber \\
& \approx &  \frac{1}{4\pi^2} 4 \sqrt{t} \int_{-\infty}^{\infty } dx 
(\pi - \sqrt{t}|x |) e^{-3(x^2 + \frac{x ( \sum_{i=1}^{3} (\phi_i -\phi_i ') 
-i (p_i + p_i'))}{3\sqrt{t}})}
\ea
The second term is a derivative which when evaluated gives a finite result 
independent 
of $t$ times $t$ and therefore vanishes in the limit $t\to 0$. Thus we can 
continue
\ba \label{a6.6}
Int & \approx&  \frac{1}{4\pi^2} 4 \pi \sqrt{t} \int_{-\infty}^{\infty} dx  
e^{-3(x^2 + \frac{x ( \sum_{i=1}^{3} (\phi_i -\phi_i ') 
-i (p_i + p_i'))}{3\sqrt{t}})} \nonumber \\
&=&  \frac{1}{4\pi^2} 4 \pi \sqrt{t} e^{\frac{( \sum_{i=1}^{3} 
(\phi_i -\phi_i ') 
-i (p_i + p_i'))^2}{12t}} \int_{-\infty}^{\infty} dx  e^{-3x^2} \nonumber \\
&=&  \frac{1}{\sqrt{3\pi}}  \sqrt{t} e^{\frac{( \sum_{i=1}^{3} 
(\phi_i -\phi_i ') 
-i (p_i + p_i'))^2}{12t}} 
\ea
Putting everything together we get the following expression 
\be \label{a6.7}
j^t_\g(\vg,\vgp^{\vh}) =  \frac{ (\frac{\pi}{t})^{3/2} 
\sqrt{\frac{t}{3\pi}} 
e^{\frac{( \sum_{i=1}^{3} (\phi_i -\phi_i ') -i (p_i + p_i'))^2}{12t}}  
e^{\frac{-\frac{1}{2} \sum_{i=1}^{3} (\phi_i - \phi_i')^2 }{2t}} 
e^{\frac{i \sum_{i=1}^{3} (\phi_i -
\phi_i') (p_i +p_i') }{2t}} e^{\frac{\frac{1}{2} \sum_{i=1}^{3} 
(p_i +p_i')^2 }{2t}} 
(1+\tilde{K}_t ) }{ (\frac{\pi}{t} )^{3/2} e^{\frac{\sum_{i=1}^3 (p_i^2 
+ (p_i')^2 )}{2t}} (1-K_t)(1-K_t')}
\ee
for 
$$
J^t_\g(\vg,\vgp):=\int d\mu_H(h_1) d\mu_H(h_2) j^t_\g(\vg,\vgp^{\vh})
$$
This result is of course the same as for the 
integral over $j^t_\g(\vg,\vgp^{\vhp})$, so the 
numerator of $I^t_\g(\vg,\vgp)$ is just given by the norm square of 
(\ref{a6.7}) (ignoring the norms which cancel with those in the denominator.) 
The $j^t_\g$ terms 
in the denominator can be obtained from (\ref{a6.7}) by setting $\phi_i =
\phi_i' $ 
and $p_i = p_i'$. Putting everything together and after some lengthy 
algebra we obtain 
as the final result :
\be \label{a6.8}
I^t_{\g} (\vg,\vgp) = e^{-\frac{\frac{1}{3} \sum_{i<j} (\phi_i - \phi_i' 
+ \phi_j' 
-\phi_j )^2}{2t}} e^{-\frac{\sum_{i=1}^3 (p_i - p_i')^2}{2t}} (1+ \bar{K}_t)
\ee
where $\bar{K}_t$ is a constant that goes to $0$ for $t\to 0$ exponentially
fast. This expression 
obviously has the same peakedness property as in the gauge-variant 
case, but the 
$\phi_i$ appear in a gauge-invariant way as should be expected. The $p_i$ 
are of course
gauge-invariant by themselves. We still want to make the relation between the 
gauge-variant and the gauge-invariant overlap functions a bit clearer. 
To do so we 
choose as a set of independent variables the following combinations of the 
$\phi_i$:
\ba \label{a6.9}
  && x_1 = \phi_2 - \phi_1 \nonumber \\
  && x_2 = \phi_3 - \phi_1 \nonumber \\
  &&  \phi_1
\ea 
and similar for the primed angles. It follows that 
$\dl \phi_2 = \phi_2 - \phi_2' = 
\dl x_1+\dl \phi_1$ and $\dl \phi_3 = \dl x_2 +\dl \phi_1$. The exponent for 
$i^t_{\g}$ in the new variables reads then
\ba \label{a6.10}
(\dl \phi_1)^2+ (\dl \phi_2)^2+ (\dl \phi_3)^2 &=& (\dl \phi_1)^2 + (\dl x_1 + 
\dl \phi_1)^2 + (\dl x_2 + \dl \phi_1)^2 \nonumber \\
&=& 3(\dl \phi_1 +\frac{\dl x_1 +\dl x_2}{3})^2 +\frac{2}{3}((\dl x_1)^2 + 
(\dl x_2 )^2 -\dl x_1 \dl x_2) \nonumber \\
&& {}
\ea
where we left out some simple manipulations. To see the peakedness of the expression 
we have to render the last term into a quadratic form. The ansatz
\be \label{a6.11}
(\dl x_1)^2 + (\dl x_2 )^2 -\dl x_1 \dl x_2 = \beta 
[(\dl x_1 - \alpha \dl x_2)^2 + (\dl x_2 - \alpha \dl x_1)^2 ]
\ee
leads to $\beta = \frac{1}{4\alpha}$ and $\alpha=2\pm \sqrt{3}$. Thus we find
\be \label{a6.12}
(\dl \phi_1)^2+ (\dl \phi_2)^2+ (\dl \phi_3)^2 = 3(\dl \phi_1 +\frac{\dl x_1 
+\dl x_2}{3})^2 +\frac{2}{3} \frac{1}{4\alpha}  [(\dl x_1 - \alpha \dl x_2)^2 
+ (\dl x_2 - \alpha \dl x_1)^2 ]
\ee
which is peaked at $\dl x_1 = \dl x_2 =\dl \phi_1 =0$ as $\alpha\not= 1$. 
Performing similar manipulations on the exponent of $I^t_{\g}$ leads to
\be \label{a6.13}
\sum_{i<j} (\phi_i - \phi_i' +\phi_j - \phi_j' )^2 = 2 \frac{1}{4\alpha } 
[(\dl x_1 - \alpha \dl x_2)^2 + (\dl x_2 - \alpha \dl x_1)^2 ]
\ee
In conclusion,
\ba \label{a6.14}
i^t_\g &  = & \exp( -\frac{1}{3t} \{ \frac{1}{4\alpha} [(\dl x_1 - \alpha 
\dl x_2)^2 + (\dl x_2 - \alpha \dl x_1)^2 ] + \frac{1}{2} (3\dl \phi_1 
+ \dl x_1 +\dl x_2 )^2 \} ) \times \nonumber \\ 
& & \times e^{-\frac{\sum_{i=1}^3 (p_i - p_i')^2}{2t}} (1+ K_t) \nonumber \\
I^t_\g &  = & \exp( -\frac{1}{3t}  
\frac{1}{4\alpha} [(\dl x_1 - \alpha \dl x_2)^2 + 
(\dl x_2 - \alpha \dl x_1)^2 ]  ) 
\times \nonumber\\ 
&& \times e^{-\frac{\sum_{i=1}^3 (p_i - p_i')^2}{2t}} (1+ \bar{K}_t)
\ea
So we see that in the gauge $\dl\phi_1 +(\dl x_1 +\dl x_2)/3 =0$ both 
expressions 
become equal (as usual understood in the limit $t\to 0$). \\
Remark :\\
Notice that $p_i,\phi_i$ and $p'_i,\phi'_i$ respectively are defined by 
$g_i=g_{e_i}((A,E))$ and $g'_i=g_{e_i}((A',E'))$ respectively, that is,
they come from two {\it different} points in the phase space. Nevertheless,
they transform under the same gauge transformation function and so it
seems surprising that we have two independent sets of gauge functions 
$\theta_I,\theta_I',\;I=1,2$. The reason is simply that there are two
independent gauge group integrals appearing in (\ref{4.13}).

\subsubsection{Peakedness in the Connection Representation}
\label{sa.2.2}

In this subsection we want to illustrate the peakedness of the gauge-invariant 
probability density given by
\ba \label{a8.1}
P^t_{\g,\vg}(\vh) & =& 
\frac{|\Psi^t_{\g,\vg}(\vh)|^2}{||\Psi^t_{\g,\vg}||^2} \nonumber \\
&=& \frac{||\psi^t_{\g,\vg}||^2 \prod_{v\in V(\g)} \int_G d\mu_H (u_v) 
d\mu_H (u_v') b^t_{\g,\vg} (\vh^{\vec{u}}) 
\overline{ b^t_{\g,\vg} (\vh^{\vec{u}'}) } }{||\Psi^t_{\g, \vg}||^2}
\ea
$||\Psi^t_{\g, \vg}||^2 $ is nothing but $||\psi^t_{\g, \vg}||^2  
\prod_{v\in V(\g)} \int_G d\mu_H (u_v) j^t_{\g }(\vg,\vg^{\vu}) $ 
which we 
calculated already in the last section. It remains to determine 
$b^t_{\g,\vg}(\vh^{\vec{u}})$. As seen earlier the norm of 
$\psi^t_{\g,\vg} $ 
is given by
\be \label{a8.2}
||\psi^t_{\g,\vg}|| = (\frac{\pi}{t})^{3/4} e^{\frac{1}{4t} 
\sum_{i=1}^3 p_i^2} 
(1+K_t)
\ee
while
\be \label{a8.3}
\psi^t_{\g,\vg} (\vh^{\vec{u}}) = \prod_{i=1}^3 \sum_{n_i} 
e^{-\frac{t}{2} n_i^2} 
e^{in_i\phi_i +n_ip_i} e^{-i n_i\alpha_i +in_i \dl \theta}
\ee
where the $\alpha_i$ parameterize the $h_e$, $\dl \theta$ is again the 
difference of the gauge angles and $K_t$ etc. denote constants esponentially 
decaying to zero as $t\to 0$. After 
applying the Poisson summation 
formula and keeping only the relevant terms in an explicit form we obtain
\ba \label{a8.4}
\psi^t_{\g,\vg} (\vh^{\vec{u}}) &=& (\frac{2\pi}{t})^{3/2} 
e^{-\frac{\frac{3}{2} 
(\dl \theta)^2}{t}} e^{+\frac{i\dl \theta \sum_i p_i}{t}} 
e^{-\frac{\dl \theta \sum_i 
(\phi_i - \alpha_i )}{t}} \times \nonumber \\
& & e^{-\frac{\frac{1}{2}\sum_i(\phi_i-\alpha_i)^2}{t}}  
e^{\frac{i\sum_i(\phi_i-\alpha_i) p_i}{t}} e^{\frac{\frac{1}{2} 
\sum_i p_i^2}{t}} 
(1+\tilde{K}_t)
\ea
We see that the $\theta-$dependent terms are nearly the same as in the last 
section, 
the only difference being that $\phi_i' $ is substituted by $\alpha_i$ and 
$2t$ in the 
exponent by $t$. Thus we can take over the final result with minor changes only.
We obtain as the final result:
\be \label{a8.5}
P^t_{\g,\vg}(\vh) = (2\sqrt{\frac{\pi}{t}})^3 [\sqrt{\frac{t}{3\pi}}] 
e^{-\frac{\sum_{i<j} 
(\phi_i - \alpha_i +\alpha_j - \phi_j )^2}{2t}} (1+ \tilde{K}_t )
\ee
where the volume factor $V_\g=[\sqrt{\frac{t}{3\pi}}]$ has popped 
out. As expected in section \ref{s4}, it is proportional to $\sqrt{t}$.
The result is obviously gauge invariant. To compare it with 
the gauge-variant case 
one should set the $\{ \alpha_i \}$ to $0$, as we calculated peakness for 
$\psi^t_g(1)$ there. A further analysis of their relation can be done 
analogously to 
the last section. We will not repeat this here.

\subsubsection{Peakedness in the Electric Field Representation}
\label{sa.2.3}

We recall from section \ref{s4.3} the form of a gauge-invariant coherent 
state in the 
electric field representation
\be \label{a7.1}
\tilde{\Psi}^t_{\g \vg}(\vec{j}, \vec{J}) = 
e^{-\frac{t}{2}\sum_{e\in E(\g)} j_e(j_e+1) } T_{\g \vec{j} \vec{J}}(\vec{g})
\ee
which in our case becomes
\be \label{a7.2}
\tilde{\Psi}^t_{\g \vg}(\vec{n}) = e^{-\frac{t}{2}\sum_i n_i^2} 
T_{\g \vec{n} }(\vec{g}) =  e^{-\frac{t}{2}\sum_i n_i^2} e^{i\sum_i n_i 
\phi_i } e^{\sum_i n_i p_i} 
\ee
with the additional condition that only those $\{n_i\}$ are allowed for which 
$\sum_i n_i = 0$. Analogously, the Gauss constraint requires that 
$\sum_i p_i=0$.
The aim is now to prove peakedness for the gauge-invariant probability 
amplitude given 
by
\be \label{a7.3}
P^t_{\g, \vg }(\vec{n}) = 
\frac{|\tilde{\Psi}^t_{\g,\vg}
(\vec{n})|^2}{||\tilde{\Psi}^t_{\g,\vg}||^2} \ee
The numerator is immediately obvious from above:
\be \label{a7.4}
||\tilde{\Psi}^t_{\g \vg}(\vec{n})|^2 =  e^{-t\sum_i n_i^2} e^{2\sum_i 
n_i p_i} 
\ee
On the other hand we verify as in the main text that
\be \label
||\tilde{\Psi}^t_{\g \vg}||^2 = ||\tilde{\psi}^t_{\g \vg}||^2 
\cdot V_{\g} 
\ee
which finally leads to
\be \label{a7.5}
P^t_{\g,\vg }(\vec{n}) = \frac{(\frac{t}{\pi})^{3/2} 
e^{-\frac{\sum_i (tn_i -p_i)^2}{t}}}{V_{\g}}
\ee
where still the additional condition $\sum_i n_i =0$ holds. 
This condition can be interpreted as the remains of the Clebsch-Gordan 
coefficients which appear in the $SU(2)$ case.

\clearpage
\newpage

\section{Graphical Illustration of Peakedness}
\label{sb}

This appendix contains extensive graphics, illustrating the
peakedness properties of the heat kernel coherent states for 
$G=U(1),SU(2)$. Except for the
first four figures, all graphics were calculated by Mathematica on the basis 
of a numerical approximation of the respective Poisson transformed
formulas. 

\subsection{The $U(1)$ case}
\label{sb.1}

\subsubsection{Peakedness in the Connection Representation}
\label{sb.1.1}

We numerically compute the function $\frac{| \psi^t_g (h) |}{\|
\psi^t_g \| } $, with the parameterization $g = e^p e^{i \phi}, \;
h=e^{i \phi_0 } $. Without restriction, one can choose $\phi_0 = 0$ and 
plot only $p$ and $\phi$. For $p$ we choose the range
$[-5,5]$ while $\phi$ is varied over its full range $[-\pi,\pi]$. The 
first four plots are obtained without Poisson transformation.
We contrast those with figure 5 obtained after Poisson summation, 
demonstrating the drastic difference in the convergence behaviour.

\begin{figure}
\includegraphics[width=100mm]{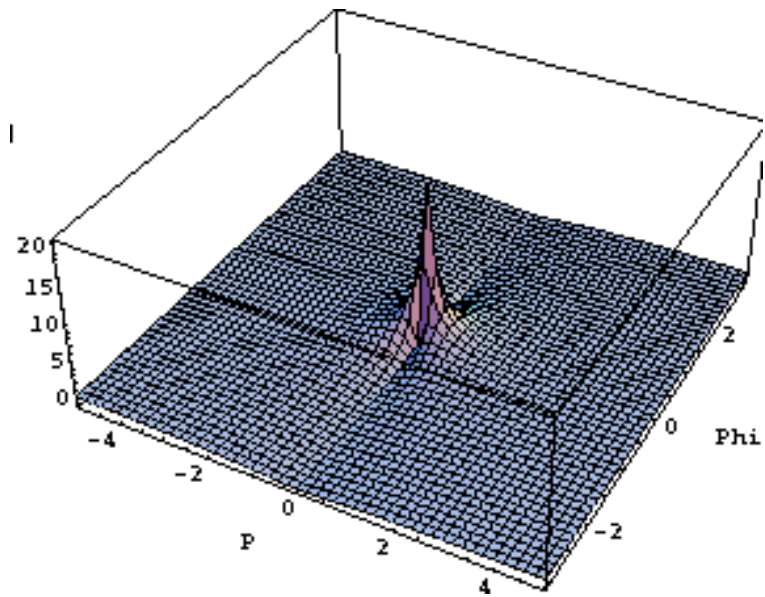}
\caption{$t=0.001, N=10$}
\end{figure} 

\begin{figure}
\includegraphics[width=100mm]{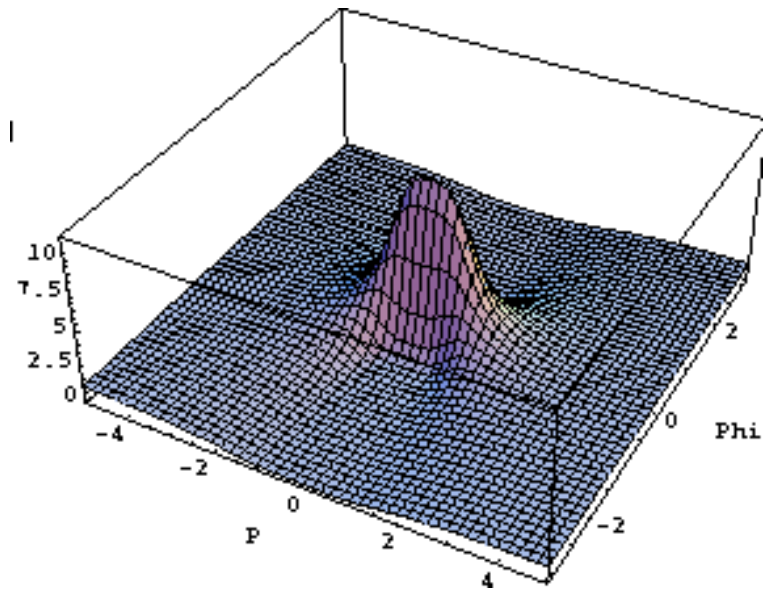}
\caption{$t=0.1, N=10$}
\end{figure} 

\clearpage
\newpage

\begin{figure}
\includegraphics[width=100mm]{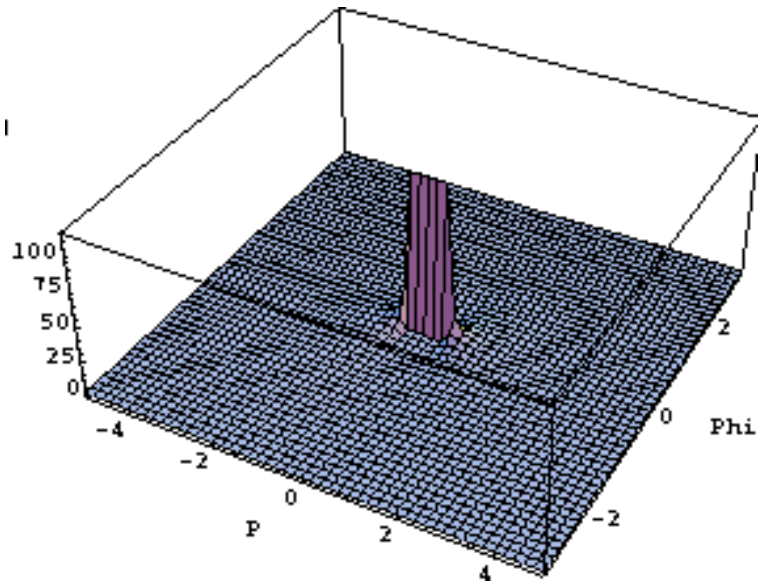}
\caption{$t=0.001, N=500$}
\end{figure} 

\begin{figure}
\includegraphics[width=100mm]{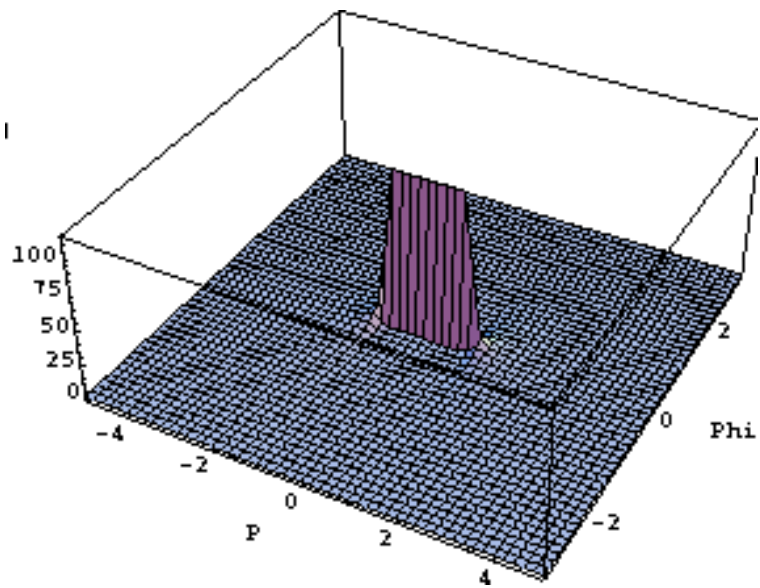}
\caption{$t=0.001, N=1000$}
\end{figure} 

\begin{figure}
\includegraphics[width=100mm]{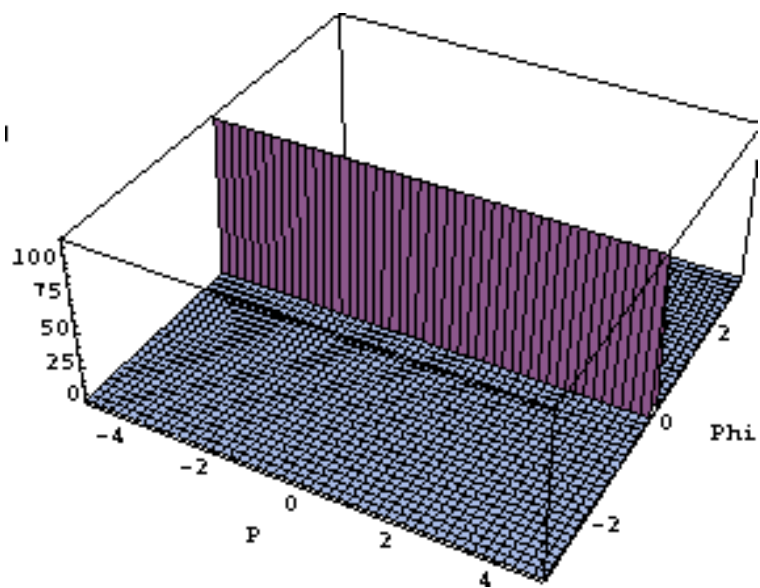}
\caption{$t=0.001, N=10$}
\end{figure} 

\clearpage
\newpage

Figures 1 and 2 reveal the bad convergence behaviour of the 
non-transformed series resulting in grossly misleading plots : 
taking the same number $N$ of summation terms in the non-transformed series, 
the one with the 
{\it higher} value of $t$ (figure 2) is the better approximation to the
actual situation (figure 5). To improve on the non-transformed
series, one has to considerably increase $N$, as shown in figures 
3 and 4.

\subsubsection{Peakedness of the Overlap Function}
\label{sb.1.2}

We compute the function $|<\psi^t_g,\psi^t_{g'}>|/
(||\psi^t_g||\;||\psi^t_{g'}||$ with the
parameterizations $g=e^{p_1} e^{i \phi } , \, g' = e^{p_2} e^{i \phi
'}$.
W.l.g. we set $\phi ' =0$, choose the range of $p_1$
to be $[p_2 - 5,p_2 +5]$ and take $\phi\in [-\pi,\pi]$. We
compute plots for the values $p_2=0,1,2,3,4$ in figures 6 through 10.

\begin{figure}
\includegraphics[width=100mm]{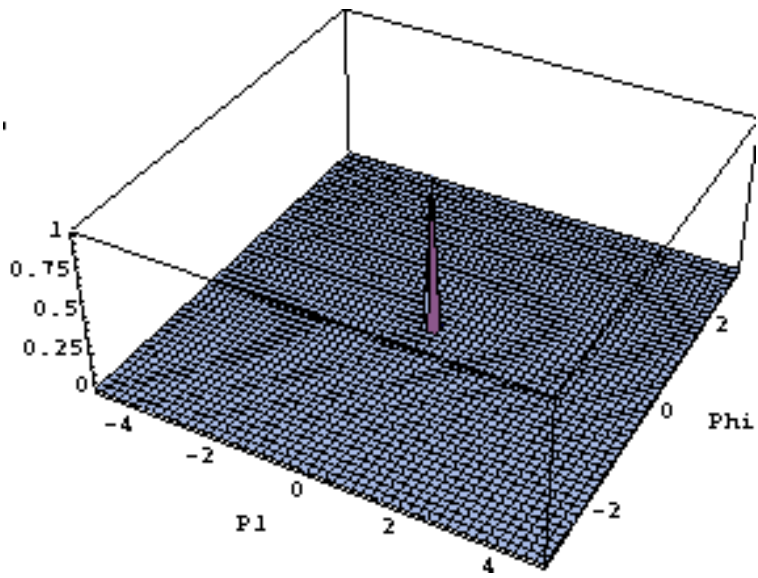}
\caption{$t=0.001, N=10, p_2 = 0$}
\end{figure} 

\begin{figure}
\includegraphics[width=100mm]{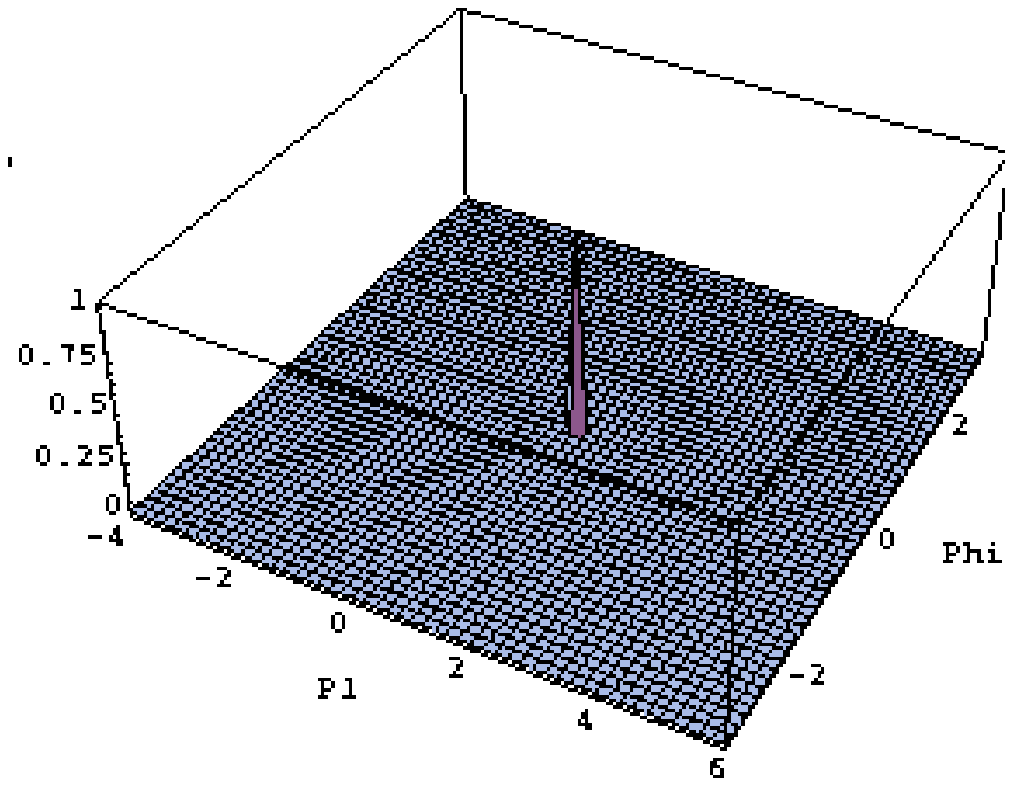}
\caption{$t=0.001, N=10, p_2 = 1$}
\end{figure} 

\clearpage
\newpage

\begin{figure}
\includegraphics[width=100mm]{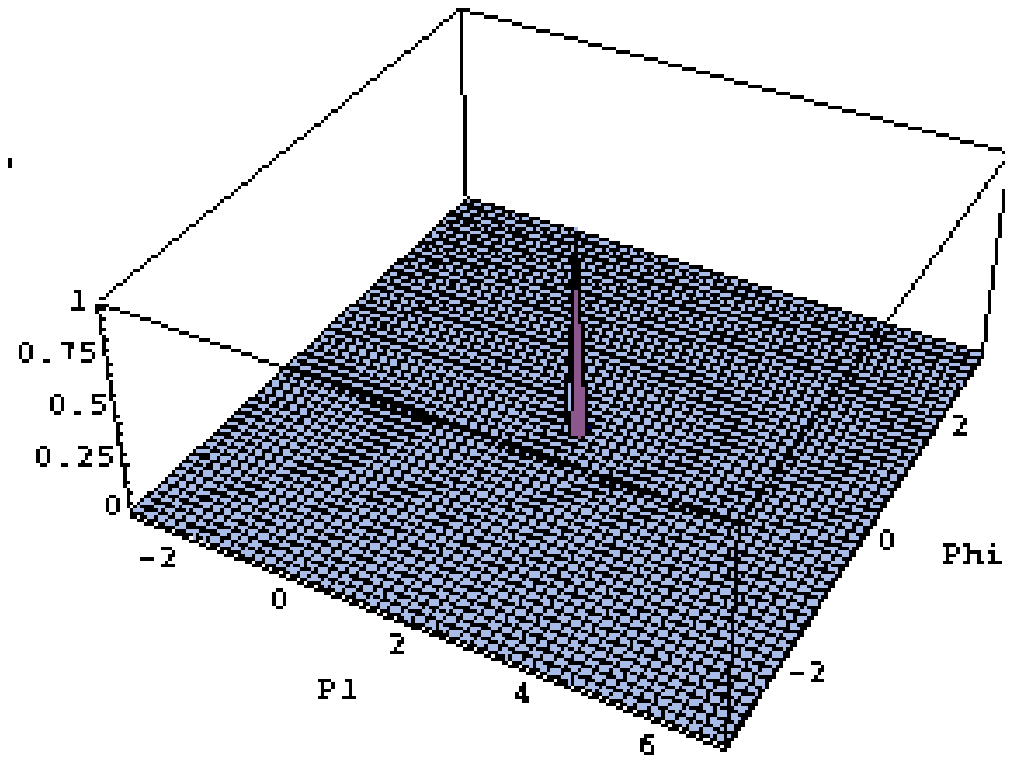}
\caption{$t=0.001, N=10, p_2 = 2$}
\end{figure} 

\begin{figure}
\includegraphics[width=100mm]{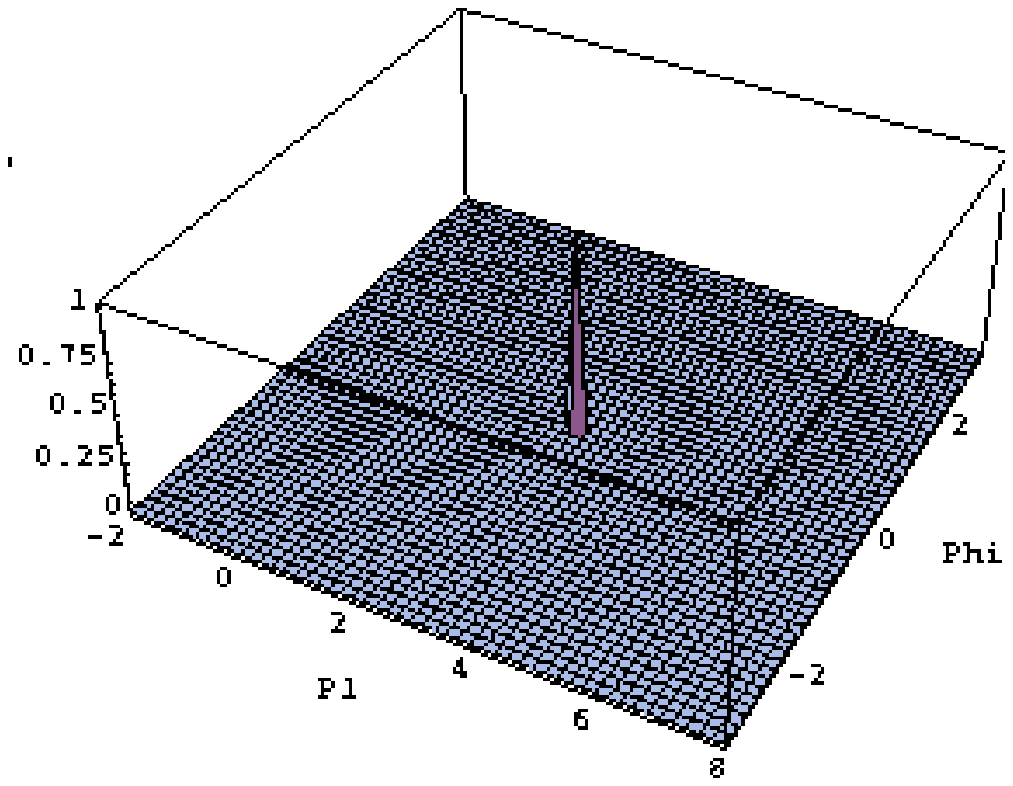}
\caption{$t=0.001, N=10, p_2 = 3$}
\end{figure} 

\begin{figure}
\includegraphics[width=100mm]{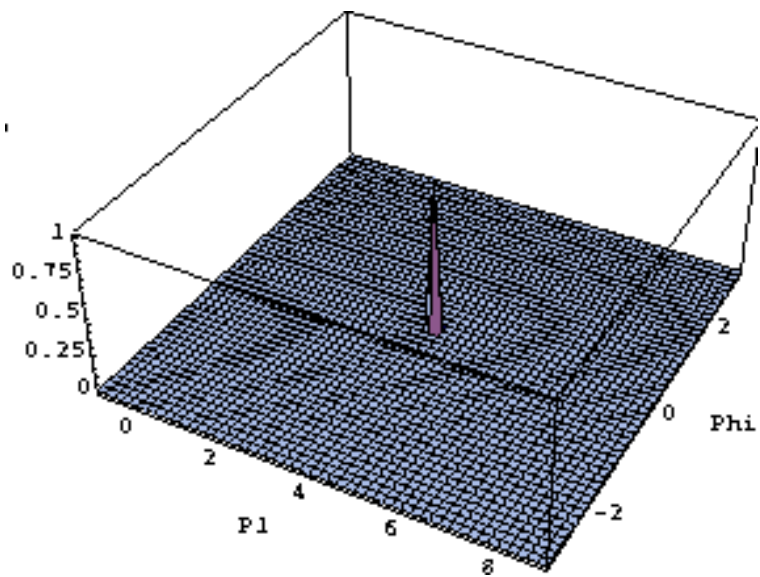}
\caption{$t=0.001, N=10, p_2 =4$}
\end{figure} 

\clearpage
\newpage

Finally, figure 11 resolves the exact nature of the peak, exhibiting
its Gauss-like structure. Its decay width is compatible with the expected 
value $\sqrt{t}\approx 0.03$.

\begin{figure}
\includegraphics[width=100mm]{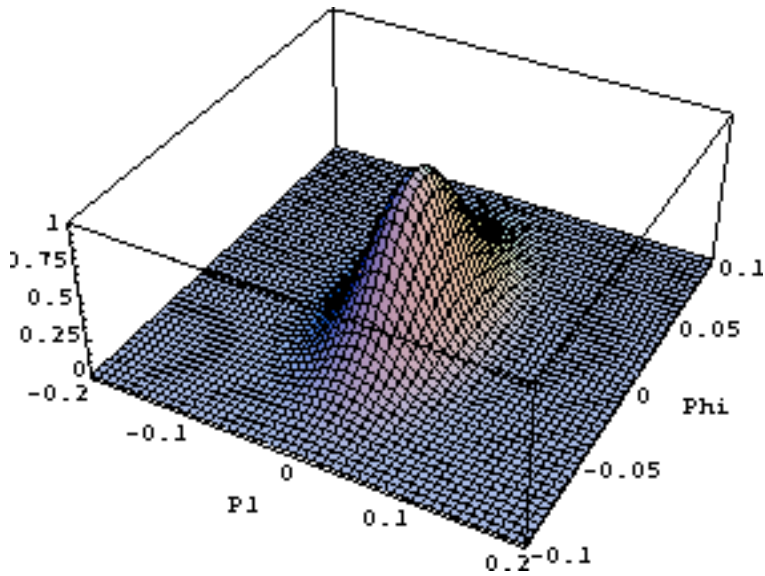}
\caption{$t=0.001, N=10, p_2 = 0$}
\end{figure} 
  
\subsection{The $SU(2)$ case}
\label{sb.2}

\subsubsection{Peakedness in the Connection Representation}
\label{sb.2.1}

We numerically compute the function $\frac{| \psi^t_g (h) |}{\|
\psi^t_g \| } $, with the parametrization $g = e^{-i p^j \tau_j/2 } 
e^{ \theta^j \tau_j}$. Without restriction, one can chosse $h = 1$.
For the parameterization vectors we take $p^j =(0,0,p)$ and $\theta^j
= \theta (\sin(\chi),\cos(\phi),\sin(\chi)\sin(\phi),\cos(\chi))$.
The variables for each graphic are $\theta \in [0,\pi ] $ and $\phi
\in [-\pi , \pi]$. Graphics were calculated for all combinations of
the values $p = \pm 1,\pm 2$ and $\chi = 0,\pi /4,\pi /2,3\pi /4,
\pi$. However, since they look completely identical in our resolution 
we display only the plots with $p=\pm 2$ in figures 12 through 21.

\begin{figure}
\includegraphics[width=100mm]{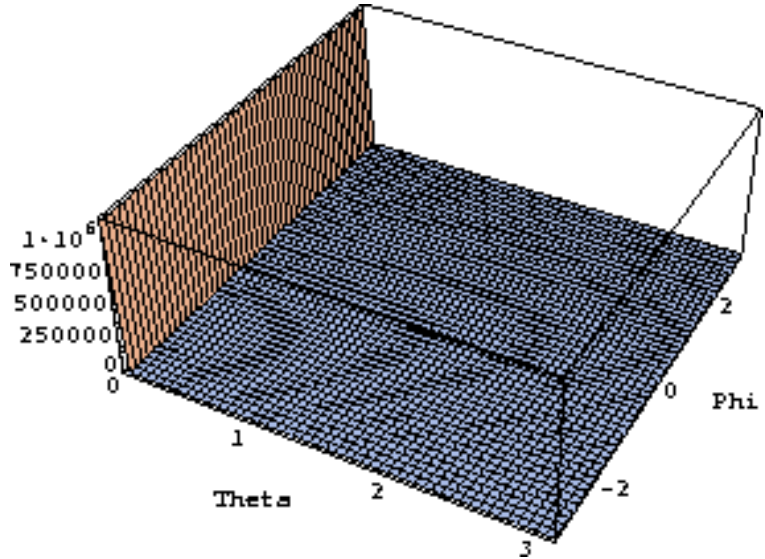}
\caption{$t=0.001, N=10, p=2, \chi=0$}
\end{figure} 

\clearpage
\newpage

\begin{figure}
\includegraphics[width=100mm]{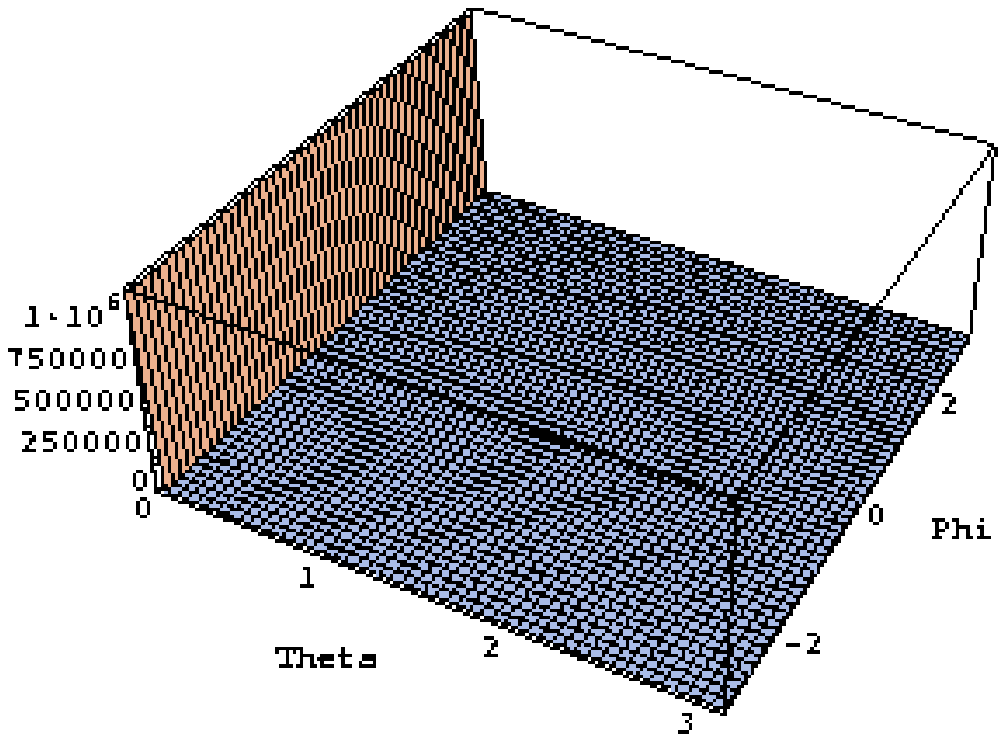}
\caption{$t=0.001, N=10, p=2, \chi= \pi/4$}
\end{figure} 

\begin{figure}
\includegraphics[width=100mm]{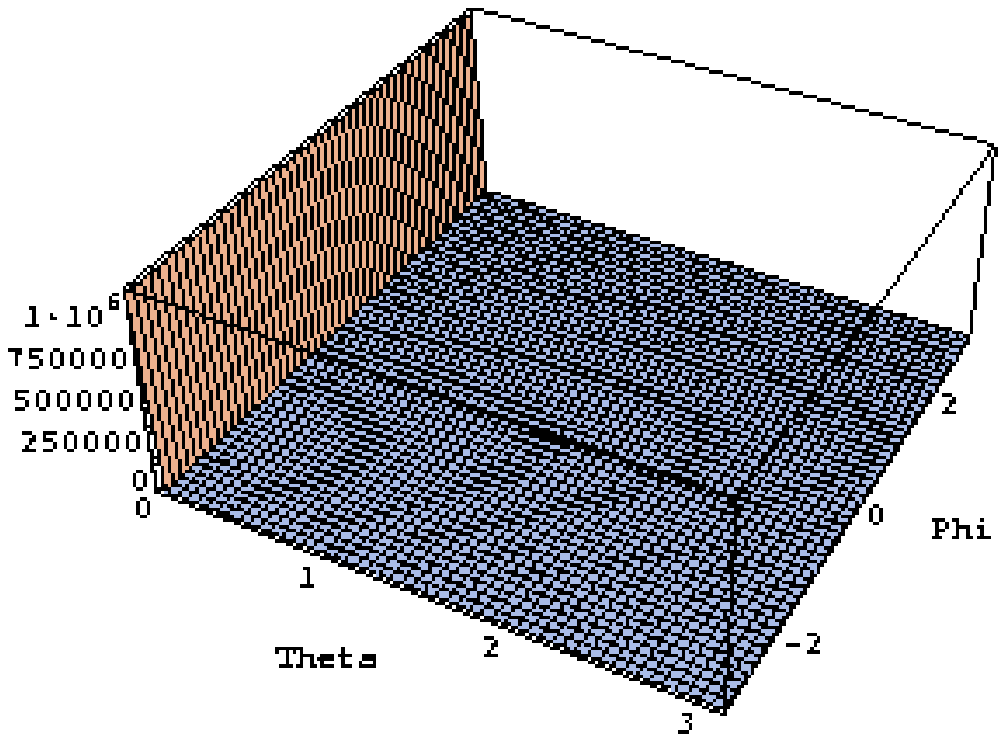}
\caption{$t=0.001, N=10, p=2, \chi= \pi/2$}
\end{figure} 

\begin{figure}
\includegraphics[width=100mm]{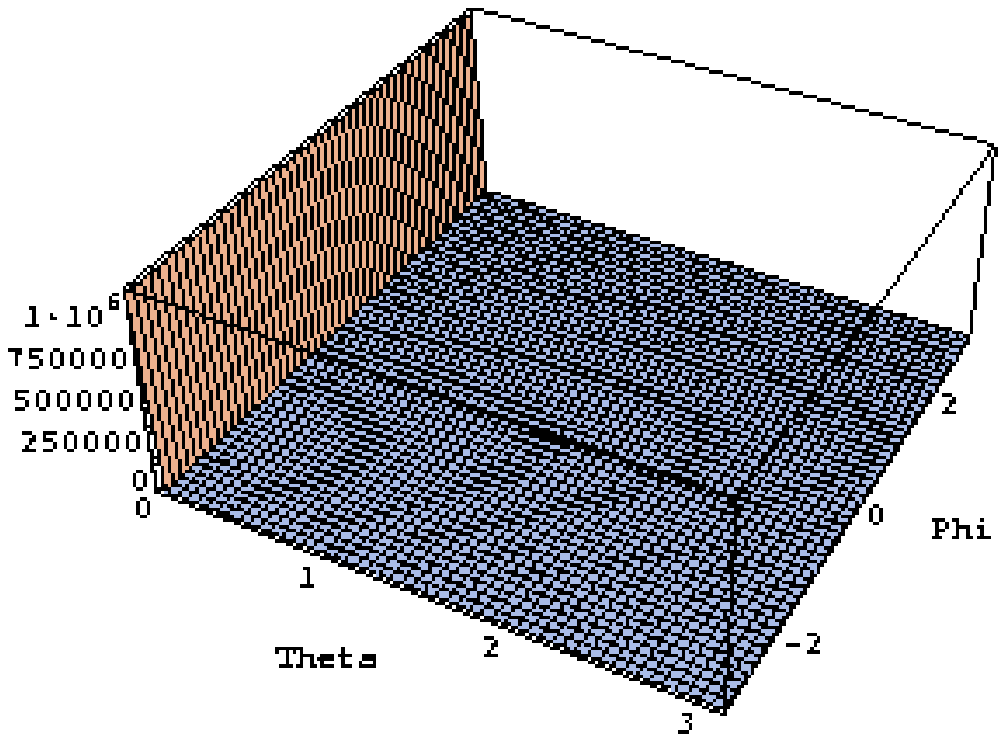}
\caption{$t=0.001, N=10, p=2, \chi= 3\pi /4$}
\end{figure} 

\clearpage
\newpage

\begin{figure}
\includegraphics[width=100mm]{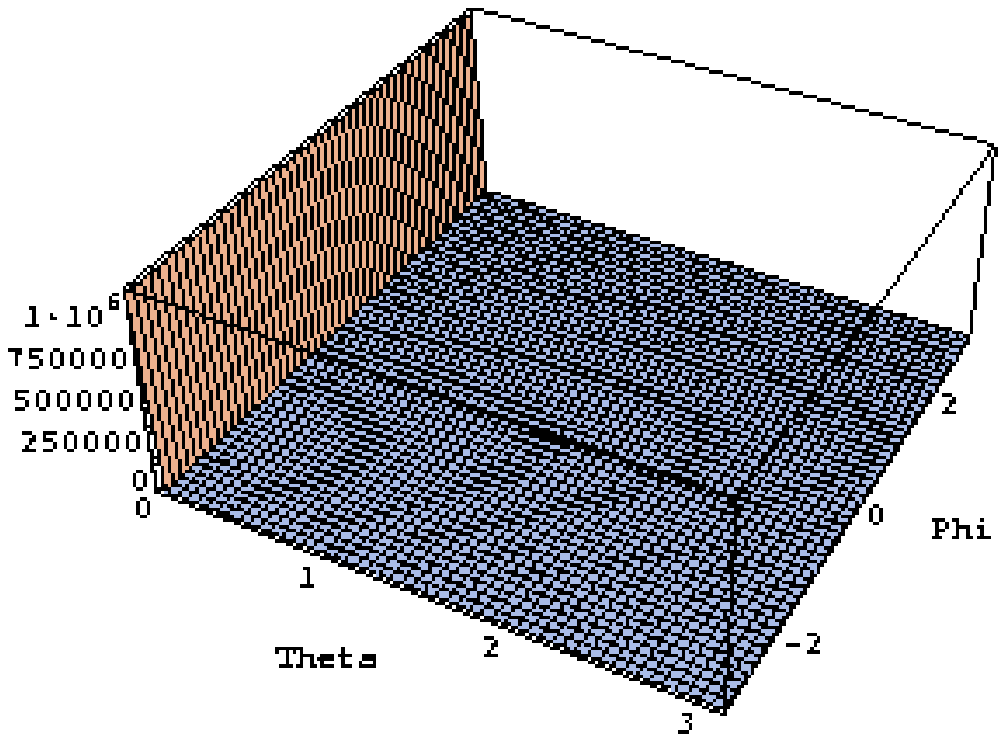}
\caption{$t=0.001, N=10, p=2, \chi =\pi$}
\end{figure} 

\begin{figure}
\includegraphics[width=100mm]{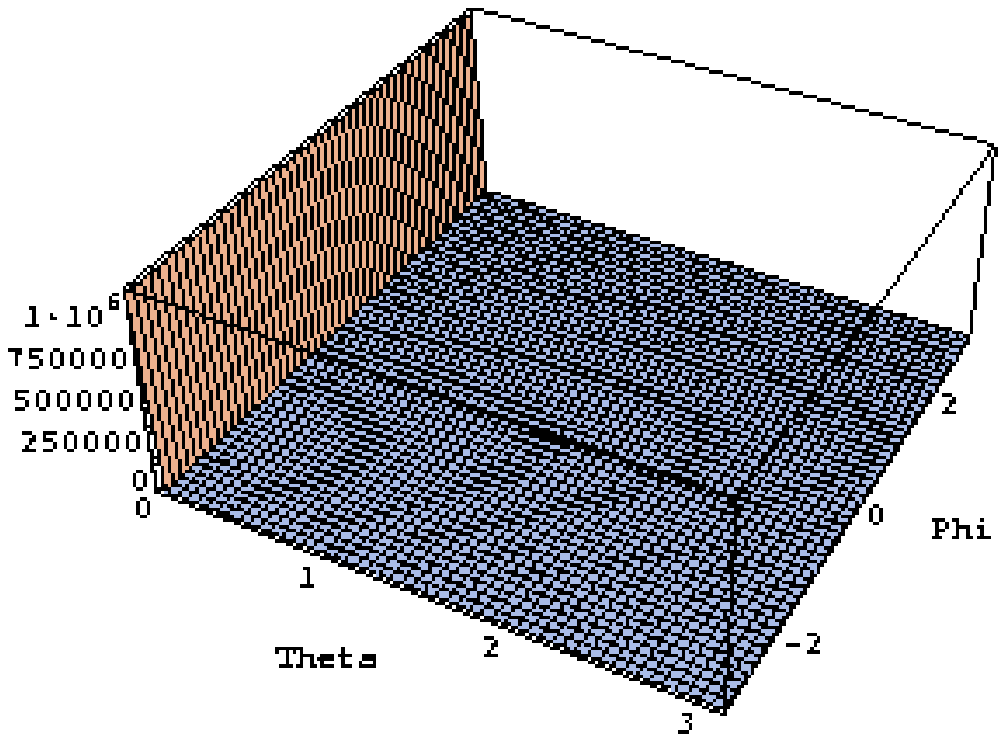}
\caption{$t=0.001, N=10, p=-2, \chi = 0$}
\end{figure} 

\begin{figure}
\includegraphics[width=100mm]{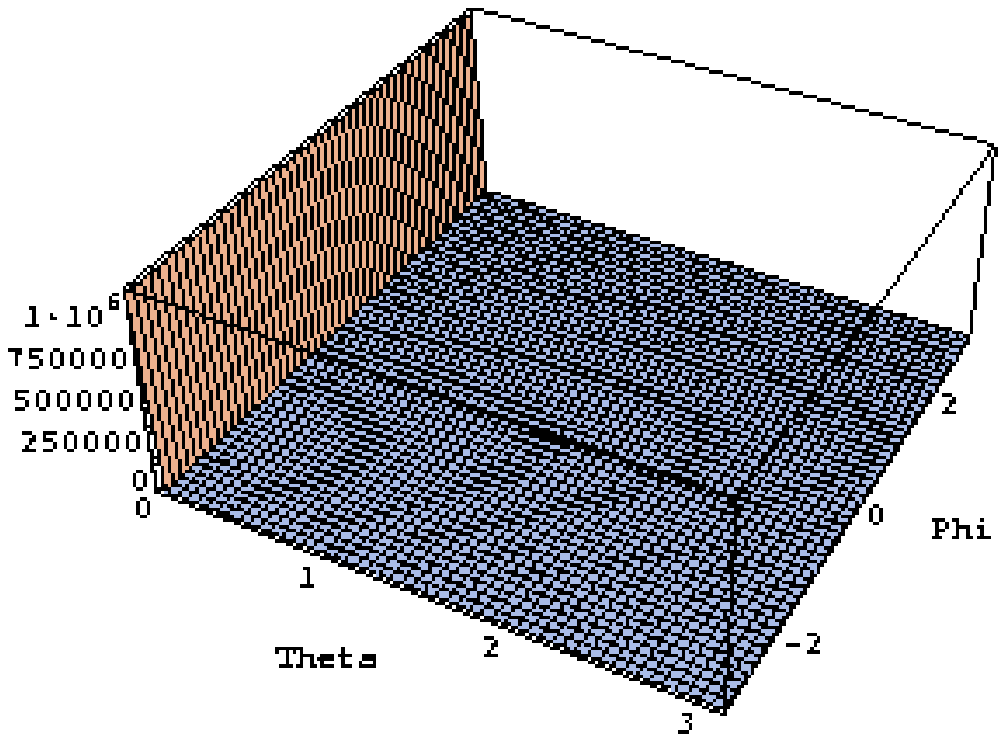}
\caption{$t=0.001, N=10, p=-2, \chi = \pi/4$}
\end{figure} 

\clearpage
\newpage

\begin{figure}
\includegraphics[width=100mm]{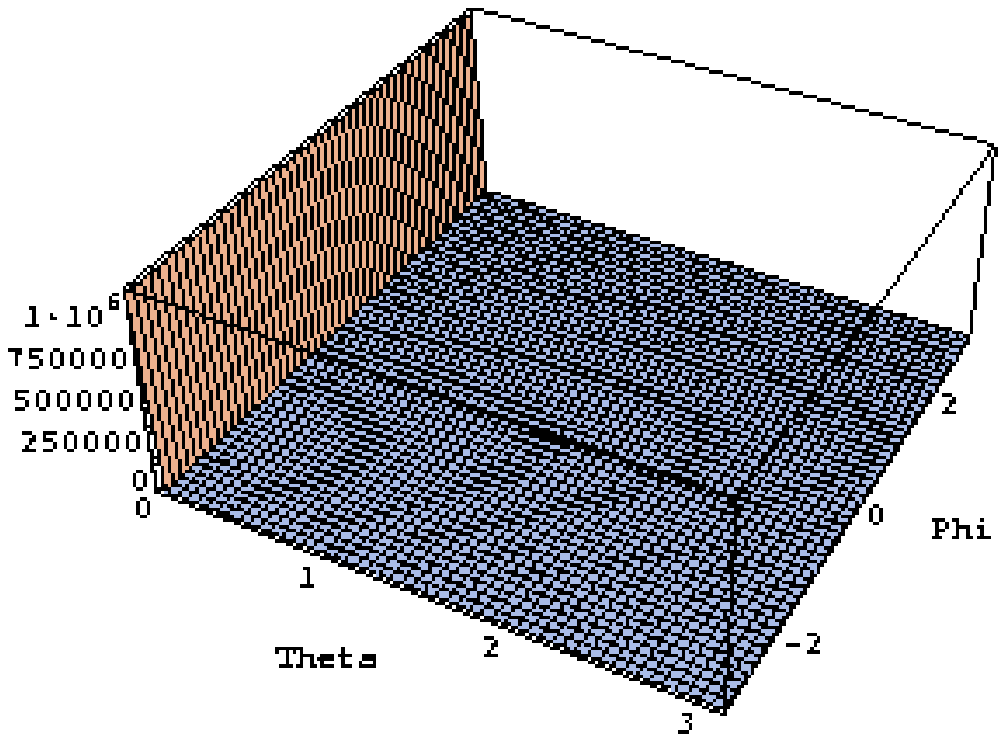}
\caption{$t=0.001, N=10, p=-2, \chi = \pi/2$}
\end{figure} 

\begin{figure}
\includegraphics[width=100mm]{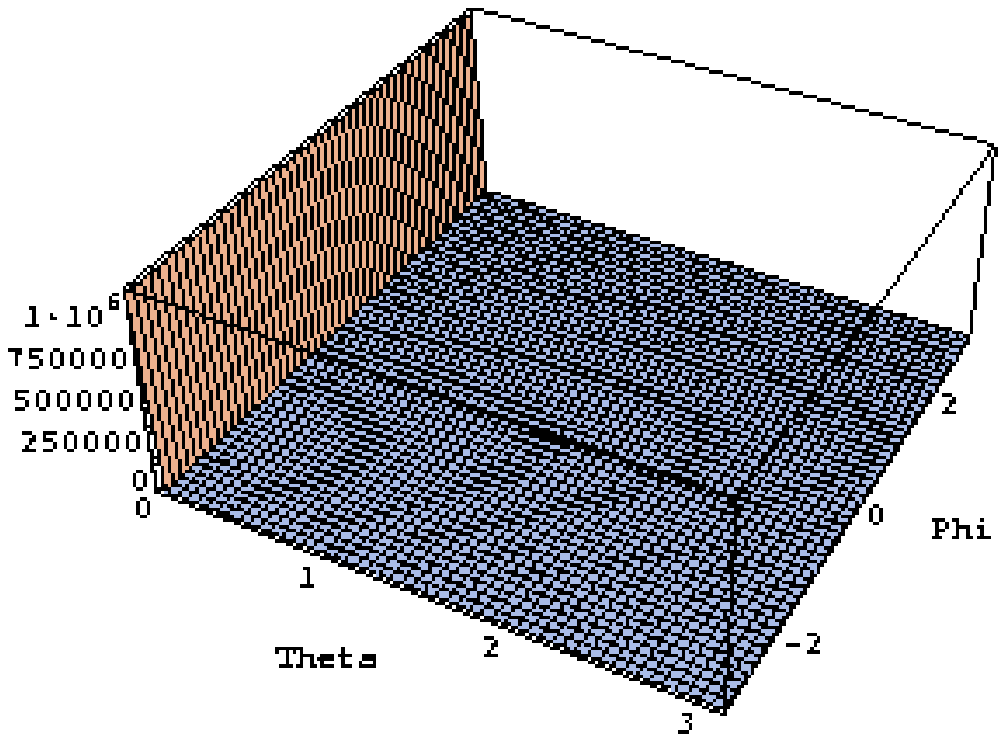}
\caption{$t=0.001, N=10, p=-2, \chi = 3\pi/4$}
\end{figure} 

\begin{figure}
\includegraphics[width=100mm]{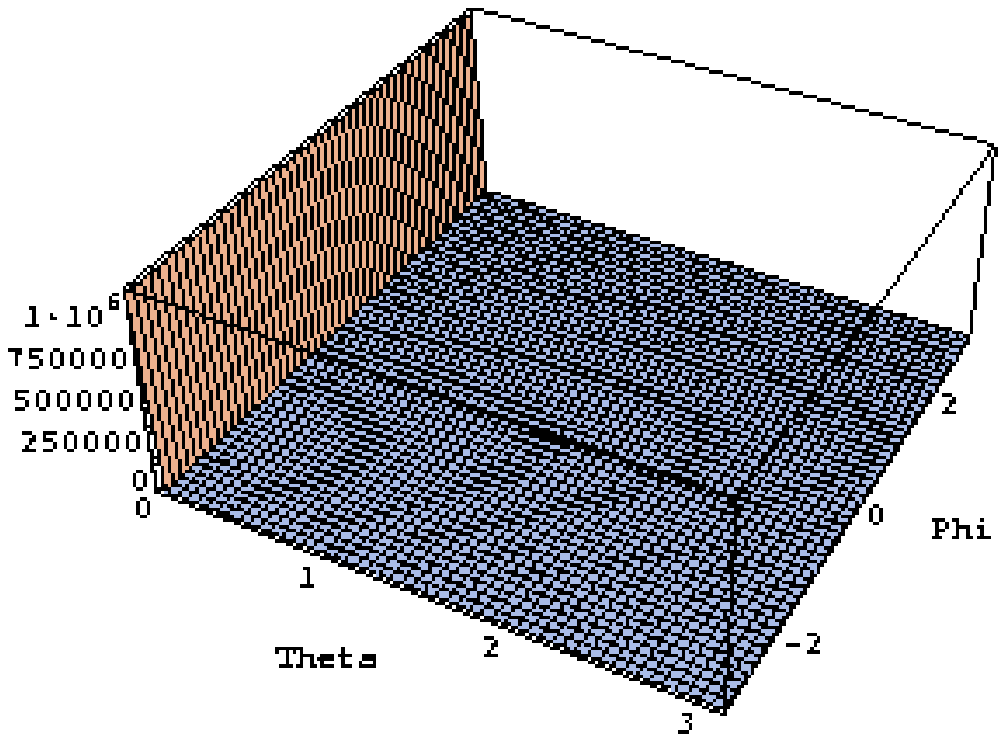}
\caption{$t=0.001, N=10, p=-2, \chi = \pi$}
\end{figure} 

\clearpage
\newpage

\subsubsection{Peakedness of the Overlap Function}
\label{sb.2.2}

We numerically compute the function $\frac{| \langle \psi^t_g ,
\psi^t_{g'} \rangle  |}{\| \psi^t_g \|\; \|
\psi^t_{g'} \| } $, with the parametrizations $g = e^{-i p^j \tau_j/2}
e^{ \theta^j \tau_j}$ $g = e^{-i (p')^j \tau_j/2}
e^{ (\theta ')^j \tau_j}$. Without restriction, one can choose
$\theta ' = 0$.
For the parameterization vector we take $\theta^j
= \theta_0 (\sin(\chi)\cos(\phi),\sin(\chi)\sin(\phi),\cos(\chi))$.
The variables for each graphic are $\theta_0 \in [0,\pi ] $ and $p
\in [p' -5 , p' +5]$. Graphics were calculated for all combinations
of the values $p' = \pm 3$, $\chi = 0,\pi /2,\pi$ and $\phi = 0, \pm
\pi/2, \pm \pi$. However, since again all of them look completely identical
we display only the plots with $\phi=0$ in figures 22 through 27 for 
parallel vectors and in figure 28 for orthogonal ones.

\underline{Case A: Parallel vectors}
We choose $p^j = (0,0,p)$ and $(p')^j = (0,0,p')$, that is, the
momentum vectors are parallel and peakedness is therefore to be
expected for $p=p'$, independent of the values for $\chi$ and $\phi$.

\begin{figure}
\includegraphics[width=100mm]{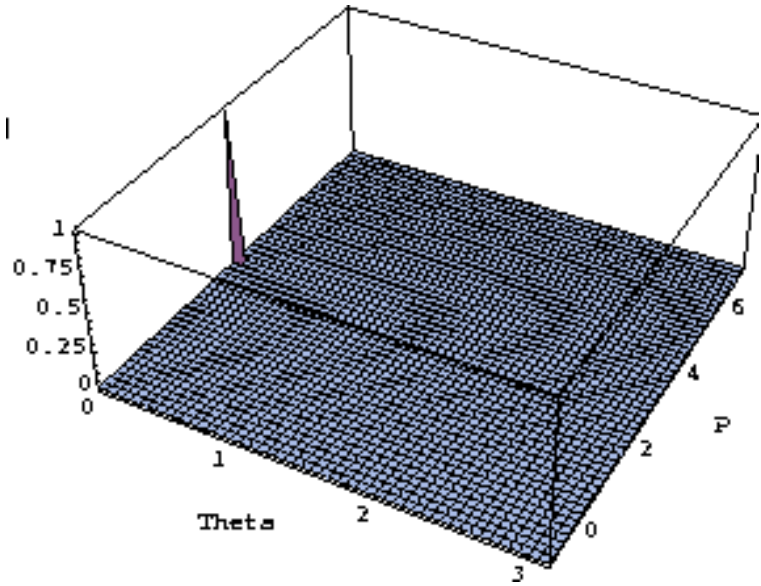}
\caption{$t=0.001, N=10, p'=3, \phi = 0, \chi=0$}
\end{figure} 

\begin{figure}
\includegraphics[width=100mm]{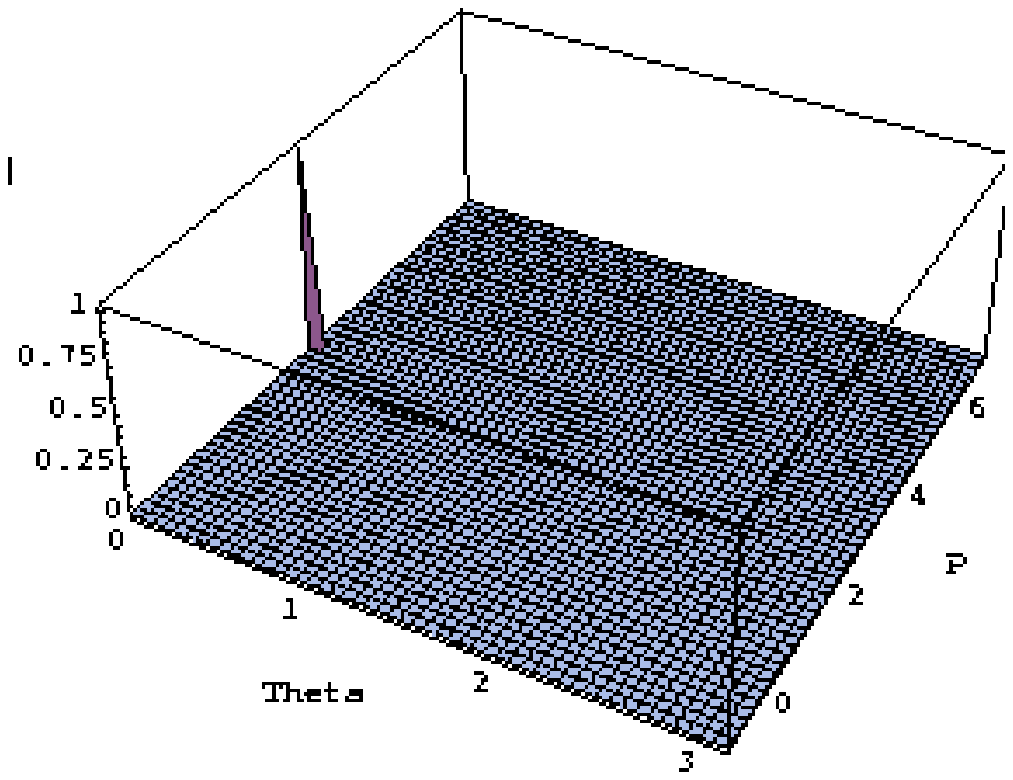}
\caption{$t=0.001, N=10, p'=3, \phi = 0, \chi= \pi/2$}
\end{figure} 

\clearpage
\newpage

\begin{figure}
\includegraphics[width=100mm]{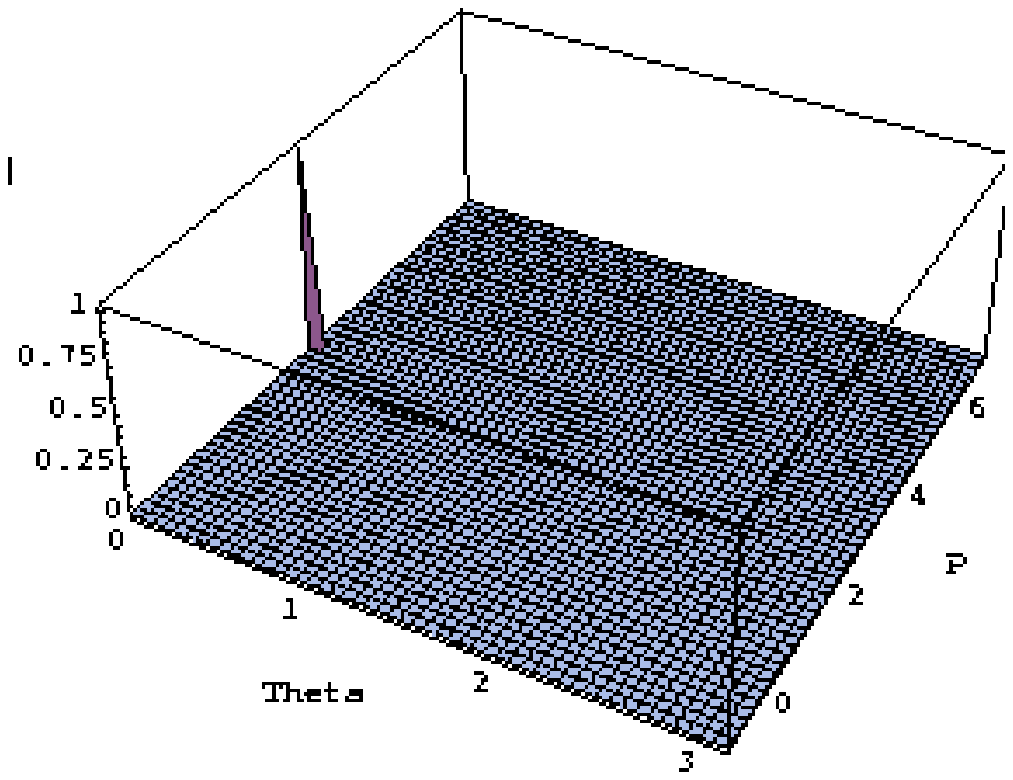}
\caption{$t=0.001, N=10, p'=3, \phi = 0, \chi= \pi$}
\end{figure} 

\begin{figure}
\includegraphics[width=100mm]{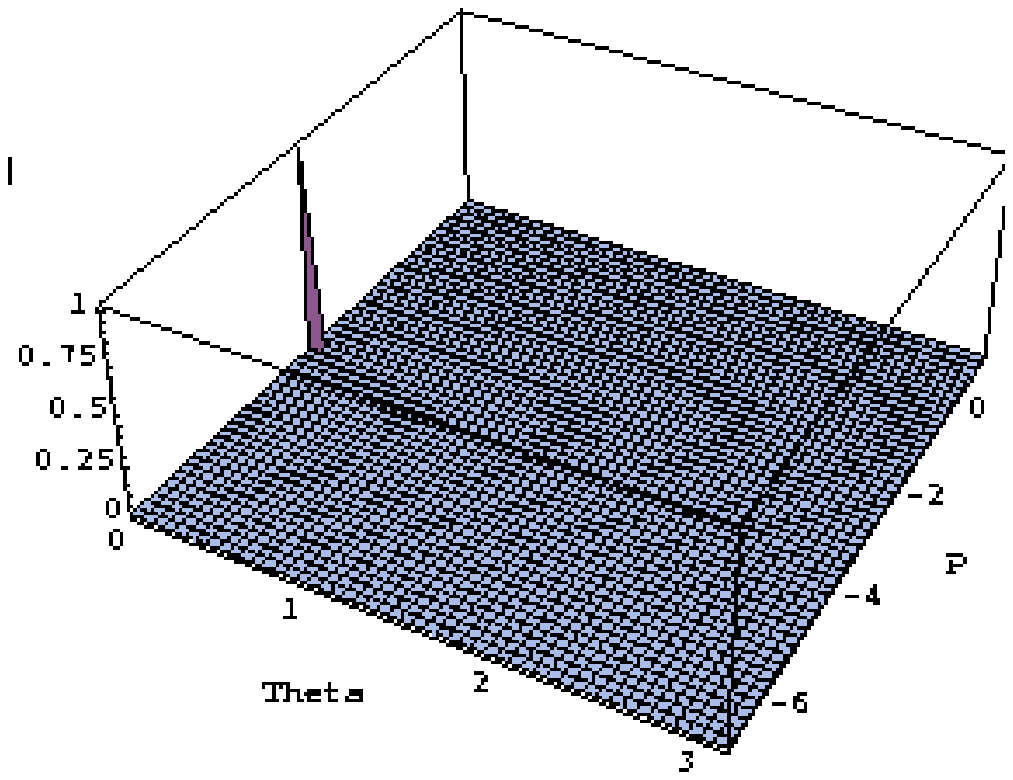}
\caption{$t=0.001, N=10, p'=-3, \phi = 0, \chi = 0$}
\end{figure} 

\begin{figure}
\includegraphics[width=100mm]{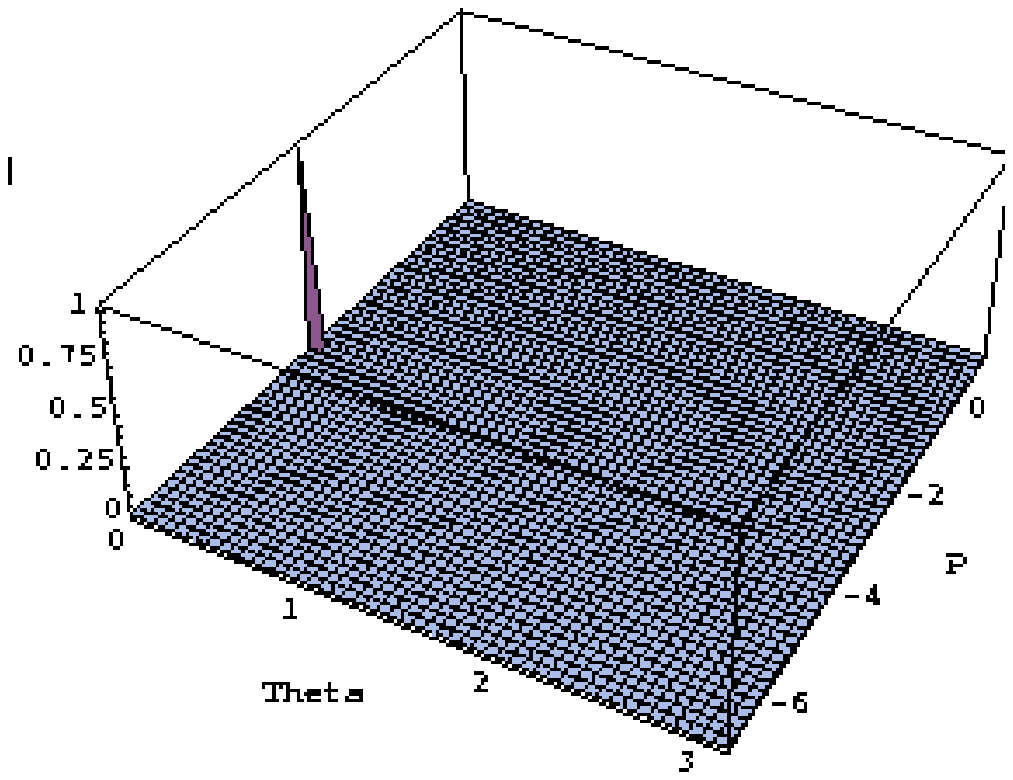}
\caption{$t=0.001, N=10, p'=-3, \phi = 0, \chi = \pi/2$}
\end{figure} 

\clearpage
\newpage

\begin{figure}
\includegraphics[width=100mm]{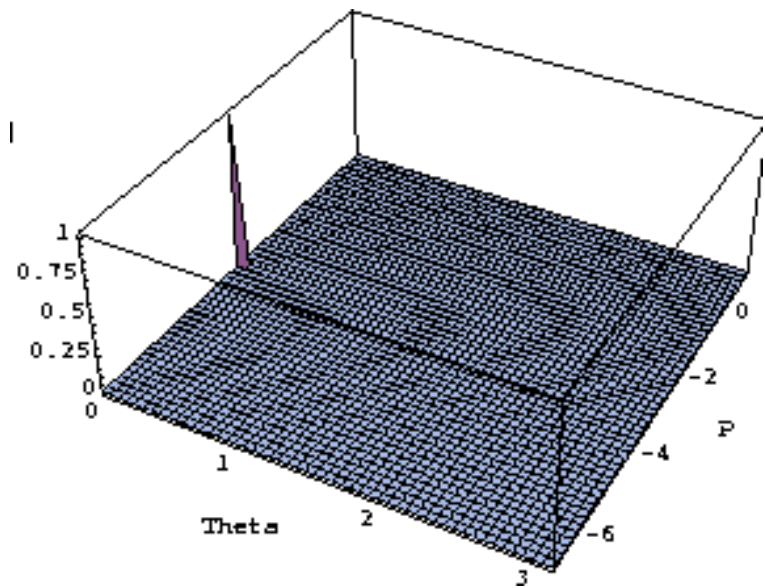}
\caption{$t=0.001, N=10, p'=-3, \phi = 0, \chi = \pi$}
\end{figure}

\underline{Case B: Orthogonal vectors}
We now choose $p^j = (0,0,p)$ and $(p')^j = (0,p',0)$, that is, the
vectors are orthogonal, and therefore the overlap function should
approximately vanish for all values of $\chi $ and $\phi $. We have
calculated all 30 figures as for case A but will display only one
here (figure 28), as they look all the same and expectedly rather boring.

\begin{figure}
\includegraphics[width=100mm]{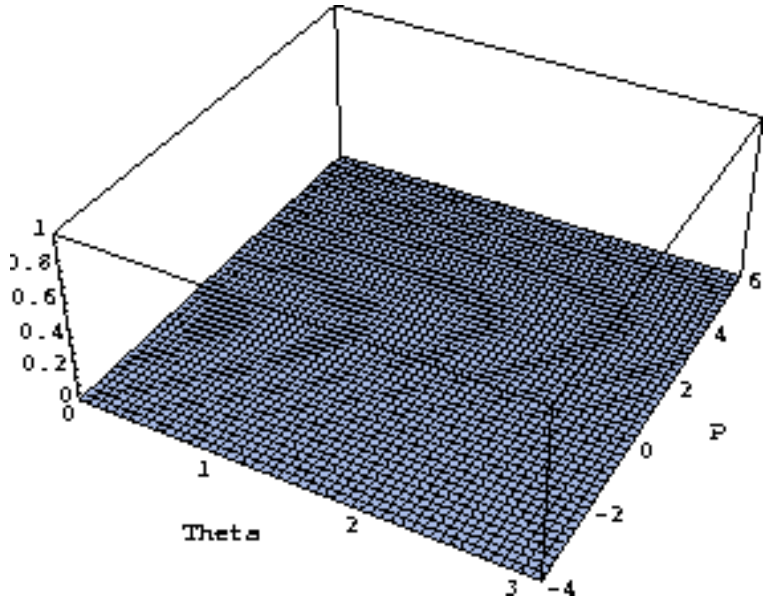}
\caption{$t=0.001, N=10, p'= 1, \phi = \pi/2, \chi = \pi$}
\end{figure} 

\clearpage
\newpage

Finally, figure 29 resolves the exact nature of the peak, exhibiting
its Gauss-like structure. Notice that the decay width of the peak
is indeed of the order of the expected value $\sqrt{t}\approx 0.03$.
This gives an idea of how drastically semi-classical these states will
be in applications to quantum gravity in $D=3$ where $t=10^{-64}$ 
for $a=1$cm !

\begin{figure}
\includegraphics[width=100mm]{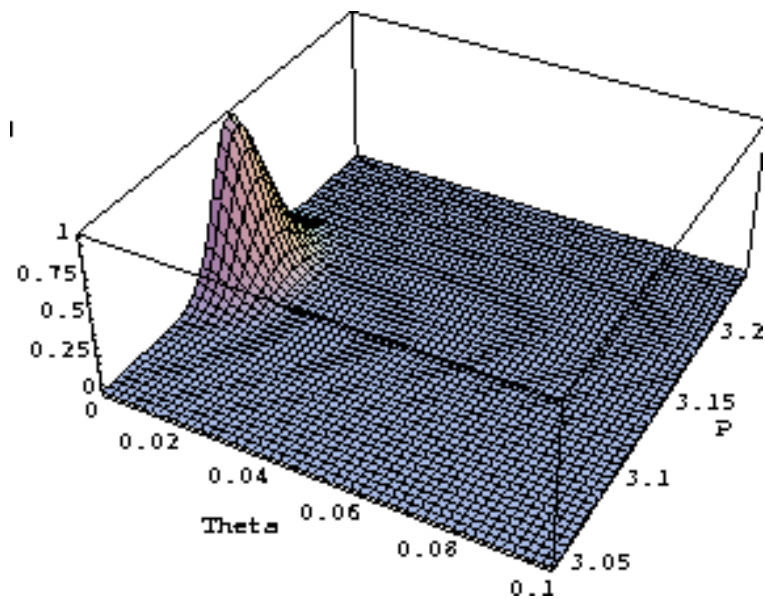}
\caption{$t=0.001, N=10, p' = 3.14, \phi=0, \chi=0$}
\end{figure} 

\end{appendix}

\end{document}